\documentclass[fleqn,usenatbib]{mnras}

\usepackage{newtxtext,newtxmath}

\usepackage[T1]{fontenc}

\DeclareRobustCommand{\VAN}[3]{#2}
\let\VANthebibliography\thebibliography
\def\thebibliography{\DeclareRobustCommand{\VAN}[3]{##3}\VANthebibliography}

\usepackage{graphicx}	
\usepackage{amsmath}	
\usepackage{bm}
\usepackage{subcaption}
\usepackage{adjustbox}
\usepackage{CJKutf8}

\newcommand*{\typewriter}{\fontfamily{lmtt}\selectfont{}}
\newcommand*{\sigreg}{\sigma_{\rm{reg}}}
\newcommand*{\taueq}{\tau_{\rm{eq}}}
\newcommand*{\tauin}{\tau_{\rm{in}}}
\newcommand*{\taudyn}{\tau_{\rm{dyn}}}
\newcommand*{\sigdyn}{\sigma_{\rm{dyn}}}
\newcommand*{\Msun}{\rm{M}_\odot}

\newcommand*{\Mstar}{\mathrm{M}_*}

\newcommand*{\tage}{t_{\rm{age}}}

\title[Stochastic SFH prior]{Stochastic prior for non-parametric star-formation histories}

\author[Wan et al.]{
Jenny T. Wan,$^{1,2,3,4}$\thanks{E-mail: jennywan@stanford.edu}
Sandro Tacchella,$^{1,2}$
Benjamin D. Johnson,$^{5}$
Kartheik G. Iyer,$^{6}$
\newauthor
Joshua S. Speagle (\begin{CJK*}{UTF8}{gbsn}沈佳士\ignorespacesafterend\end{CJK*}),$^{7,8,9,10}$
Roberto Maiolino$^{1,2}$
\\
$^{1}$Kavli Institute for Cosmology, University of Cambridge, Madingley Road, Cambridge, CB3 0HA, UK\\
$^{2}$Cavendish Laboratory, University of Cambridge, 19 JJ Thomson Avenue, Cambridge, CB3 0HE, UK\\
$^{3}$Department of Physics, Stanford University, 382 Via Pueblo Mall, Stanford, CA 94305, USA\\
$^{4}$Kavli Institute for Particle Astrophysics \& Cosmology, P.O. Box 2450, Stanford University, Stanford, CA 94305, USA\\
$^{5}$Center for Astrophysics $|$ Harvard \& Smithsonian, 60 Garden St., Cambridge, MA 02138, USA\\
$^{6}$Columbia Astrophysics Laboratory, Columbia University, 550 West 120th Street, New York, NY 10027, USA\\
$^{7}$Department of Statistical Sciences, University of Toronto, 9th Floor, Ontario Power Building, 700 University Avenue, Toronto, ON, M5G 1Z5, Canada\\
$^{8}$David A. Dunlap Department of Astronomy \& Astrophysics, University of Toronto, 50 St George Street, Toronto, ON, M5S 3H4, Canada\\
$^{9}$Dunlap Institute for Astronomy \& Astrophysics, University of Toronto, 50 St George Street, Toronto, ON, M5S 3H4, Canada\\
$^{10}$Data Sciences Institute, University of Toronto, 17th Floor, Ontario Power Building, 700 University Avenue, Toronto, ON, M5G 1Z5, Canada
}

\date{Accepted XXX. Received YYY; in original form ZZZ}

\pubyear{2023}

\begin{document}
\label{firstpage}
\pagerange{\pageref{firstpage}--\pageref{lastpage}}
\maketitle

\begin{abstract}
The amount of power contained in the variations in galaxy star-formation histories (SFHs) across a range of timescales encodes key information about the physical processes which modulate star formation. Modelling the SFHs of galaxies as stochastic processes allows the relative importance of different timescales to be quantified via the power spectral density (PSD). In this paper, we build upon the PSD framework and develop a physically-motivated, ``stochastic'' prior for non-parametric SFHs in the spectral energy distribution (SED)-modelling code {\typewriter Prospector}. We test this prior in two different regimes: 1) massive, $z = 0.7$ galaxies with both photometry and spectra, analogous to those observed with the LEGA-C survey, and 2) $z = 8$ galaxies with photometry only, analogous to those observed with NIRCam on JWST. We find that it is able to recover key galaxy parameters (e.g. stellar mass, stellar metallicity) to the same level of fidelity as the commonly-used continuity prior. Furthermore, the realistic variability information incorporated by the stochastic SFH model allows it to fit the SFHs of galaxies more accurately and precisely than traditional non-parametric models. In fact, the stochastic prior is $\gtrsim 2\times$ more accurate than the continuity prior in measuring the recent star-formation rates (log SFR$_{100}$ and log SFR$_{10}$) of both the $z = 0.7$ and $z = 8$ mock systems. While the PSD parameters of individual galaxies are difficult to constrain, the stochastic prior implementation presented in this work allows for the development hierarchical models in the future, i.e. simultaneous SED-modelling of an ensemble of galaxies to measure their underlying PSD.
\end{abstract}




\section{Introduction}

Galaxies are a remarkably diverse class of objects, in every sense of the word. From massive, red ellipticals retired from their heyday of star-forming vigor, to blue, star-studded spirals with entourages of tightly wound arms, to irregular galaxies with sweeping tidal tails and clusters of newborn stars, galaxies span the impossibly large parameter space set by the laws of physics. Explaining this observed diversity in the galaxy population through a robust physical framework remains one of the biggest challenges in galaxy evolution studies today.

Currently, the main limiting factor in our understanding of how galaxies form and evolve is the uncertainty regarding the physics of star formation and feedback \citep{NaabOstriker2017}. The star formation activity within galaxies is regulated by physical processes that operate over many orders of magnitude in both temporal and spatial scales, yet it is only possible to observe their \textit{cumulative} effects on the galaxy population at any given epoch. How, then, can we deduce any information about the multitude of pathways through which a galaxy can build its stars?

Despite the immense diversity between individual systems, scaling relations do exist across the galaxy population, one of the most important being the star-forming main sequence (SFMS). The SFMS describes the correlation that exists between the stellar masses ($\Mstar$) and the star formation rates (SFR) of star-forming galaxies from present day up to at least redshift $\sim 6$ \citep{Brinchmann2004, Daddi2007, Elbaz2007, Whitaker2012, Speagle2014, Schreiber2015, Boogaard2018, Leja2022}. The SFMS is observed to have a scatter of $\sim0.2-0.4$ dex, implying that galaxies along the main sequence ridgeline self-regulate their star formation through the interplay between gas inflow, outflow, and consumption processes \citep{Bouche2010, Lilly2013, Rodriguez2016, Tacchella2016, Matthee2019}. 

These star formation-driving processes act over different timescales in a galaxy's lifetime, and therefore leave distinct temporal fingerprints on the star-formation histories (SFHs) of galaxies \citep[e.g.][]{Iyer2020, TFC2020}. The fluctuations in the SFRs of galaxies are tied to mechanisms ranging from the local creation and destruction of giant molecular clouds (GMCs) on short timescales \citep[$\lesssim$ 100 Myr;][]{Scalo1984, Krumholz2015, Orr2019}, to galaxy-galaxy mergers and galactic winds from stellar and AGN feedback on intermediate timescales \citep[$\sim 0.1 - 1$ Gyr;][]{GunnGott1972, Hernquist1989, WangLilly2020, Matthee2019, Shin2023}, to dark matter halo accretion rates and environmental large-scale structure on the longest timescales \citep[$\gtrsim$ 1 Gyr;][]{Rodriguez2016, Tacchella2018, Behroozi2019}. Assuming that a given physical process acts only over a specific timescale, analyzing the variability in galaxy SFHs across a wide range of timescales will allow us to distinguish between these processes and constrain their relative importance to galaxy growth.

Various signatures in a galaxy's observed spectral energy distribution (SED) act as measures of its SFR averaged over different timescales. For example, the nebular H$\alpha$ line resulting from the ionizing radiation of the most massive (O-type) stars provides a near-instantaneous tracer of star-formation activity, and the combined rest-frame UV+IR spectral energy distribution of a galaxy (contributed to by O and B-type stars) provides a reliable estimate of the star-formation rate up to $\sim 100$ Myr \citep[e.g.][]{Flores2021, Tacchella2022}. Additionally, Balmer absorption features track star formation over intermediate timescales \citep{Worthey1997, WangLilly2020a}, and the strength of the $4000 \Angstrom$ break, caused by the accumulation of metal absorption lines from older stellar populations, provides an estimate of the total time over which a galaxy has been forming stars \citep{Barbaro1997, Kauffmann2003}. The combination of these spectral features can be employed to approximate the variability, or ``burstiness'', of a galaxy's SFH on timescales ranging from $\sim 5$ Myr to many Gyr.

For a single galaxy, these features can only be observed at one point in time, so the challenge lies in deducing all the ups and downs in its SFH from a single spectral snapshot. To address this complicated problem, \cite{CT2019} introduces a framework for modelling the SFHs of galaxies as a stochastic time series. This stochastic process is defined through a power spectral density (PSD) with a broken power law functional form, providing an effective way to quantify the amount of power contained in the SFR fluctuations (i.e. burstiness) in galaxy SFHs on a given timescale. \cite{TFC2020} links this PSD framework to the physical processes driving variations in the SFR on different timescales by connecting the gas regulator model of \citet{Lilly2013} to the gas inflow rate and the life cycle of GMCs, creating the Extended Regulator model. Furthermore, for a stochastic process, the frequency-domain PSD can be straightforwardly transformed into the time-domain auto-covariance function (ACF) via the Weiner-Khinchin theorem. \cite{Iyer2022} leverages this relation between the PSD and ACF to incorporate the Extended Regulator model into a Gaussian process \citep[GP; see, e.g.][]{Aigrain2022} framework and explores how changes in model stochasticity affects spectral signatures across various galaxy populations.

However, it remains difficult to break the degeneracy between individual star-formation rate indicators and confounding variables such as the assumed initial mass function (IMF), variations in dust attenuation, and metallicity effects \citep[e.g.][]{Shivaei2018, Tacchella2022Halo7D}. SED-fitting methods are especially important for this reason -- they estimate the SFHs of galaxies using the full range of photometric and/or spectroscopic information available, while exploring these degeneracies and robustly estimating uncertanties. \citet{Iyer2022}'s {\typewriter GP-SFH} \footnote{\url{https://github.com/kartheikiyer/GP-SFH}} framework provides an avenue to integrate the physical covariance structure of the Extended Regulator model into SED modelling tools.

In this paper, we adapt the Extended Regulator model developed in \cite{CT2019} and \cite{TFC2020}, in conjunction with the GP implementation of \citet{Iyer2022}, into a physically-motivated prior for non-parametric SFHs in the SED-modelling code {\typewriter Prospector} \footnote{\url{https://github.com/bd-j/prospector}} \citep{Leja2017, Johnson2021, JohnsonLeja2017}. We test our stochastic SFH prior on mock galaxy observations in two different regimes -- massive galaxies at intermediate ($z = 0.7$) redshift with both photometric and spectroscopic data, and high-redshift ($z = 8$) galaxies with only photometric data. We demonstrate that the the stochastic model is able to accurately measure the basic galaxy parameters. Furthermore, it is able to recover galaxy SFHs with a higher level of fidelity than the more-commonly used continuity prior because of the physical variability information incorporated into the model.

In Section \ref{sec:modelling sfh variability}, we summarize the Extended Regulator PSD formalism and explain the construction of the stochastic SFH prior. Section \ref{sec:prior flexibility} demonstrates how the PSD formalism translates into priors on physical quantities and illustrates its flexibility in modelling galaxy SFHs. We generate mock observations of galaxies comparable to those from the Large Early Galaxy Astrophysics Census \citep[LEGA-C;][]{LEGAC2021} in Section \ref{sec:legac recovery}, and show that the stochastic SFH prior is able to recover their key properties well. In Section \ref{sec:highz recovery}, we test the performance of the stochastic prior on a sample of mock high-redshift galaxies, designed for comparison to systems observed with NIRCam on JWST. We discuss our results in Section \ref{sec:discussion} and conclude in Section \ref{sec:conclusion}. Throughout this work, we assume a flat $\Lambda$CDM cosmology with $\Omega_m = 0.3$, $\Omega_\Lambda = 0.7$, and $H_0 = 70$ km s$^{-1}$ Mpc$^{-1}$.

\section{Modelling star formation history variability}
\label{sec:modelling sfh variability}

The variability in the SFHs of galaxies across a range of timescales holds a wealth of information about the physical processes which regulate star formation. The power spectral density, or PSD, of galaxies' SFHs has proven to be an extremely effective way to model the stochastic behavior of galaxy star formation. \cite{CT2019}, \cite{TFC2020}, \cite{Iyer2020}, and \cite{Iyer2022} describe in detail the theoretical development of the PSD framework and lay the groundwork for its implementation into SED-modelling tools. For completeness, a brief summary is given in Section \ref{sec:psd}. We describe how this framework is adapted into a prior for binned SFHs in Section \ref{sec:stoch prior}.

\subsection{The PSD formalism \& Extended Regulator model}
\label{sec:psd}

Following \citet{CT2019}, we model the time-dependence of star formation and the movement of star-forming galaxies along the SFMS ridgeline as a purely (stationary) stochastic processes. \textcolor{black}{Building upon this premise, \citet{TFC2020} numerically modelled the gas cycle in galaxies to characterize how the stochastic processes that drive
\begin{enumerate}
    \item gas inflow into galaxies,
    \item cycling between atomic and molecular gas in equilibrium, and
    \item[(iii)] the formation and disruption of giant molecular clouds (GMCs),
\end{enumerate}
affect the variability in $\ln \mathrm{SFR}$.} These processes act over distinct timescale ranges and can be characterized in the frequency domain using a PSD, which quantifies the amount of power contained in fluctuations at any given frequency for a stochastic process. In other words, the PSD is a way of estimating the relative importance of different frequency modes (or timescales) for a given process.

Natural processes tend to de-correlate after some amount of time to behave like white noise. This property can be captured by a PSD in the form of a broken power-law:
\begin{equation}
    \label{eq:broken power law}
    \mathrm{PSD}(f) = \frac{s^2}{1 + (2\pi\tau_{\mathrm{dec}})^2 f^2},
\end{equation}
where \textcolor{black}{$f$ is the frequency, $s^2$ is the absolute normalization (scatter squared) at $f = 0$}, and $\tau_{\mathrm{dec}}$ is the de-correlation timescale of the process (or where the PSD ``breaks''). \footnote{A white noise process has equal power on all frequencies, i.e. has a constant PSD. We see in Eq. \ref{eq:broken power law} that on timescales much larger than $\tau_{\mathrm{dec}}$ ($f \ll 1/\tau_{\mathrm{dec}}$), PSD $\rightarrow s^2$, satisfying this requirement.}

\citet{TFC2020} (see also \citealt{WangLilly2020}) demonstrate that this form of the PSD appears if we consider star formation through the lens of the gas regulator model described in \citet{Lilly2013}. In this model, the SFR of a galaxy is regulated by the gas mass present its gas reservoir. However, it has been observationally established that star formation within a galaxy is sustained by a population of GMCs rather than the gas reservoir as a whole \citep[e.g.][]{Myers1986, ScovilleGood1989}. Thus, the creation and disruption of GMCs, along with local dynamical processes such as spiral arm instabilities and bars \citep[e.g.][]{Krumholz2015, Yu2021, Shin2023}, also play an integral role in modulating a galaxy's SFR.

Given that the stochastic processes regulating star formation within galaxies (gas inflow/outflow, GMC formation and disruption, etc.) are well-described by broken power-law PSDs \citep{TFC2020, Iyer2020}, and assuming 
\begin{enumerate}
    \item gas inflow into the galaxy is coupled to the equilibrium gas cycling and
    \item the short-term dynamical processes are independent of the large-scale behavior of the gas reservoir,
\end{enumerate}
we arrive at the Extended Regulator model of \citet{TFC2020}. \textcolor{black}{Specifically, in this model, the inflow rate directly relates to the gas mass and the gas cycling, while the dynamical term (i.e. GMC formation/disruption) is an independent process. Therefore, the stochasticity of the inflow and gas cycling are related (i.e. multiplicative), while the dynamical term is an additional and independent source of stochasticity. (For details, see \citealt{TFC2020} and Section 2.2 of \citealt{Iyer2022}.)} Thus, the full PSD of the Extended Regulator model can be written as
\begin{equation}
    \begin{aligned}
        \mathrm{PSD}_{\mathrm{ExReg}}(f) = 
        & ~s_{\mathrm{reg}}^2 ~\Big{(} \frac{1}{1 + (2\pi \tau_{\rm{in}})^2 f^2} \times \frac{1}{1 + (2\pi \tau_{\rm{eq}})^2 f^2} \Big{)} \\
        & + ~\frac{s^2_{\rm{dyn}}}{1 + (2\pi \tau_{\rm{dyn}})^2 f^2},
    \end{aligned}
\end{equation}
where $\tauin$, $\taueq$, and $\taudyn$ are timescales associated with gas inflow, equilibrium gas cycling, and GMC formation/disruption, respectively. Additionally, $s_{\mathrm{reg}}^2$ is the overall variability in the gas component and $s_{\mathrm{dyn}}^2$ is the variability in the dynamical component. \textcolor{black}{It's important to emphasize that while this model does not explicitly incorporate all the possible physical processes that induce variability in a galaxy's SFR, the timescales $\tauin$, $\taueq$, and $\taudyn$ are ``effective'' timescales that average over a wide range of processes. $\tauin$ accounts for changes in the gas reservoir due to pristine inflows, outflows, mergers, and the recapture of previously-ejected gas; $\taueq$ encompasses the transformations that gas within a galaxy undergoes due to stellar and AGN feedback; and $\taudyn$ is the timescale associated with the dynamical processes that modulate the life cycle of GMCs (e.g. stellar feedback, gas compression by spiral arms, and compaction-induced starbursts).} 

To explore the correlation structure of the stochastic star-formation process, we can define the associated auto-covariance function, or ACF. The ACF is the Fourier pair of the PSD, and can be computed straightforwardly from the PSD using the Wiener-Kinchin theorem. The ACF of the Extended Regulator model is
\begin{equation}
\label{eq:AFC_exreg}
    \begin{aligned}
        \mathrm{ACF}_{\mathrm{ExReg}}(\tau) = 
        & ~\sigreg^2 \times \frac{\tau_{\rm{in}} e^{-|\tau|/\tau_{\rm{in}}} - \tau_{\rm{eq}} e^{-|\tau|/\tau_{\rm{eq}}}}{\tau_{\rm{in}} - \tau_{\rm{eq}}} \\
        & + ~\sigma_{\rm{dyn}}^2 \times e^{-|\tau| / \tau_{\rm{dyn}}},
    \end{aligned}
\end{equation}
where $\tau$ is the timescale associated with the frequency $f$. Simply put, this ACF contains information about how correlated the SFR of a galaxy is over some timescale $\tau$. We note that due to the nature of the Fourier transform, $\sigreg \neq s_{\mathrm{reg}}$ and $\sigdyn \neq s_{\mathrm{dyn}}$; however, both $s$ and $\sigma$ contain the same information about the intrinsic stochasticity of the processes regulating star-formation in our model. \footnote{$s_{\rm{reg}}^2 = 2 \sigreg^2 (\tauin + \taueq)$ and $s_{\rm{dyn}}^2 = 2 \sigdyn^2 \taudyn$.} For a more in-depth treatment of the connection between the PSD and ACF in the Extended Regulator model, see Appendix A of \citet{Iyer2022}.

\subsection{Prior for non-parametric SFHs}
\label{sec:stoch prior}

The Extended Regulator model correlation structure can be adapted for use as a prior for non-parametric, or binned, SFHs in SED-fitting codes, where ``non-parametric'' just means that the shape of the SFH is not pre-determined. Rather, the SFH is split into discrete time bins, and the SFR within each bin is fit, such that any reasonable function in SFH space can be approximated. In this paper, we focus on non-parametric SFHs within the SED-modelling framework of {\typewriter Prospector} \citep{Leja2017, JohnsonLeja2017, Johnson2021}. 

\subsubsection{Continuity prior}

\textcolor{black}{A common method} for determining the SFRs associated with non-parametric SFHs in {\typewriter Prospector} uses the continuity prior \citep{Leja2019}. The continuity prior fits for the log of the ratio between SFRs in adjacent time bins, $\Delta \log \mathrm{SFR} = \log (\mathrm{SFR}_n / \mathrm{SFR}_{n+1})$. A Student's $t$-distribution prior is placed on $\Delta \log \mathrm{SFR}$, with $\nu = 2$ degrees of freedom and a scale of $\sigma = 0.3$ \citep[based on comparisons to the Illustris simulations;][]{Leja2019}.

The continuity prior enforces a constant expectation value for $\mathrm{SFR}(t)$. Furthermore, it weights against sharp transitions in the SFR between bins, meaning that bursty SFHs are disfavored. However, the burstiness level of the continuity prior can be adjusted by changing the values chosen for $\sigma$ \citep{Tacchella2022-bursty-sfh-prior}. For example, $\sigma = 1.0$ will result in a broader SFH distribution which allows for SFHs with stronger variations. Additionally, because the Student's $t$-distribution has heavier tails than the normal distribution, sudden significant changes (e.g. a starburst or rapid quenching event) are not completely excluded from the solution space. Thus, it remains flexible enough to describe both maximally old and starburst galaxies.

\subsubsection{Stochastic prior}

While the continuity prior is reasonably successful at recovering the SFHs of galaxies, \textcolor{black}{and its width can be somewhat arbitrarily tuned to allow for more or less ``bursty'' star formation behavior, it does not directly extract any information about the amount of variability on different timescales.} Here, we construct a new prior for non-parametric SFHs which incorporates the physically-motivated correlations between SFRs quantified in $\mathrm{ACF}_{\mathrm{ExReg}} (\tau)$ and allows for the straightforward constraint of parameters related to the strength of SFH variability across various timescales. 

Following \citet{Iyer2022}, we assume that a galaxy's set of star formation rates $\bm{\mathrm{x}}_N \equiv \{ \log \mathrm{SFR}_{t_n} \}_{n=1,...,N}$ \footnote{Note that while the Extended Regulator model describes the variability in the natural log of the star formation rate, $\ln \mathrm{SFR}(t)$, the Gaussian Process implementation of \citet{Iyer2022} ({\typewriter GP-SFH}) converts $\ln \mathrm{SFR}(t)$ into the base-10 logarithm $\log \mathrm{SFR}(t)$, which facilitates comparisons with other quantities in the literature, as well as straightforward usage in {\typewriter Prospector}.} in the time bins $t_n = t_1,...,t_N$ follows a multivariate normal distribution
\begin{equation}
    \bm{\mathrm{x}}_N \sim \mathcal{N} \big{(} \bm{\mu}_N, \bm{\mathrm{C}}(\bm{\tau}) \big{)},
\end{equation}
where $\bm{\mu}_N$ is the $N$-vector of mean $\log \mathrm{SFR}$ values and $\bm{\mathrm{C}}(\bm{\tau})$ is the corresponding $N \times N$ covariance matrix. To incorporate the Extended Regulator model covariance values, we simply take $\bm{\mathrm{C}}(\bm{\tau})$ to be $\mathrm{ACF}_{\rm{ExReg}}(\bm{\mathrm{\tau}})$. Specifically, $\bm{\tau} \equiv \{ t - t' \}$ for $t, t' = t_1,...,t_N$, where $t_1,...,t_N$ are the centers of the respective $N$ time bins in our non-parametric SFH. 

However, to be directly comparable to the continuity prior, we want to constrain the log of the ratios between SFRs in adjacent time bins rather than the individual SFRs themselves. \textcolor{black}{The set of log SFR ratios of a galaxy can be written as
\begin{equation}
    \begin{aligned}
        \bm{\mathrm{y}}_{A/B} &\equiv \big{\{} \log (\mathrm{SFR}_{t_n}/\mathrm{SFR}_{t_{n+1}}) \big{\}}_{n = 1,...,N-1} \\
        &= \big{\{} \log \mathrm{SFR}_{t_n} -  \log \mathrm{SFR}_{t_{n+1}} \big{\}}_{n = 1,...,N-1} \\
        &=\bm{\mathrm{x}}_A - \bm{\mathrm{x}}_{B},
    \end{aligned}
\end{equation}
where $\bm{\mathrm{x}}_A$ is the set of log SFRs of a galaxy in the time bins $t_1$ to $t_{N-1}$, and $\bm{\mathrm{x}}_B$ is the set of log SFRs from $t_2$ to $t_N$.} As $\bm{\mathrm{x}}_A$ and $\bm{\mathrm{x}}_B$ are both multivariate normally-distributed variables, their difference also follows a multivariate normal distribution. Therefore, we can describe also the $\log \mathrm{SFR}$ ratios with an multivariate normal distribution and use the property that
\begin{equation}
    \begin{aligned}
        \mathrm{Cov}(x_i - x_{i+1}, x_j - x_{j+1}) = & ~ \mathrm{Cov}(x_i, x_j) -  \mathrm{Cov}(x_i, x_{j+1}) \\
        &-  \mathrm{Cov}(x_{i+1}, x_j) +  \mathrm{Cov}(x_{i+1}, x_{j+1})
    \end{aligned}
\end{equation}
to transform the $N \times N$ covariance matrix $\mathrm{ACF}_{\rm{ExReg}} (\bm{\tau})$ of \{log SFR\} into the $(N-1) \times (N-1)$ covariance matrix $\mathrm{ACF}_{\rm{SFH}} (\bm{\tau})$ of \{log SFR ratios\}. Thus, the set of $\log \mathrm{SFR}$ ratios $\bm{\mathrm{y}}_{A/B}$ for a galaxy follows the distribution:
\begin{equation}
    \bm{\mathrm{y}}_{A/B} \sim \mathcal{N} \big{(} \bm{\mu}_{A/B}, \mathrm{ACF}_{\rm{SFH}}(\bm{\tau}) \big{)},
\end{equation}
where $\bm{\mu}_{A/B}$ is the mean vector of log SFR ratios, i.e. the assumed base SFH.

The stochastic prior maintains many of the useful properties of the continuity prior, such as the expectation of a flat baseline $\mathrm{SFR}(t)$. The key difference is that the stochastic prior correlates the SFR between SFH bins based on a physical model for the stochastic behavior of the processes working within galaxies: gas inflow, the cycling of gas within the reservoir, and the star-formation process (i.e. life cycle of GMCs).

\section{Flexibility of the stochastic SFH prior}
\label{sec:prior flexibility}

In this section, we aim to build physical intuition for the stochastic prior and demonstrate its flexibility in modelling galaxy SFHs. \textcolor{black}{In Section \ref{sec:fixed psd params}, we examine} the effect that the ``PSD parameters'' -- $\sigreg$, $\taueq$, $\tauin$, $\sigdyn$, and $\taudyn$ -- have on the resulting specific star-formation rate (sSFR) and mass-weighted age ($t_{\rm{age}}$) prior distributions for a few fixed set of PSD parameters. In each case, we take 10,000 draws from the prior and calculate sSFR for each prior call. We display the sSFR prior distribution rather than the SFR prior distribution in order to remove the effect of the broad, flat stellar mass prior. The specific star-formation rate is calculated as sSFR = SFR/M$_*$, and we show the SFR averaged over two different timescales, 100 Myr and 5 Myr. 
We assume a SFH with 10 bins specified in look-back time, with the first two bins fixed at ($1 - 5$) Myr and ($5 - 10$) Myr, and the remaining bins are equally spaced in logarithmic time between $10$ Myr and $0.95 t_H$ Gyr. In Section \ref{sec:free psd params}, we then consider the full distribution of PSD parameters and their associated priors, focusing on five specific evolutionary regimes.

\subsection{Fixed PSD parameters}
\label{sec:fixed psd params}

\begin{figure*}
    \centering
    \begin{subfigure}[b]{0.48\textwidth}
        \centering
        \includegraphics[width=\textwidth]{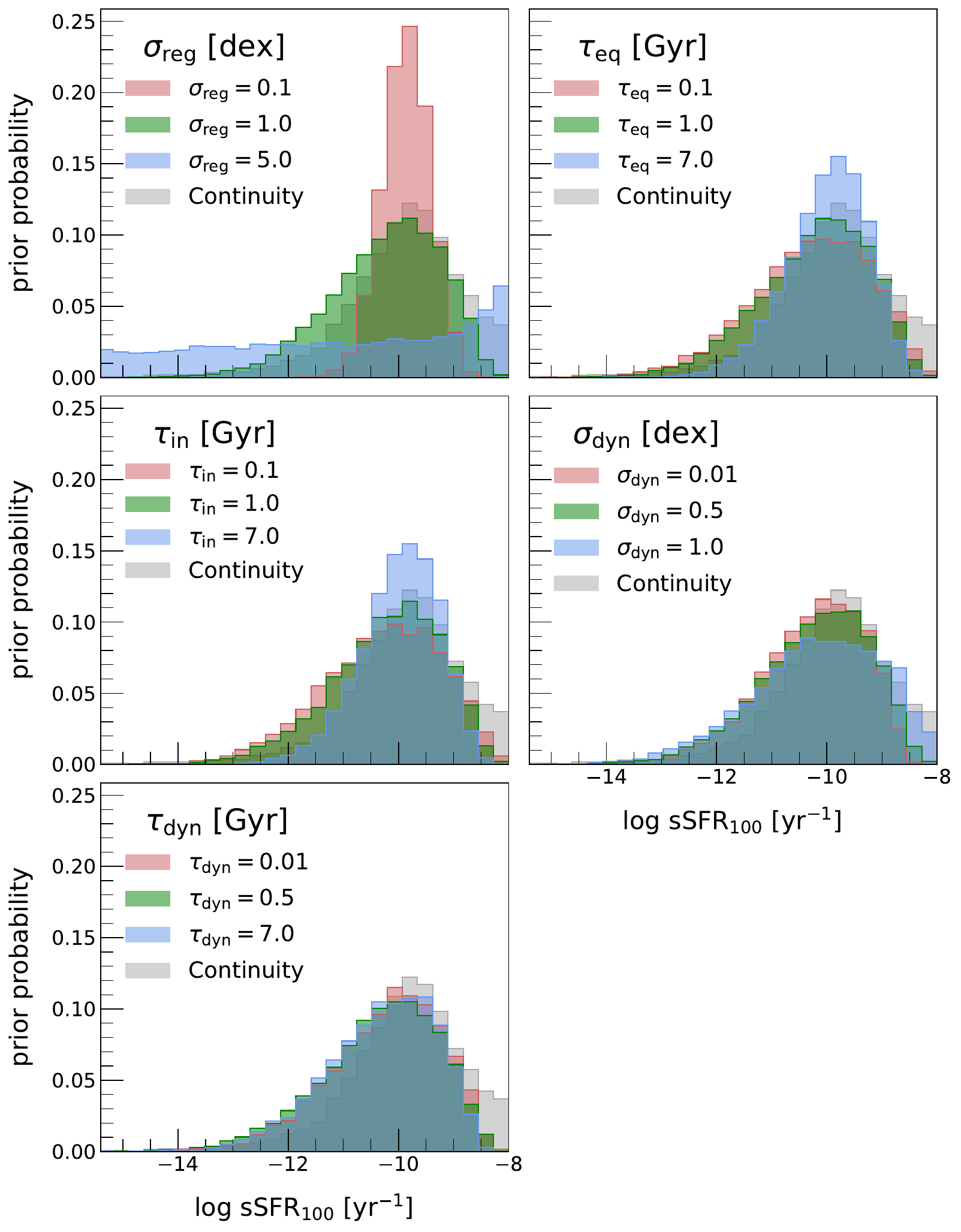}
    \end{subfigure}
    \hfill
    \begin{subfigure}[b]{0.48\textwidth}
        \centering
        \includegraphics[width=\textwidth]{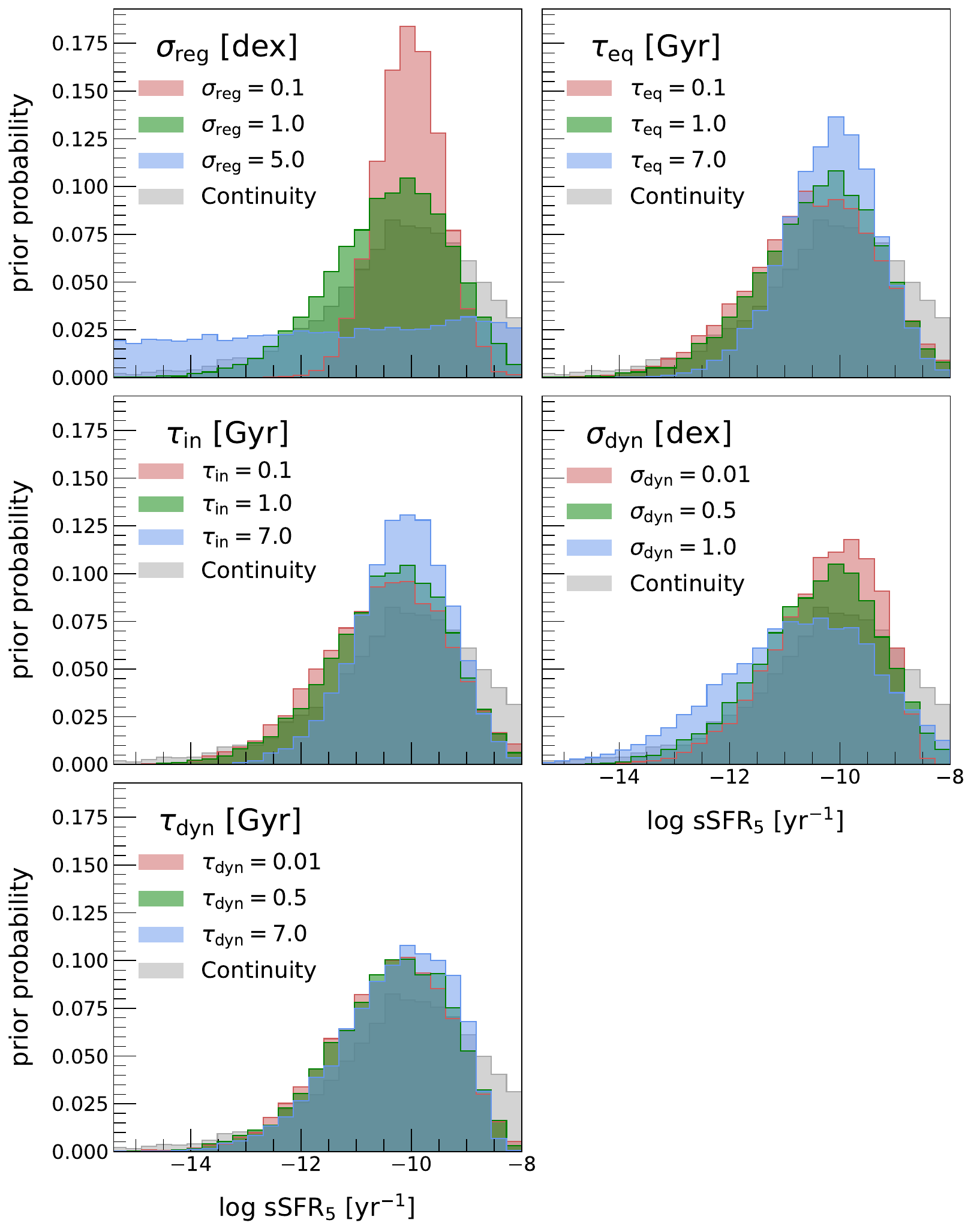}
    \end{subfigure}
    \caption{Histograms of the specific star-formation rate (sSFR), averaged over the most recent 100 Myr (\textit{left}) and 5 Myr (\textit{right}), obtained from 10,000 draws of the prior distribution for sets of fixed PSD parameters. The top row shows how the sSFR prior distribution changes for different values of $\sigreg$ and $\taueq$; the middle row shows the effect of $\tauin$ and $\sigdyn$; and the bottom row $\taudyn$. The PSD parameters are set to $(\sigreg/\rm{dex}, \taueq/\rm{Gyr}, \tauin/\rm{Gyr}, \sigdyn/\rm{dex}, \taudyn/\rm{Gyr}) = (1.0,~1.0,~1.0,~0.5,~0.01)$ unless explicitly specified. The sSFR prior distribution from the continuity SFH model is shown in gray for comparison.}
    \label{fig:fixed PSD sSFR prior}
\end{figure*}

The left half of Figure \ref{fig:fixed PSD sSFR prior} shows the sSFR (averaged over 100 Myr) prior distributions derived from sets of fixed PSD parameters with $z = 0.7$ (the redshift dependence of the prior is shown in the subsequent section). The PSD parameters are set to $(\sigreg/\rm{dex}, \taueq/\rm{Gyr}, \tauin/\rm{Gyr}, \sigdyn/\rm{dex}, \taudyn/\rm{Gyr}) = (1.0,~1.0,~1.0,~0.5,~0.01)$ where parameter values are not explicitly specified. We change the value of the PSD parameters individually to isolate each parameter's impact on the sSFR prior. \textcolor{black}{For comparison, we also show the sSFR prior distribution obtained from the continuity prior in gray.}

It is clear that $\sigreg$ has the most significant impact on the behavior of the sSFR prior distribution. Specifically, when $\sigreg$ is small, the sSFR prior is narrow, and becomes broader as $\sigreg$ increases. Galaxies with a high level of variability in their gas inflow and cycling processes (i.e. large $\sigreg$) will experience a wider range of star-formation activity; therefore, a broader sSFR prior distribution is a necessary product of the model to describe such systems. 

$\taueq$ and $\tauin$ have effectively the same impact on the resulting sSFR prior. Changing either parameter alters the width of the sSFR prior distribution, albeit to a lesser degree than $\sigreg$. When $\taueq$ is long, the sSFR distribution is tighter than when it is short. Galaxies with long $\taueq$ values will have SFHs which are correlated over long periods of time, which limits the range of possible SFRs such systems can experience. Conversely, galaxies with short $\taueq$'s have SFHs which de-correlate more quickly, meaning they have the ability to fluctuate more significantly in their SFRs. Thus, the sSFR prior distribution is narrower for long $\taueq$ than short $\taueq$. The same applies to $\tauin$.

\textcolor{black}{We expect dynamical processes to leave a smaller imprint on the PSD than gas inflow and cycling in modulating the SFHs of galaxies \citep{TFC2020}, and to occur on shorter timescales.} Thus, we choose values for $\sigdyn$ and $\taudyn$ (i.e. the dynamical PSD parameters) which are smaller than that of $\sigreg$, $\taueq$, and $\tauin$ (i.e. the gas PSD parameters). This by construction limits the effect that $\sigdyn$ and $\taudyn$ can have on the sSFR prior distribution. However, we see that, similar to $\sigreg$, increasing $\sigdyn$ serves to broaden the sSFR prior distribution. Changing $\taudyn$ has effectively no impact on the sSFR prior over 100 Myr timescales.

The choice of averaging timescale is important to whether or not the dynamical PSD parameters visibly alter the sSFR prior distribution. Because we choose $\taudyn$ to be relatively short, any SFH fluctuations resulting from the dynamical component are washed out when the SFR is averaged over 100 Myr. Thus, we also show the sSFR prior distributions calculated by averaging the SFR over the most recent 5 Myr (right half of Figure \ref{fig:fixed PSD sSFR prior}). In this case, the behavior of the gas PSD parameters remains largely unchanged; however, the effect of the dynamical parameters becomes more apparent. The sSFR prior distribution broadens more as $\sigdyn$ increases when the averaging timescale is 5 Myr versus 100 Myr. It is now also possible to see small differences in the sSFR prior due to changes in $\taudyn$ -- as $\taudyn$ becomes larger, the distribution becomes narrower.

\begin{figure}
    \centering
    \includegraphics[width=0.48\textwidth]{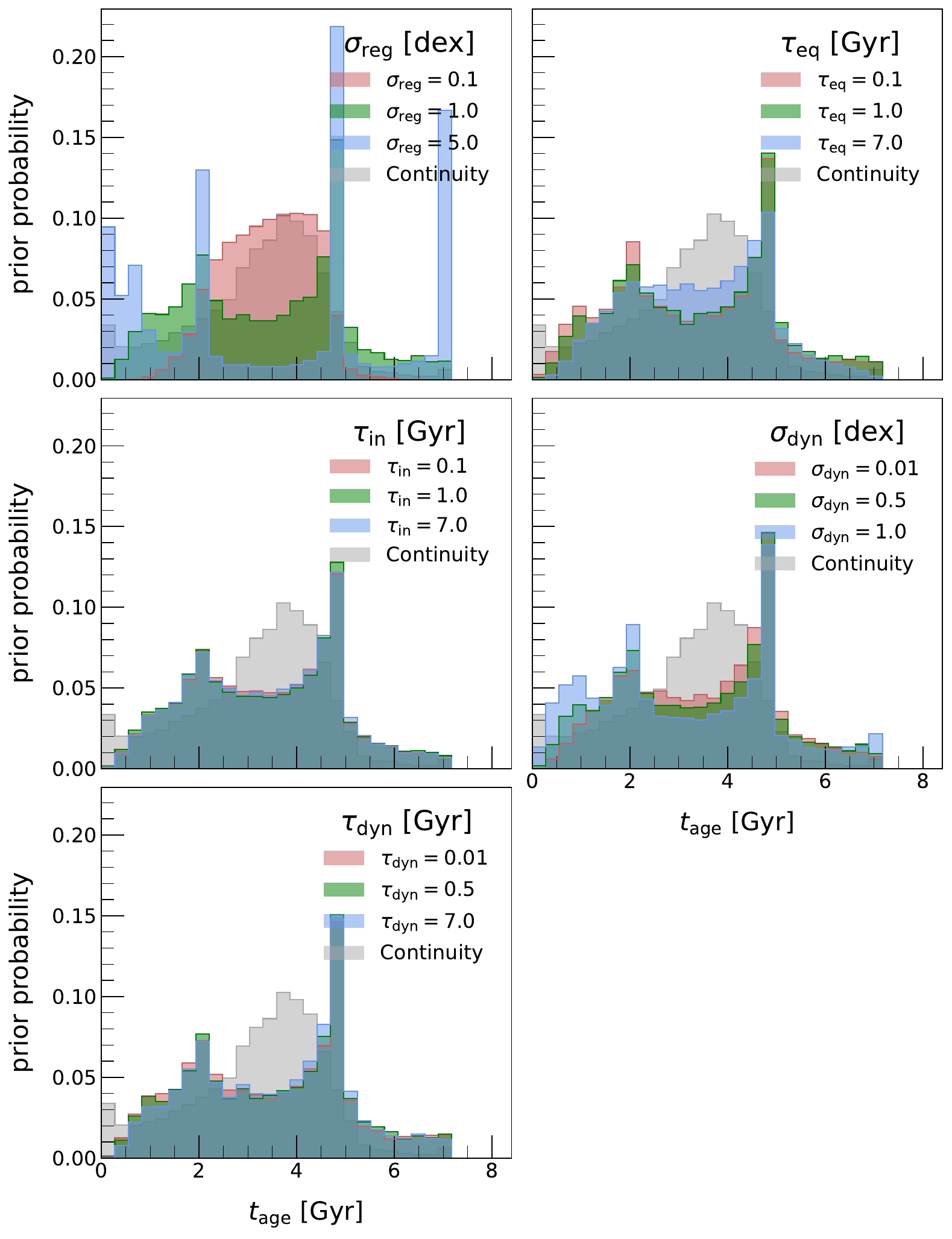}
    \caption{Histograms of the mass-weighted age obtained from 10,000 draws of the prior distribution for sets of fixed PSD parameters. The top row shows how the sSFR prior distribution changes for different values of $\sigreg$ and $\taueq$; the middle row shows the effect of $\tauin$ and $\sigdyn$; and the bottom row $\taudyn$. The PSD parameters are set to $(\sigreg/\rm{dex}, \taueq/\rm{Gyr}, \tauin/\rm{Gyr}, \sigdyn/\rm{dex}, \taudyn/\rm{Gyr}) = (1.0,~1.0,~1.0,~0.5,~0.01)$ unless specified. The $t_{\rm{age}}$ prior distribution from the continuity SFH model is shown in gray for comparison. The multi-modal structure that appears in the derived $\tage$ prior distributions at large values of $\sigreg$ is an artefact of binning -- when $\sigreg$ is large, it is more likely that a galaxy will form a significant amount of mass in the bin (or set of bins) whose lengths are on the order of $\taueq$ and $\tauin$, hence boosting the prior probability at specific values of $\tage$.}
    \label{fig:fixed PSD mwa prior}
\end{figure}

We repeat the exercise of changing one PSD parameter at a time for the $t_{\rm{age}}$ (mass-weighted age) prior distribution, the result of which is shown in Figure \ref{fig:fixed PSD mwa prior}. As with the sSFR prior distributions, $\sigreg$ has the largest influence on the $t_{\rm{age}}$ prior. The most prominent feature of the $\tage$ prior distribution is that as $\sigreg$ increases, the $\tage$ prior not only broadens, but also becomes multi-modal. The broadening can be explained intuitively -- when $\sigreg$ is large, long-timescale SFH fluctuations contain more power and a wider range of SFHs is allowed, meaning that a wider range of ages is allowed. 

To explain the multi-modal structure is slightly more nuanced. Because $\sigreg$ describes the level variability in the gas processes of a galaxy, $\taueq$ and $\tauin$ (which are both set 1 Gyr by default) are the timescales of importance here (refer to Eq. \ref{eq:AFC_exreg}). A small $\sigreg$ corresponds to SFHs which are less correlated over 1 Gyr timescales and contain less power in Gyr-length fluctuations. In other words, the resulting SFHs are likely to experience low-amplitude variations around the baseline SFR. Such more-or-less constant SFHs lead to mass-weighted ages which average out to half the age of the universe ($\sim 3.6$ Gyr at a redshift of $z = 0.7$), which is what we see from the $\sigreg = 0.1$ dex iteration of the $\tage$ prior distribution. 

On the other hand, when $\sigreg$ is adequately large (e.g. in the $\sigreg = 1.0$ dex case), SFHs become more correlated over 1 Gyr intervals and will experience larger amplitude fluctuations. This means that SFHs can remain significantly above (or below) the baseline over 1 Gyr intervals, i.e. there is an increased prior probability that a galaxy would have formed a significant amount of its mass in a 1 Gyr ``burst''. Here, our choice of SFH time bins becomes key -- the bins (or set of bins) in which such a burst could occur will be those that span $\sim1$ Gyr, resulting in a mass-weighted age approximately equal to the lookback time of that bin. In our setup, bins $1-6$ (collectively), and bins 8 and 9 (individually), are on the order of 1 Gyr. The midpoints of bins 8 and 9 are $\sim2.0$ Gyr and $\sim4.9$ Gyr, respectively, which manifest as the two peaks present in the $\sigreg = 1.0$ dex iteration of the $\tage$ prior. The impact of bins $1-6$ shows itself as a bump in the $\tage$ prior distribution towards young ages (i.e. the prior probability of $\tage$ does not smoothly decline towards young ages). This bimodal structure can be seen in most of the $\tage$ prior distributions in Figure \ref{fig:fixed PSD mwa prior} since $\sigreg = 1.0$ dex is the default value we choose when varying the other PSD parameters.

When $\sigreg$ becomes \textit{very} large (e.g. $\sigreg = 5.0$ dex), or when $\sigdyn$ begins to contribute significantly (e.g. $\sigdyn = 1.0$ dex), a four-pronged distribution appears. The two peaks at the midpoints of SFH bins 8 and 9 remain, and we find two new peaks at approximately the maximally old and maximally young values of $\tage$. Because the overall covariance matrix of the SFH prior distribution is set by both $\sigreg$ and $\sigdyn$, when the combination of $\sigreg$ and $\sigdyn$ is large, the correlation between short-timescale SFH bins (i.e. bins of width $\sim \taudyn$) gets boosted along with the long-timescale bins. This increases the prior probability of SFR bursts occurring in the short-timescale bins, namely the first four bins and the last bin, leading to the $\tage$ prior distribution peaks at very young and very old ages.

Additionally, we find that when $\taueq = 7$ Gyr, the bimodality of the $\tage$ prior distribution becomes less severe. This is because a $\taueq$ of 7 Gyr serves to correlate the SFR of a galaxy over nearly its entire lifetime at this redshift, effectively smoothing over its SFH. In doing so, it increases the prior probability of a galaxy's age being $\sim0.5 t_H$ (where $t_H$ is the age of the universe). However, $\tauin$ remains at 1 Gyr, so the double-peaked structure does not entirely disappear.

The multi-modality of the $\tage$ prior distribution is clearly not ideal. We do not want to inform the model that galaxies are more likely to be 2 Gyr and 5 Gyr old as a baseline assumption because there is no physical reason for that to be true. If we want to wash out this effect while maintaining a broad enough SFH prior to model galaxies that span a wide range of star-formation activity, allowing the PSD parameters to remain free parameters in the fitting procedure is key -- particularly $\sigreg$. We elaborate on this in the following section.

\subsection{Free PSD parameters}
\label{sec:free psd params}

We can add an additional layer of flexibility to the model by allowing the PSD parameters to remain free in the SED-fitting procedure. By varying the prior distributions assigned to each PSD parameter, the stochastic SFH prior is able to capture the SFHs of a wide range of different galaxy types, from local Milky Way-like galaxies to galaxies in the high-redshift universe. We look at examples of sSFR and $t_{\rm{age}}$ prior distributions representative of five different classes of galaxies, modifying the priors of the PSD parameters to fit the galaxy regime of interest. These priors are listed in Table \ref{tab:diff sfh priors}.

\begin{table*}
    \centering
    \caption{Priors on the stellar mass and PSD parameters tailored for different populations of galaxies and used to generate the sSFR prior distributions in Figure \ref{fig:prior diff regimes}. In all cases, the prior for the log SFR ratios follows a multivariate-normal distribution with $\bm{\mu} = 0$ and $\Sigma = \mathrm{ACF}_{\mathrm{SFH}}(\sigma_{\mathrm{reg}}, \taueq, \tauin, \sigdyn, \taudyn)$.}
    \label{tab:diff sfh priors}
    \begin{tabular}{lcccccc}
        \hline
        \hline
        Galaxy regime & $\sigreg$ & $\tau_{\mathrm{eq}}$ & $\tau_{\mathrm{in}}$ & $\sigma_{\mathrm{dyn}}$ & $\tau_{\mathrm{dyn}}$ \\
        & [dex] & [Gyr] & [Gyr] & [dex] & [Gyr] \\
        \hline
        Local MW analog & log-U~(0.1, 5.0) & U~(0.01, $t_H$) & U~(0.01, $t_H$) & log-U~(0.001, 0.1) & Clipped-$\mathcal{N}$~(min $= 0.005$, max $= 0.2$, \\
& & & & & \hspace{1cm} $\mu = 0.01$, $\sigma = 0.02$) \\
$z = 1$ MW analog & log-U~(0.1, 5.0) & U~(0.01, $t_H$) & U~(0.01, $t_H$) & log-U~(0.001, 0.1) & Clipped-$\mathcal{N}$~(min $= 0.005$, max $= 0.2$, \\
& & & & & \hspace{1cm} $\mu = 0.01$, $\sigma = 0.02$) \\
Local dwarf galaxy & log-U~(0.1, 5.0) & log-U~(0.01, 1.0) & log-U~(0.01, $t_H$) & log-U~(0.01, 0.5) & log-U~(0.01, 0.1) \\
Cosmic noon ($z = 2$) & log-U~(0.1, 5.0) & U~(0.01, 2.0) & U~(0.01, 2.0) & log-U~(0.01, 0.5) & U~(0.01, 2.0) \\
High-$z$ ($z = 6$) & log-U~(0.1, 5.0) & log-U~(0.001, 0.02) & log-U~(0.001, 0.02) & log-U~(0.01, 1.0) & log-U~(0.001, 0.02)
 \\
        \hline
    \end{tabular}
\end{table*}

\begin{figure}
    \centering
        \begin{subfigure}[b]{0.48\textwidth}
        \centering
        \includegraphics[width=\textwidth]{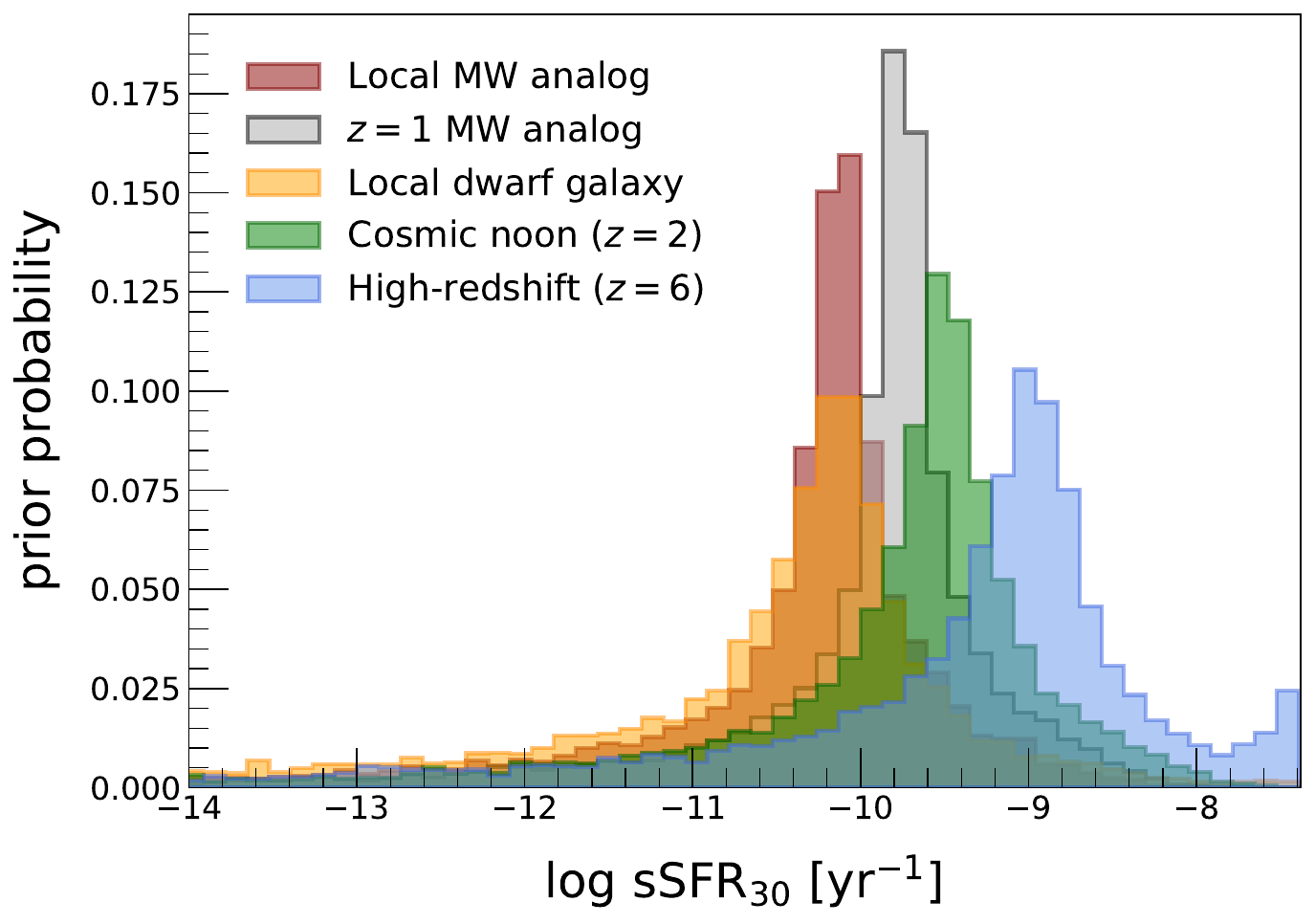}
    \end{subfigure}
    \hfill
    \begin{subfigure}[b]{0.48\textwidth}
        \centering
        \includegraphics[width=\textwidth]{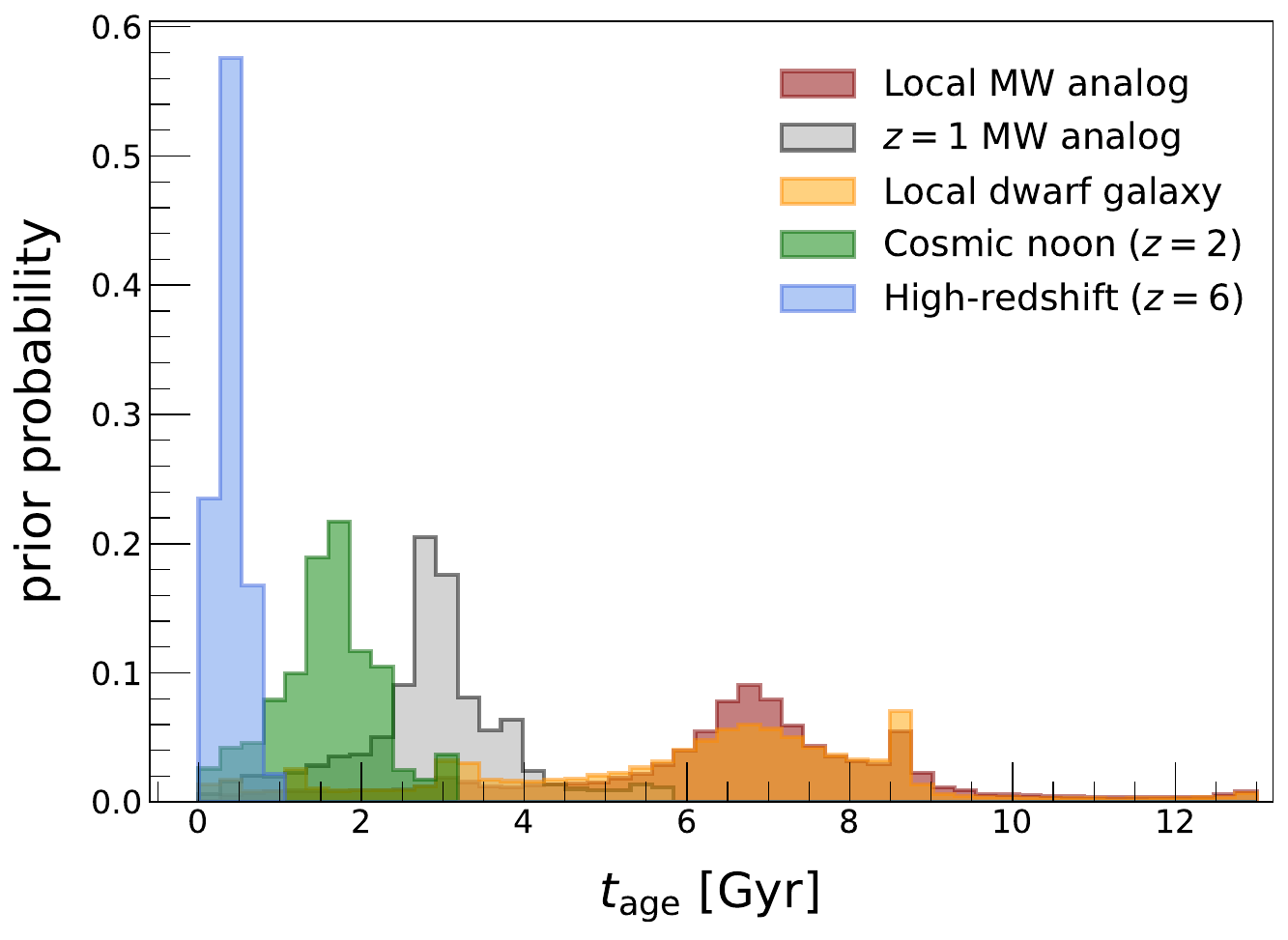}
    \end{subfigure}
    \caption{Histograms of the specific star-formation rate (top) and mass-weighted age (bottom) obtained from 10,000 draws of the prior distribution for the stochastic SFH model, tailored for Milky Way-type galaxies at $z = 0$(red) and $z = 1$ (gray), local dwarf galaxies (orange), galaxies at cosmic noon (green), and high-redshift galaxies (blue). The overall values of the sSFR distributions increase with redshift, and galaxy regimes with burstier SFHs result in broader priors. The mass-weighted age prior distributions broaden and shift towards older ages with decreasing redshift. Galaxy regimes with burstier SFHs lead to broader $\tage$ prior distributions as well.}
    \label{fig:prior diff regimes}
\end{figure}

The top panel of Figure \ref{fig:prior diff regimes} shows the sSFR prior distributions adjusted for five different galaxy evolution regimes -- local MW, $z=1$ MW, local dwarf, cosmic noon, and high-$z$ galaxies. Firstly, increasing the redshift pushes the distribution towards higher sSFRs. This effect is a natural result of $t_H$  decreasing with increasing redshift. Firstly, because $t_H$ is shorter at higher redshifts, for galaxies of a given stellar mass, the same amount of mass must be formed in a shorter amount of time, which shifts the entire sSFR distribution towards higher rates. This mimics redshift-dependent behavior of the star-forming MS, whose normalization increases with increasing redshift. Furthermore, the broadest sSFR priors are seen in the dwarf and high-$z$ cases, which accommodates for the burstier star-formation behavior of these types of galaxies. This broadening is primarily a result of the log-uniform priors on $\taueq$, $\tauin$, and $\taudyn$ which (i) confine these timescales to smaller values, and (ii) up-weight the prior importance of shorter timescales.

The bottom panel of Figure \ref{fig:prior diff regimes} illustrates the $\tage$ prior distributions for the five evolutionary stages. As before, the primary effect comes from redshift. In both cases, the distribution peaks at the half the age of the universe at the given redshift, which is the expected mass-weighted age for a system which forms stars at a constant rate over the course of its SFH. As redshift increases, the prior distribution of $\tage$ becomes tighter and shifts towards younger ages. This behavior is expected, since the universe is younger at higher redshifts, restricting the range of possible ages a galaxy can be.

Additionally, if we compare the local MW analog to the local dwarf galaxy priors (which are both at $z = 0$), we notice that the local dwarf $\tage$ prior distribution is slightly broader. For dwarf-like galaxies with our assumed model, gas cycling is much more rapid than in MW-type galaxies; therefore, dwarf galaxies experience much more variable SFHs, resulting in a greater likelihood that such systems built up a significant portion of their stellar masses in a burst of star-formation activity. Depending on when the burst occurs, this would lead to mass-weighted ages that are either younger or older than the baseline expectation of half the age of the universe, which is captured by the broadening of the $\tage$ prior distribution.

\subsection{Recommended usage}
\label{sec:recommendation}

We do not expect to be able to measure the PSD for individual galaxies. \textcolor{black}{Galaxy SFHs are effectively ``draws'' from an overarching PSD, meaning that one galaxy's SFH could in principle have been produced by any number of different PSDs. However, if one has an ensemble of galaxies (from the same population), each individual galaxy acts as a draw from the overall PSD, and across $\sim$hundreds of galaxies, it should become more feasible to reconstruct an approximation the PSD.}

However, given that {\typewriter Prospector} does not currently have the machinery to perform SED-fitting on multiple galaxies concurrently, to maximize the amount of information extracted by the stochastic SFH prior for a single galaxy, we suggest a combination of the methods discussed in Sections \ref{sec:fixed psd params} and \ref{sec:free psd params}. In other words, our recommendation is to fix one or two PSD parameters to \textcolor{black}{physically-motivated} values and adopt informative priors on the remaining parameters. \textcolor{black}{For example, gas inflow rates are likely correlated with the accretion history of galaxies' parent halos. These halos evolve over timescales comparable to the age of the universe; thus, it might be reasonable to fix $\tauin$ to a timescale on the order of the Hubble time (at the epoch of observation). Besides hard physical limits, one could also use other physical models (e.g. cosmological simulations) to set a narrow, informed prior or to fix parameters. This has been done as part of the continuity prior, for example, which was set such that the SFHs of Illustris simulated galaxies can be produced.} This prevents the SFH prior from having more freedom than the data is equipped to handle, while simultaneously allowing us to either estimate the parameters of interest or marginalize over their uncertainties. We test this method of using the stochastic prior in the subsequent sections.

\section{Recovery tests I: photometry + spectroscopy at \texorpdfstring{\MakeLowercase{\textit{z} = 0.7}}{z = 0.7}}
\label{sec:legac recovery}

In this section, we validate the performance of the stochastic prior when used to model (mock) massive, intermediate-redshift galaxies. The tests detailed here are designed for direct comparison to the photometry and high S/N ($\sim15-20 \Angstrom^{-1}$) spectra of the massive, $z = 0.6 - 1.0$ galaxies which comprise the LEGA-C survey, and act as a showcase for SDSS data, the upcoming MOONS, and other similar surveys. In Section \ref{sec:model_and_priors}, we detail the physical model used to generate and model our mock systems. We describe the method used to generate our mock observations in Section \ref{sec:mock_data}. We present the results of our validation tests in Section \ref{sec:recovery tests}. 

We note that the mock galaxies presented in this section are predominantly star-forming systems. We test our prior explicitly on a mock sample of $z = 0.7$ quiescent galaxies in Appendix \ref{sec:quiescent recovery}.

\subsection{Physical SED model and priors}
\label{sec:model_and_priors}

We present the physical galaxy SED model in {\typewriter Prospector} used to generate and model mock galaxy observations in order to test the performance of the stochastic SFH prior. Our overall model setup is similar to that of \citet{Tacchella2022Halo7D}. We use the dynamic nested sampling package {\typewriter dynesty} \citep{Speagle2020} to sample the posterior probability distribution. Table \ref{tab:priors} summarizes the free parameters and priors used in our physical model, and a brief description is given below.

\begin{table*}
    \centering
    \caption{Description of the free parameters and corresponding priors used to jointly fit the photometry and spectra of the intermediate-redshift mock galaxy sample in {\typewriter Prospector}.}
    \label{tab:priors}
    \begin{tabular}{lll}
        \hline
        \hline
        Parameter & Description & Prior \\
        \hline
        $z$ & Redshift & Uniform~($z_{\mathrm{spec}} - 0.005$, $z_{\mathrm{spec}} + 0.005$), where $z_{\mathrm{spec}}$ is the spectroscopic redshift \\

$\sigma_{*}$/(km s$^{-1}$) & Stellar velocity dispersion & Uniform~(40.0, 400.0) \\ 

log(M$_*$ / M$_\odot$) & Stellar mass & Uniform~(9.5, 12.0) \\

log(Z$_*$ / Z$_\odot$) & Stellar metallicity & Uniform~($-1.0$, 0.19) \\

log SFR ratios & Ratio of SFRs in adjacent time bins & Multivariate-Normal~($\bm{\mu} = 0$, $\Sigma = \mathrm{ACF}_{\mathrm{SFH}}(\sigma_{\mathrm{reg}}, \taueq, \tauin, \sigdyn, \taudyn)$) \\

$\sigma_{\rm{reg}} / \rm{dex}$ & Overall variability in gas inflow and cycling processes & log-Uniform~(0.1, 1.0) \\

$\tau_{\rm{eq}}$ / Gyr & Equilibrium timescale & Uniform~(0.01, $t_H$), where $t_H$ is the age of the universe at the redshift of the object\\

$\sigma_{\rm{dyn}} / \rm{dex}$ & Overall variability in short-timescale dynamical processes & log-Uniform~(0.001, 0.1) \\

$\tau_{\rm{dyn}}$ / Gyr & Dynamical timescale & Clipped-Normal~(min $= 0.005$, max $= 0.2$, $\mu= 0.01$, $\sigma = 0.02$) \\

$n$ & Power-law multiplicative modifier to Calzetti law & Uniform~($-1.0$, 0.4) \\

$\hat{\tau}_{\rm{dust, 2}}$ & Diffuse dust optical depth & Clipped-Normal~(min $=0.0$, max $=4.0$, $\mu=0.3$, $\sigma=1$) \\

$\hat{\tau}_{\rm{dust, 1}}$ & Birth cloud optical depth & Clipped-Normal in ($\hat{\tau}_{\rm{dust, 1}}$ / $\hat{\tau}_{\rm{dust, 2}}$) (min$=0.0$, max$=2.0$, $\mu=1.0$, $\sigma=0.3$) \\

$U_{\rm{min}}$ & Minimum radiation field strength & Clipped-Normal~(min$=0.1$, max$=15.0$, $\mu=2.0$, $\sigma=1.0$) \\

$\gamma_e$ & Fraction of dust heated at radiation intensity $U_{\rm{min}}$ & log-Uniform~($10^{-4}$, $0.1$) \\

$q_{\rm{PAH}}$ & Fraction of dust grain mass in PAHs & Uniform~(0.5, 7.0) \\

$\sigma_{\rm{gas}}$ & Velocity dispersion of gas & Uniform~($30$, $250$) \\

$\log(\mathrm{Z} / \mathrm{Z_\odot})$ & Gas-phase metallicity & Uniform~($-2.0$, $0.5$) \\

$\log(U)$ & Gas ionization parameter for nebular emission & Uniform~($-4.0$, $-1.0$) \\

$f_{\rm{out}}$ & Fraction of spectral data points considered outliers & Uniform~($10^{-5}$, $0.5$)  \\
        \hline
    \end{tabular}
\end{table*}

We use the Flexible Stellar Population Synthesis ({\typewriter FSPS}) package \citep{Conroy2009, ConroyGunn2010} with the MIST isochrones \citep{Dotter2016, Choi2016} and the empirical MILES spectral library \citep{MILES2011} to conduct stellar population synthesis. The Modules for Experiments in Stellar Astrophysics (MESA) code \citep{Paxton2011, Paxton2013, Paxton2015} is used to compute the stellar evolutionary tracks from which the MIST isochrones are constructed. We adopt the \citet{Kroupa2001} initial mass function.

The stellar mass, metallicity, and velocity dispersion remain free parameters in our fits. $\log \rm{M_*}/\rm{M_\odot}$ is allowed to vary between 9.5 and 12.0. We assume that all stars within a galaxy share a single metallicity, which is allowed to vary within a uniform prior between $-1.0$ and $0.19$. We fit for the velocity dispersion using a flat prior between 40 and 400 km s$^{-1}$. We also allow very slight adjustments to the redshift, within $\pm 0.005$ of the spectroscopic redshift.

We adopt the two-component dust attenuation model of \citet{CharlotFall2000}, which considers the attenuation of young and old stellar light separately. We assume a flat prior on the dust law slope $n$ \textcolor{black}{(the ``dust index'')}, letting it vary uniformly between $-1$ and $0.4$. A clipped-normal prior is placed on diffuse dust optical depth, which is centered at $0.3$ with a width of $1.0$ between $0.0$ and $4.0$. The birth-cloud optical depth is tied to that of the diffuse dust with an informative joint prior on the ratio $\hat{\tau}_{\mathrm{dust,1}}/\hat{\tau}_{\mathrm{dust,2}}$, a clipped-normal distribution centered at 1.0 with a width of 0.3 in the range $0.0 < \hat{\tau}_{\mathrm{dust,1}}/\hat{\tau}_{\mathrm{dust,2}} < 2.0$. Additionally, we describe the dust emission following the \citet{DraineLi2007} model. We adopt informative priors on the minimum radiation field strength and warm dust fraction, and a flat prior on the PAH mass fraction (see Table \ref{tab:priors}).

We utilize the default nebular emission model in {\typewriter FSPS}, adopting flat priors on the gas-phase metallicity ($-2.0 < \log(\mathrm{Z}/\mathrm{Z_\odot}) < 0.5$, the gas ionization parameter ($-4.0 < \log(U) < 1.0$), and the gas-phase velocity dispersion ($30$ km s$^{-1}$ $< \sigma_{\rm{gas}} < 250$ km s$^{-1}$). 
Furthermore, we include several nuisance parameters, which mitigate the effects of inaccurate calibration and outlying spectroscopic data points. We remove the continuum shape from our inference of physical parameters by multiplying the model spectrum by a order 10 Chebyshev polynomial calibration function at every likelihood call. With {\typewriter Prospector}'s pixel outlier model, we assume that a fraction $f_{\rm{out}}$ of the spectral pixels are outliers and inflate their uncertainties by $50\times$. $f_{\rm{out}}$ is a free parameter in the model, with a uniform prior between $10^{-5}$ and 0.5. 

Lastly, we describe the star formation activity of galaxies in our sample with a non-parametric SFH model. The SFR in each bin is determined using the stochastic SFH prior described in Section \ref{sec:stoch prior}, which assumes that the $N$-vector of log-SFR ratios follows a multivariate-normal distribution with a mean of 0 and a covariance matrix given by the Extended Regulator model $\mathrm{ACF}_{\rm{SFH}}$. We set $N = 10$ in our analysis, a choice which balances the need for accurate constraints on the PSD parameters with the maintenance of computational tractability. The time bins are specified in look-back time, with the first two bins fixed at ($1 - 5$) Myr and ($5 - 10$) Myr, and the remaining bins are equally spaced in logarithmic time between $10$ Myr and $0.95 t_H$ Gyr.

\subsection{Mock observations}
\label{sec:mock_data}

We create four sets of 100 mock galaxies to verify that the stochastic SFH prior recovers the SFHs of a range of galaxies, as well as their stellar population parameters, with sufficient accuracy. In our mock galaxy samples, we vary the stellar velocity dispersion ($\sigma_*$), stellar metallicities ($\log \rm{Z_*})$, dust attenuation ($n$, $\hat{\tau}_{\rm{dust,1}}$, $\hat{\tau}_{\rm{dust,2}}$) and emission ($U_{\rm{min}}$, $\gamma_e$, $q_{\rm{PAH}}$) values, and nebular emission values ($\sigma_{\rm{gas}}$, $\log \rm{Z_{gas}}$, $\log U$). For each galaxy, values of these stellar population parameters are sampled from their corresponding priors listed in Table \ref{tab:priors}. Additionally, all mock galaxies assume a \cite{Kroupa2001} IMF, a total stellar mass of $10^{10.7}~\Msun$, and a redshift of 0.7. 

SFHs for every galaxy are drawn from the Gaussian Process implemented in \cite{Iyer2022}. We focus on SFHs representative of four different PSD regimes in the galaxy population: smooth, varying, bursty, and highly bursty SFHs. These illustrative cases are based off of \citet{Iyer2022}, which we will describe briefly:

\begin{enumerate}
    \item[1.] \textit{Smooth PSD}: Based on Milky Way-type galaxies. Long-term evolutionary trends in the Milky Way point to a large $\taueq$, which will dominate the SFH.
    \item[2.] \textit{Varying PSD}: Based on galaxies found at cosmic noon. All timescales remain relevant in a typical star-forming galaxy at $z = 2$, leading to larger and more correlated SFR variations.
    \item[3.] \textit{Bursty PSD}: Based on high-$z$ galaxies. Galaxies at $z\sim 4-6$ are lower in mass and are often located in more disruptive environments, causing SFR variations over very short timescales.
    \item[4.] \textit{Highly bursty PSD}: Based on local dwarf galaxies. Gas cycling is much more rapid in these low-mass galaxies, leading to a much smaller $\taueq$ value.
\end{enumerate} 

\begin{figure}
    \centering
    \includegraphics[width=0.48\textwidth]{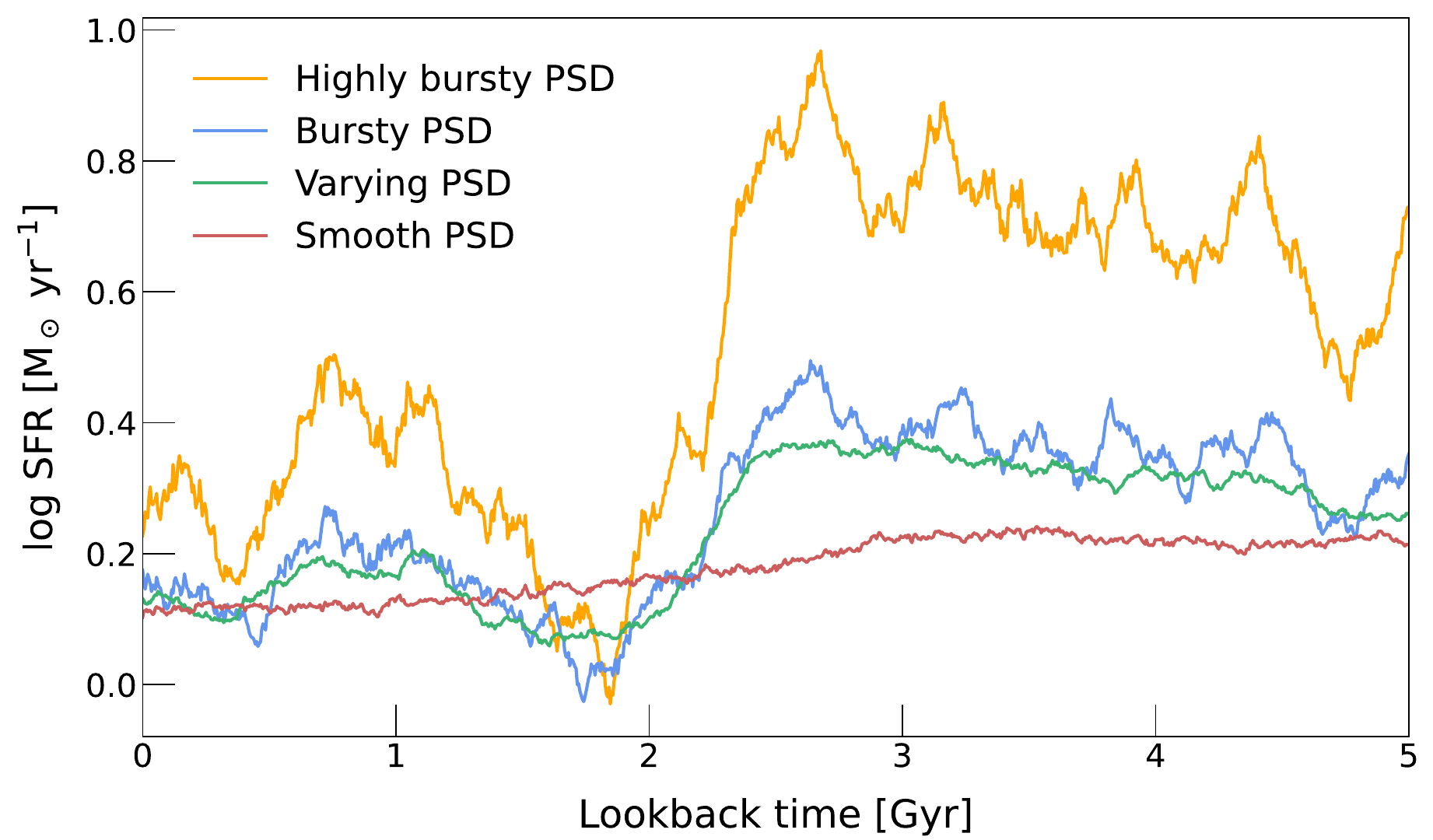}
    \caption{Examples of individual, high-resolution SFHs generated for each of the four galaxy regimes described in Section \ref{sec:mock_data} using the corresponding PSD parameter values listed in Table \ref{tab:psd_cases}. A ``smooth'' power spectrum leads to very flat SFHs, while a ``highly bursty'' PSD results in a SFH with large, rapid variations, and the ``bursty'' and ``varying'' PSDs generate SFHs somewhere in between. The time resolution of these SFHs (1000 time bins) is clearly much higher than what is achievable by the SED-fitting model (10 time bins) -- this is simply an illustration to show qualitatively how the SFHs in each PSD regime differ from one another. All four SFHs are drawn with the same seed value to facilitate direct comparison.}
    \label{fig:example sfhs}
\end{figure}

The PSD parameters associated with each regime are specified in Table \ref{tab:psd_cases}. In all four cases, we fix the inflow timescale, $\tauin$, to the age of the universe at $z = 0.7$. This choice is motivated by the fact that the gas inflow into a galaxy is largely determined by the accretion rate of its parent dark matter halo, and the growth of such halos are associated with Hubble-length timescales. Figure \ref{fig:example sfhs} provides a qualitative example of how the SFHs of galaxies in the different PSD regimes differ from one another. A ``smooth'' power spectrum leads to relatively flat, slowly-varying SFHs, while a ``highly bursty'' spectrum results in a SFH that contains rapid fluctuations on both long and short timescales. ``Bursty'' and ``varying'' PSDs generate SFHs which fall somewhere in between -- both display similar levels of variation in their star-formation activity over long timescales, but a bursty system will experience greater fluctuations over short timescales.

For each regime, we initialize an instance of the {\typewriter GP\_SFH()} class with its corresponding set of PSD parameters and the time array described in Section \ref{sec:model_and_priors}. The instance computes the covariance matrix specified by ACF at the range of times given by the time array. We then generate 100 SFH realizations by sampling a multivariate normal distribution, with $\Sigma$ equal to the covariance matrix calculated by the instance, at each time in the time array. A flat baseline SFH is assumed, i.e. the multivariate normal distribution has a mean vector $\log \mathrm{SFR_{base}}(t) = 0$. For implementation into {\typewriter Prospector}, we transform the SFH realizations into $\log \mathrm{SFR~ratios}$ by simply taking the difference between $\log \mathrm{SFRs}$ at adjacent times.

\begin{table}
    \centering
    \caption{PSD parameters passed into {\typewriter GP\_SFH()} to draw SFHs for mock galaxies in each regime.}
    \label{tab:psd_cases}
    \begin{tabular}{lccccc}
        \hline
        \hline
        PSD Regime & $\sigma_{\rm{reg}}$ & $\tau_{\mathrm{eq}}$ & $\tau_{\mathrm{in}}$ & $\sigma_{\mathrm{dyn}}$ & $\tau_{\mathrm{dyn}}$ \\
        & [dex] & [Gyr] & [Gyr] & [dex] & [Gyr] \\
        \hline
        Smooth & 0.17 & 2.5 & $t_H$ & 0.005 & 0.025 \\
Varying & 0.24 & 0.2 & $t_H$ & 0.007 & 0.05 \\
Bursty & 0.27 & 0.015 & $t_H$ & 0.008 & 0.006 \\
Highly bursty & 0.53 & 0.03 & $t_H$ & 0.016 & 0.01 \\
        \hline
    \end{tabular}
\end{table}

We create mock galaxy spectra to match the resolution of the spectra from DR3 of the LEGA-C survey. The corresponding mock photometry is based on the photometric data available through the COSMOS2020 catalog \citep{COSMOS2020} and the COSMOS Super-deblended catalog \citep{Superdeblended}. To generate spectra and photometry, we pass the SFHs and stellar population parameters of our mock galaxies through the {\typewriter predict()} function in {\typewriter Prospector}. We select one representative galaxy from the LEGA-C sample (ID 109840 \footnote{LEGA-C galaxy 109840 is a blue, star-forming galaxy at redshift $z = 0.6764$, and its measured spectrum has a average S/N of $27.9 \Angstrom^{-1}$.}) which has a match in the photometric catalogs to serve as the template galaxy for our mock observations. For a given mock galaxy, the noise of every spectral and photometric point is randomly drawn from a normal distribution with $\mu = 0$ and $\sigma$ equal to the uncertainty at each corresponding point in the template spectrum and photometric data. These noise values are subsequently added to the intrinsic fluxes generated by {\typewriter Prospector}.

\subsection{Results}
\label{sec:recovery tests}

In this section, we explore how well our model is able to constrain various galaxy parameters in our intermediate-redshift sample. We ensure that the stochastic prior is able to reproduce both the stellar population parameters (Section \ref{sec:recovery sps}) and the SFHs (Section \ref{sec:recovery sfh}) of these galaxies, and compare the efficacy of our model against the standard continuity prior. We also examine the recovery of recent SFRs and galaxy ages in Section \ref{sec:higher-order quantities}, as well as the PSD parameters of galaxies across different PSD regimes in Section \ref{sec:recovery PSD}. All reported quantities are the median of the posterior distribution, and the $1\sigma$ error bars are the 16th and 84th percentiles.

\begin{figure*}
    \centering
    \begin{subfigure}[b]{0.99\textwidth}
         \centering
         \includegraphics[width=\textwidth]{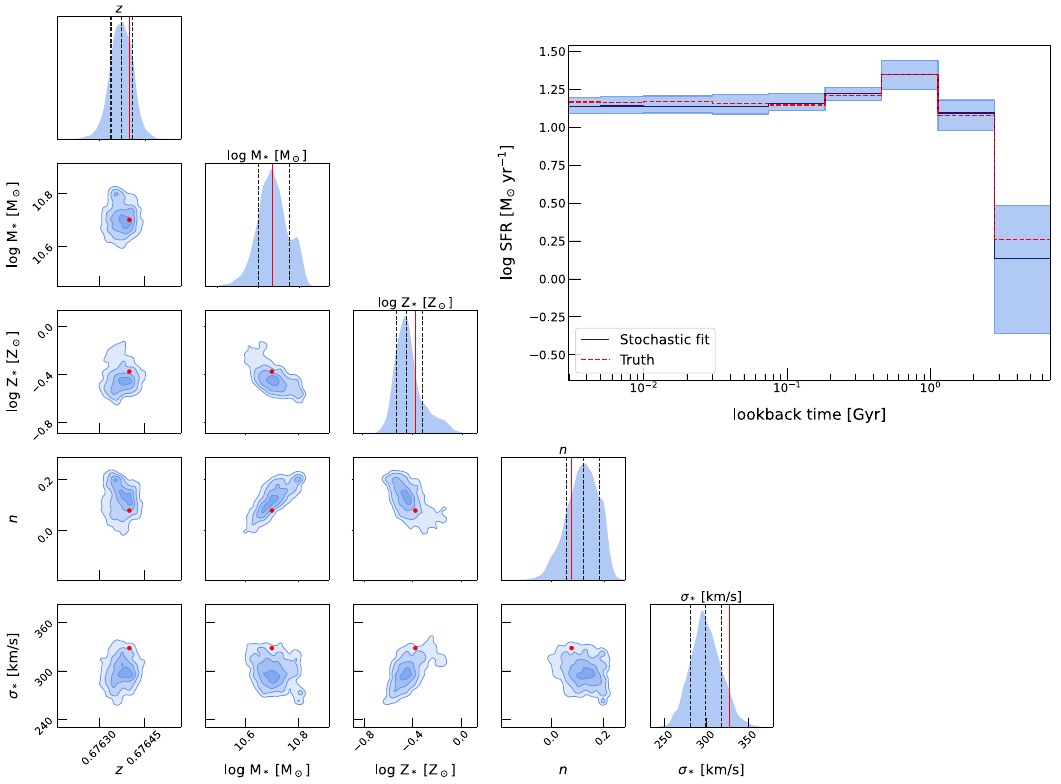}
    \end{subfigure}
    \begin{subfigure}[b]{0.99\textwidth}
         \centering
         \includegraphics[width=\textwidth]{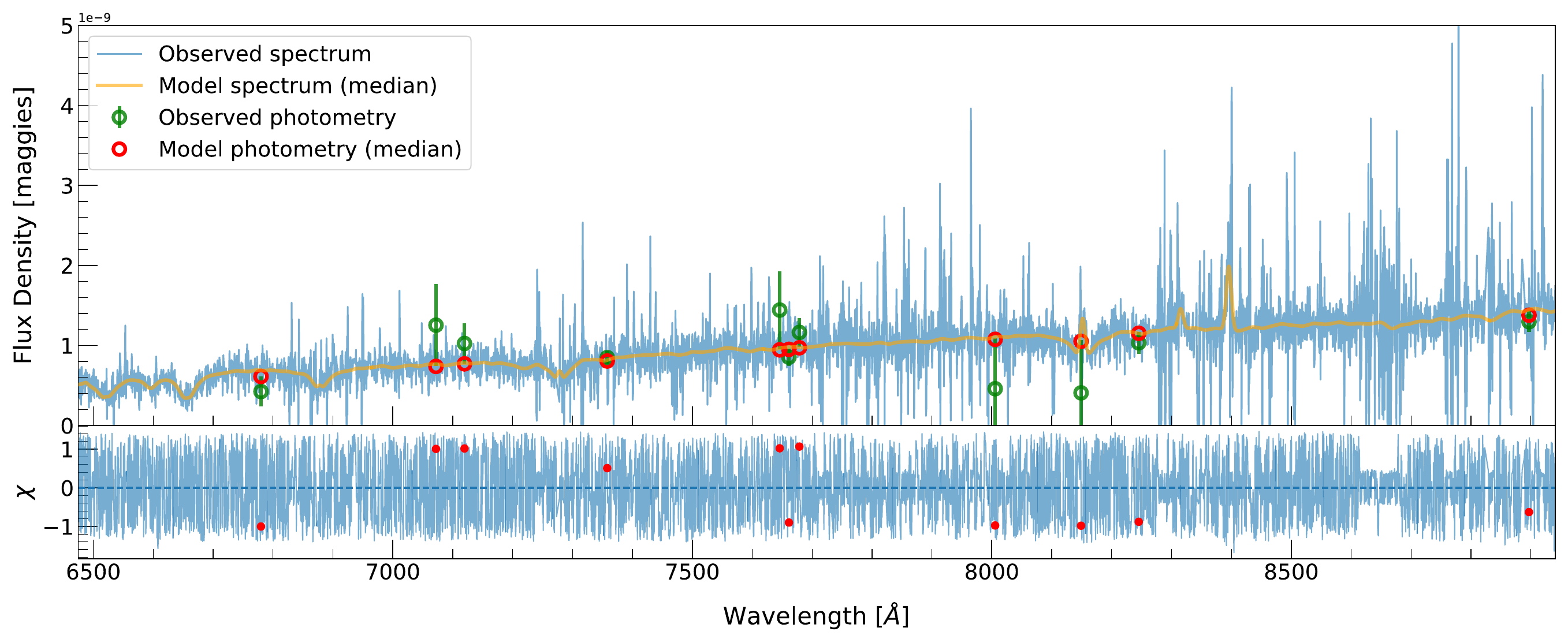}
    \end{subfigure}
    \caption{Example {\typewriter Prospector} fit for one mock intermediate-redshift galaxy. The corner plot shows the joint posterior distributions of key quantities from the galaxy fit, including redshift ($z$), stellar mass ($\mathrm{M_*}$), stellar metallicity ($\mathrm{Z_*}$), dust index ($n$), and stellar velocity dispersion ($\sigma_*$). The inset in the upper right shows the posterior SFH, with the truth over-plotted with a dashed red line. The blue shaded region shows the 16th$-$84th percentile. The bottom panel shows the photometric and spectroscopic data, along with the model, for this example galaxy. The observational data is plotted in blue (spectrum) and green (photometry), while the model fit is shown in orange and red (for the spectrum and photometry, respectively). Overall, the posteriors are well-converged and the fit to the data is good.}
    \label{fig:mock fit}
\end{figure*}

We show an example of a mock galaxy and its associated {\typewriter Prospector} fit in Figure \ref{fig:mock fit}. The model fits both the spectroscopic and photometric data well, and is able to recover key galaxy parameters (e.g. redshift, stellar mass, stellar metallicity). We verify that this is the case across the sample of mock galaxies. Further investigation into the recovery of galaxy properties can be found in the subsequent sections.

\subsubsection{Stellar population parameters}
\label{sec:recovery sps}

\begin{figure}
    \centering
    \includegraphics[width=0.48\textwidth]{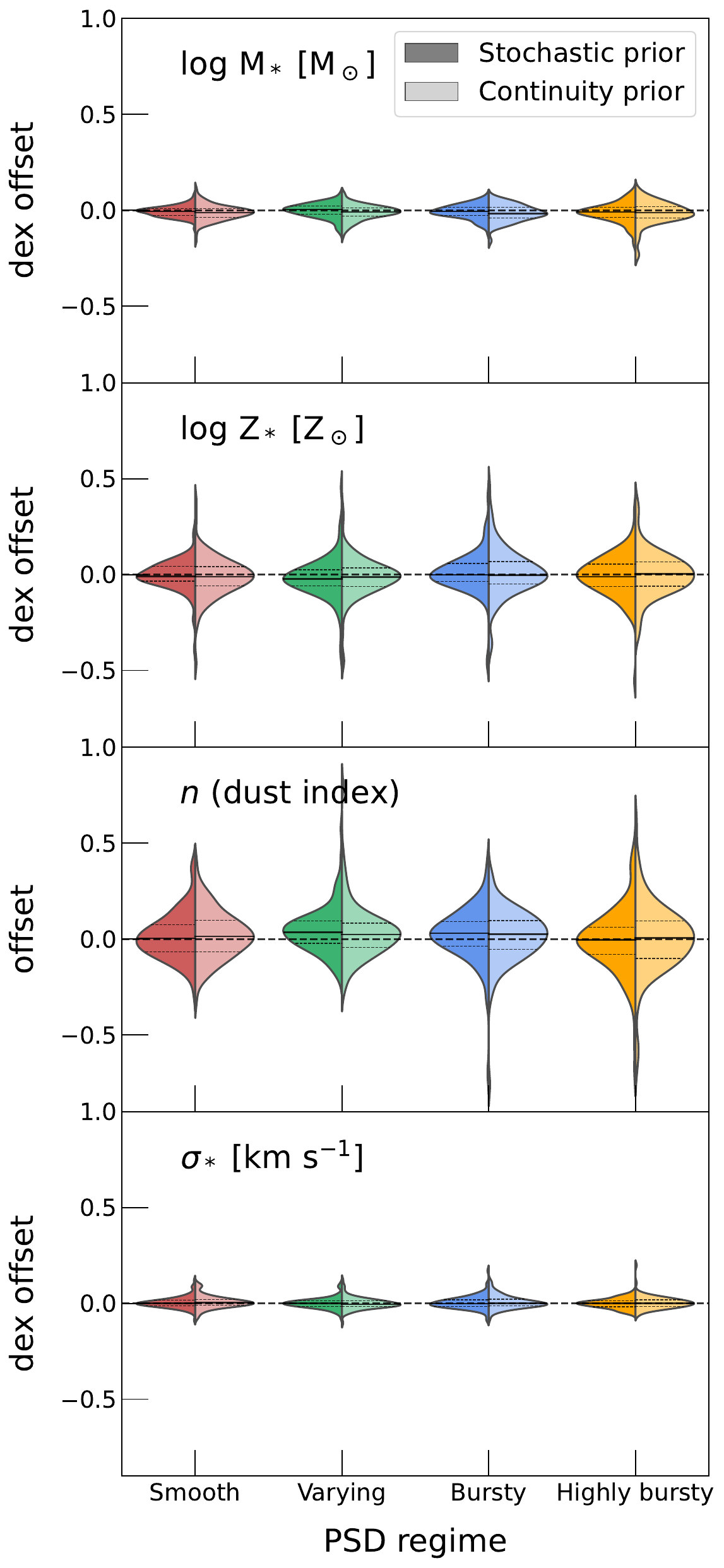}
    \caption{Histograms showing the dex offset between the median posterior recovered stellar mass, metallicity, dust index, and stellar velocity dispersion values and the true values for 100 mock galaxies in each of our four stochasticity regimes. For each regime, the dark shaded histogram (left half of each pair) corresponds to the stochastic SFH prior, and the light shaded histogram (right half) corresponds to the continuity prior. In all cases, these quantities are recovered to the same level of fidelity with the stochastic prior as with the continuity prior. It should be noted that all mock galaxies were generated to have an average S/N of 27.9 $\Angstrom^{-1}$. The recovery of the galaxy parameters shown in this figure will likely be less reliable at low S/N.}
    \label{fig:compare sps params}
\end{figure}

We test how well our fits are able to recover the basic stellar population parameters of mock galaxies, as well as their SFHs. Figure \ref{fig:compare sps params} shows histograms of the offset between the median posterior stellar mass (log $\Mstar$), metallicity (log Z$_*$), dust index ($n$), and stellar velocity dispersion ($\sigma_*$) values and the input (``true'') values for 100 mock galaxies in each of our four stochasticity regimes. The left (dark shaded) half of each histogram represents the results from fits using the stochastic SFH prior, and the right (light shaded) half represents those from the typical continuity prior.

{\typewriter Prospector}, configured with the stochastic SFH prior, is able to recover all of the basic parameters recovered with reasonable scatter and minimal bias. The stellar masses of the galaxies are accurately reproduced, with systematic offsets of $\lesssim 0.02$ dex on average, and scatters of $\lesssim 0.05$ dex in all four regimes. The metallicities are similarly recovered within $\lesssim 0.02$ dex of their true values on average. However, the scatter is larger than that of the stellar mass, at $\sim 0.12$ dex, and all regimes show a tail towards under-estimated metalliticies. The median fit values of the stellar velocity dispersions are both accurate and precise, with systematic offsets of $\lesssim 0.01$ dex and scatters of $\sim 0.05$ dex. The dust index is the least well-constrained property, with an average scatter of $\sim 0.15$ across the four regimes; nevertheless, the median offsets are still quite small, at $< 0.04$.

Additionally, we compare the results obtained when using the stochastic SFH prior against those from the continuity prior. When the continuity prior is assumed, the median offsets in the recovered $\log \mathrm{M_*}$ values average $\lesssim 0.03$ dex, and the scatter in each galaxy regime is $\lesssim 0.06$ dex, comparable to the performance of the stochastic prior. We find systematic offsets of $\lesssim 0.02$ dex and scatters of $\sim 0.13$ dex in the posterior $\log \mathrm{Z_*}$ values, which is, again, nearly identical to the stochastic prior results. As with the stochastic model, the difference between the recovered and true $\sigma_*$ is $\lesssim 0.01$ dex on average. The scatter is also similar between both priors, at $\sim 0.04$ dex. The dust index $n$ is, once again, the least precisely-recovered parameter when the continuity prior is assumed, with an average scatter of $\sim 0.17$ over all four regimes, but the systematic offsets remain small, at $\lesssim 0.02$. Overall, we find negligible differences between the two priors in how well the basic parameters are recovered across all four of the regimes.

\subsubsection{Star formation histories}
\label{sec:recovery sfh}

\begin{figure*}
    \centering
    \includegraphics[width=0.9\textwidth]{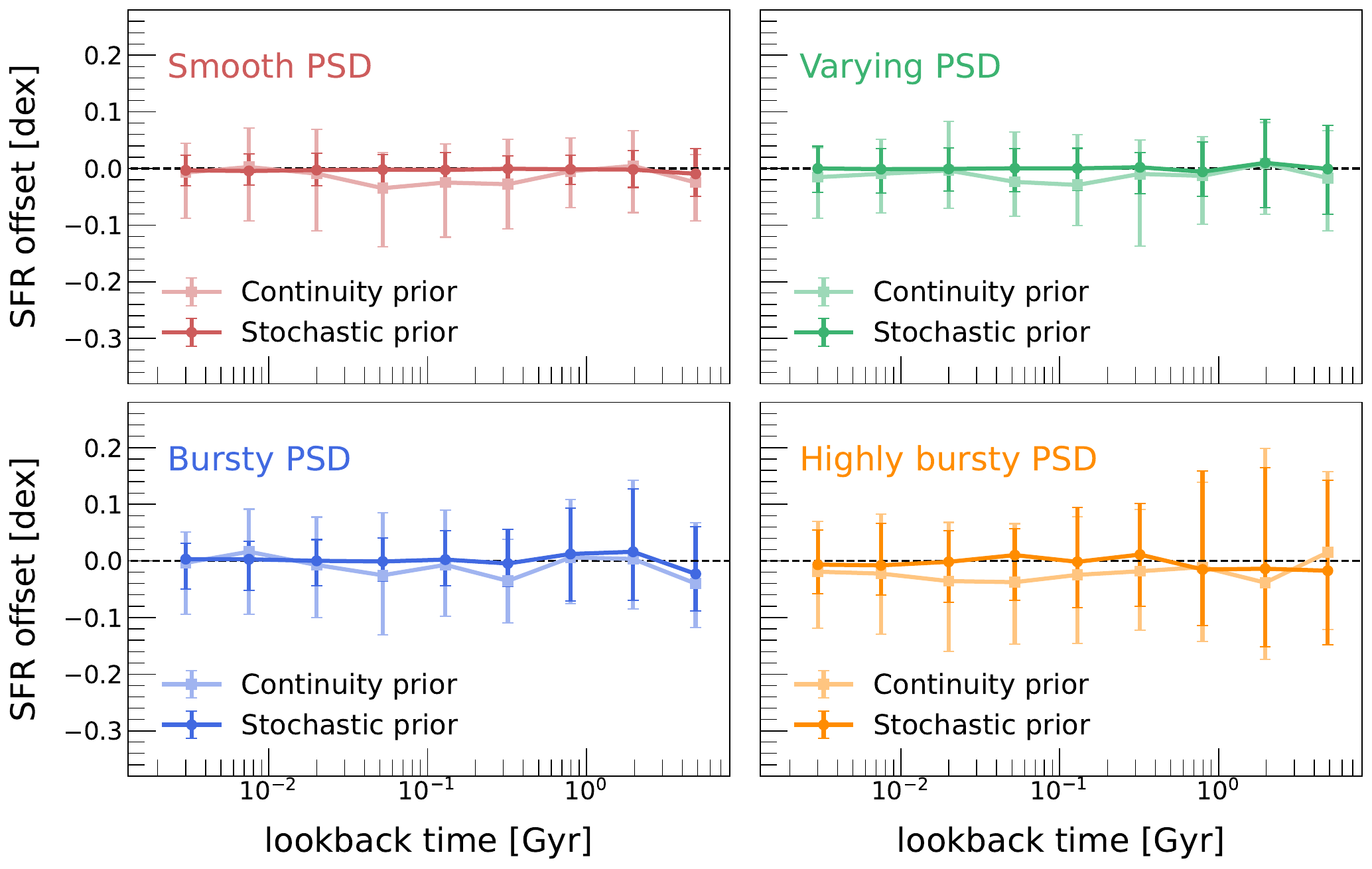}
    \caption{The dex offsets between the median posterior recovered SFRs and their true values in each time bin for the mock galaxies in each of the four stochasticity regimes. The dark-coloured points correspond to the stochastic SFH prior, and the light-coloured points correspond to the continuity prior. Error bars show the 16th$-$84th percentiles. Across all stochasticity regimes, the SFRs in each time are reproduced with minimal bias and scatter. The SFHs are recovered more precisely with the stochastic prior than the continuity prior, especially at recent lookback times. The scatters in the recovered SFRs for the first six time bins are $\sim 2\times$ smaller when the stochastic prior is used than when the continuity prior is used.}
    \label{fig:compare SFHs}
\end{figure*}

Furthermore, we ensure that the stochastic prior is able to recover the true SFHs of our mock galaxies. We look at the differences between the median posterior SFRs and their input values in each SFH time bin across each galaxy regime, and compare the stochastic prior results against the continuity prior (Figure \ref{fig:compare SFHs}).

We find no significant difference between the recovered SFHs obtained from the stochastic prior fits and the true values. The systematic offsets between the measured and true SFRs across all time bins are $< 0.01$ dex. The stochastic prior slightly out-performs the continuity prior in accuracy. Averaging over all SFR bins and all regimes, the SFHs are recovered with a median offset of $\sim 0.02$ dex when the continuity prior is used.

Additionally, the scatter in the offsets (i.e. the error bars in Fig. \ref{fig:compare SFHs}) between the recovered and input SFRs in each bin is smaller when the galaxies are modeled with the stochastic prior. In the smooth PSD regime, the SFHs are recovered with a scatter of 0.04 dex using the stochastic prior (compared to 0.10 dex with the continuity prior). In both the varying and bursty PSD regimes, this scatter is 0.09 dex (compared to 0.12 dex). Lastly, in the highly bursty regime, the scatter in SFH recovery is 0.18 dex (compared to 0.19 dex). However, regardless of which prior is used to model the SFHs, the recovery scatter increases with the burstiness level of the PSD. 

It is also interesting to note how the \textit{discrepancy} in the recovery scatter between the two SFH priors evolves from the early to late time bins in all four stochasticity regimes (here, earlier time bins equate to time bins at smaller lookback times). The scatter in the first six SFH bins is $2\times$ smaller when the stochastic prior is assumed (average scatter of $\sim0.04$ dex) than when the continuity prior is assumed (average scatter of $\sim0.08$ dex). Once we reach the seventh bin, the scatter from the stochastic SFH model ($\sim0.13$ dex) becomes comparable to that of the continuity model ($\sim0.14$ dex). This trend is a product of how the stochastic prior is constructed. The stochastic model of SFHs assumes that star-formation processes are correlated on very short timescales and decorrelate on long timescales (anywhere from $\sim 10$ Myr $- \gtrsim 1$ Gyr, depending on the regime). Thus, if the time difference between two bins is shorter than the decorrelation timescale, then the prior distribution for the log SFR ratio between those bins will be tighter. The drastic increase in recovery precision in the recent time bins, however, demonstrates the impact that a physically-motivated model for star-formation burstiness has on SFH estimation. This alone is a compelling reason to use such a model to fit galaxy SFHs.

\subsubsection{Recent SFRs and ages}
\label{sec:higher-order quantities}

\begin{figure*}
    \centering
    \includegraphics[width=0.98\textwidth]{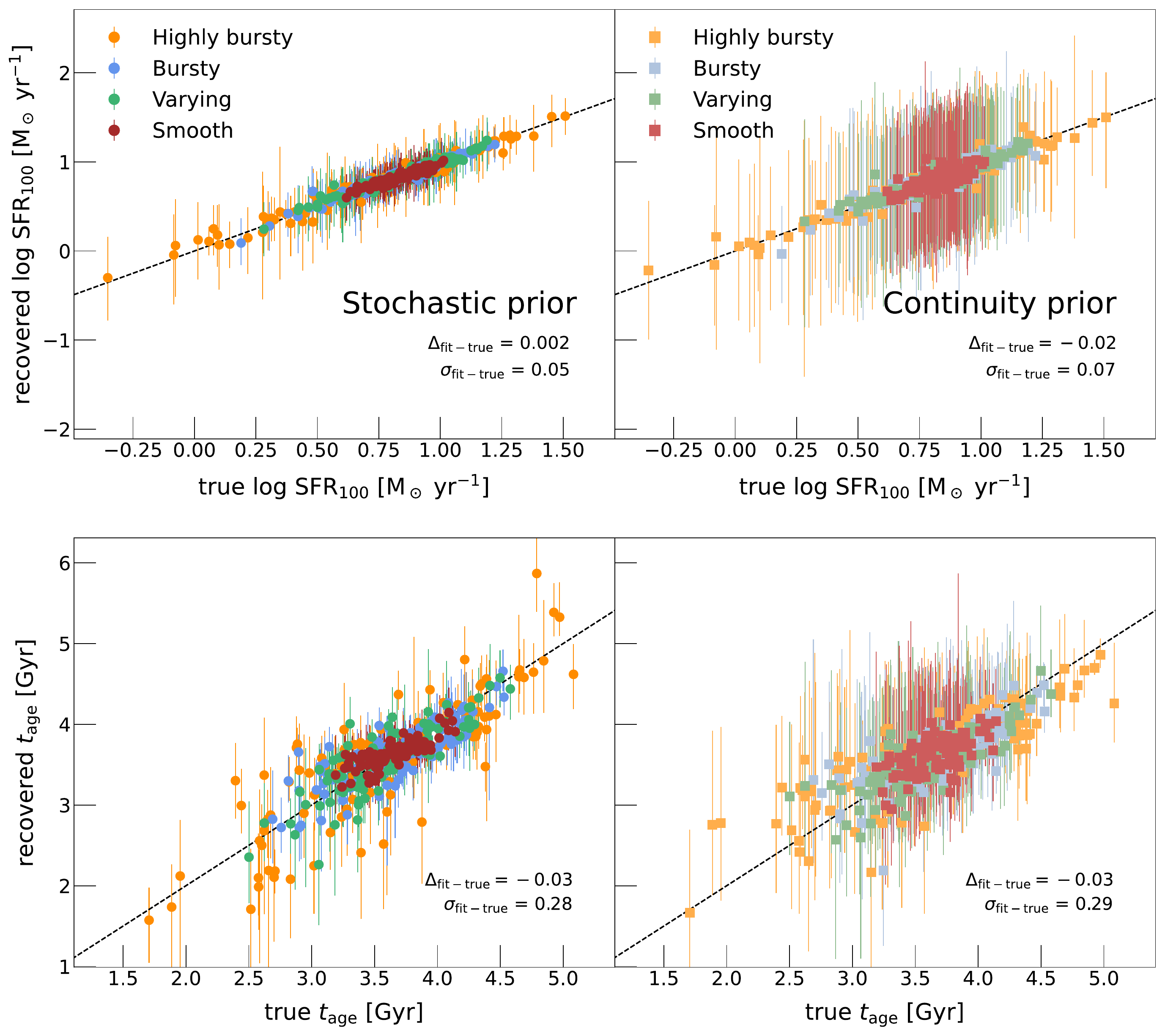}
    \caption{Recovered vs. true log SFR$_{100}$ (top) and mass-weighted age ($t_{\rm{age}}$; bottom) of the intermediate-redshift mock galaxy sample, fit with the stochastic and continuity SFH models. The perfect, one-to-one recovery scenario is denoted with a black dashed line. The $\Delta_{\rm{fit - true}}$ and $\sigma_{\rm{fit - true}}$ values reported in each panel quantify the median offset and scatter (in dex for log SFR$_{100}$ and Gyr for $\tage$) between in the recovered and input values. The stochastic SFH model recovers both the recent SFRs and the ages of the galaxies with minimal scatter and bias. There is no significant difference in how well log SFR$_{100}$ and $t_{\rm{age}}$ are recovered between the stochastic and continuity priors.}
    \label{fig:mwa_sfr}
\end{figure*}

Additionally, we explore the ongoing SFRs and mass-weighted ages recovered by the stochastic SFH prior and compare it against the continuity prior. The top panel of Figure \ref{fig:mwa_sfr} shows the recovered vs. true log SFRs averaged over the last 100 Myr (log SFR$_{100}$) for the mock galaxies in all four PSD regimes; the bottom panel shows the recovered vs. true mass-weighted age ($t_{\rm{age}}$). The left column shows the results from the stochastic prior, and the right column shows the continuity prior. The data points are colour-coded by PSD regime.

The recent SFRs for all of the mock galaxies in the intermediate-redshift sample are recovered accurately using the stochastic model, with 0.05 dex of scatter and no significant offset ($< 0.002$ dex). The mass-weighted ages are also generally well-recovered, with a scatter of 0.28 Gyr and an average offset of $< 0.03$ Gyr. When the continuity prior is used, we find that log SFR$_{100}$ is reproduced with an average offset of 0.02 dex and scatter of 0.07 dex. $t_{\rm{age}}$ is reproduced with offsets of $\sim$0.03 Gyr and a scatter of 0.29 Gyr. Overall, there is no significant difference in performance between the stochastic and continuity SFH priors in recovering the recent SFRs and mass-weighted ages of galaxies in our mock sample.

We notice that the errors on both the log SFR$_{100}$ and $t_{\rm{age}}$ values calculated from the continuity prior fits are larger than those from the stochastic prior fits. As described in Secion \ref{sec:recovery sfh}, the stochastic prior is able to constrain the log SFR ratios of the mock galaxies to a higher level of precision than the continuity prior in the earlier time bins. Notably, the scatter in the offsets between the input and recovered values for the first two log SFR ratios is nearly an order of magnitude smaller when the stochastic prior is used. It makes sense, then, that the stochastic model would lead to smaller errors on these high-order quantities, both of which depend heavily on the resulting SFHs (especially SFR$_{100}$).

\subsubsection{PSD parameters}
\label{sec:recovery PSD}

\begin{figure*}
     \centering
     \begin{subfigure}[b]{0.49\textwidth}
         \centering
         \includegraphics[width=\textwidth]{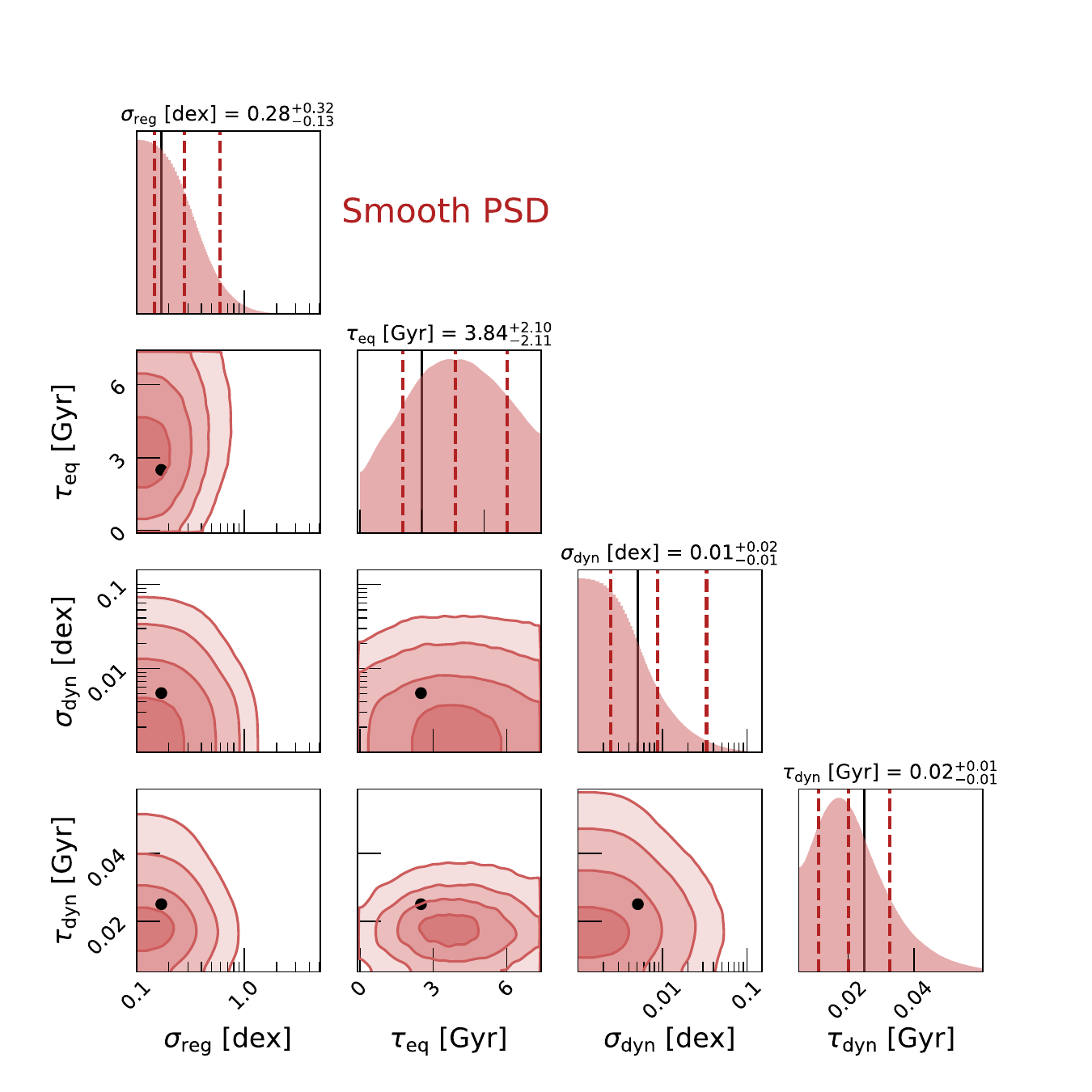}
         \label{fig:smooth PSD}
     \end{subfigure}
     \hfill
     \begin{subfigure}[b]{0.49\textwidth}
         \centering
         \includegraphics[width=\textwidth]{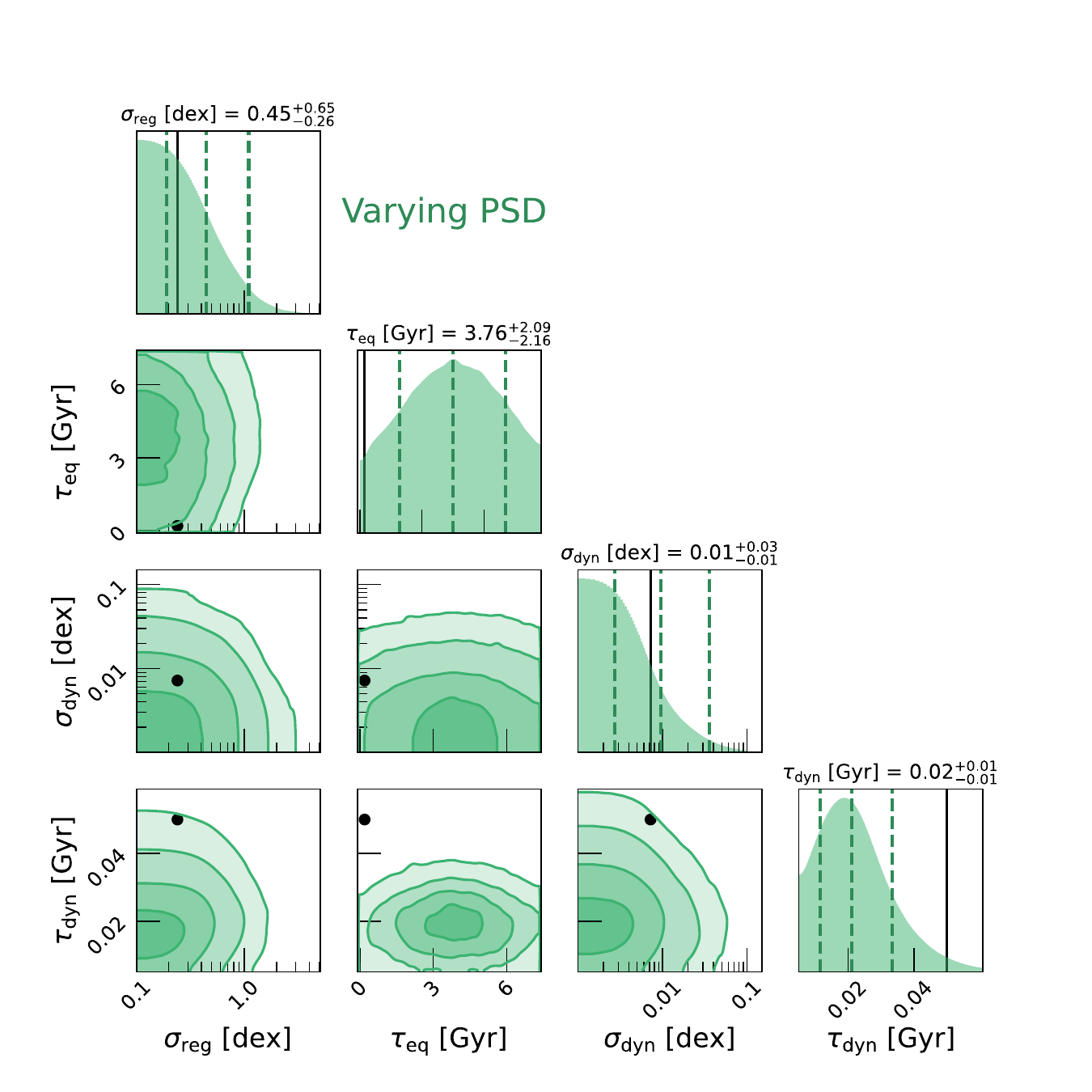}
         \label{fig:varying PSD}
     \end{subfigure}

     \begin{subfigure}[b]{0.49\textwidth}
         \centering
         \includegraphics[width=\textwidth]{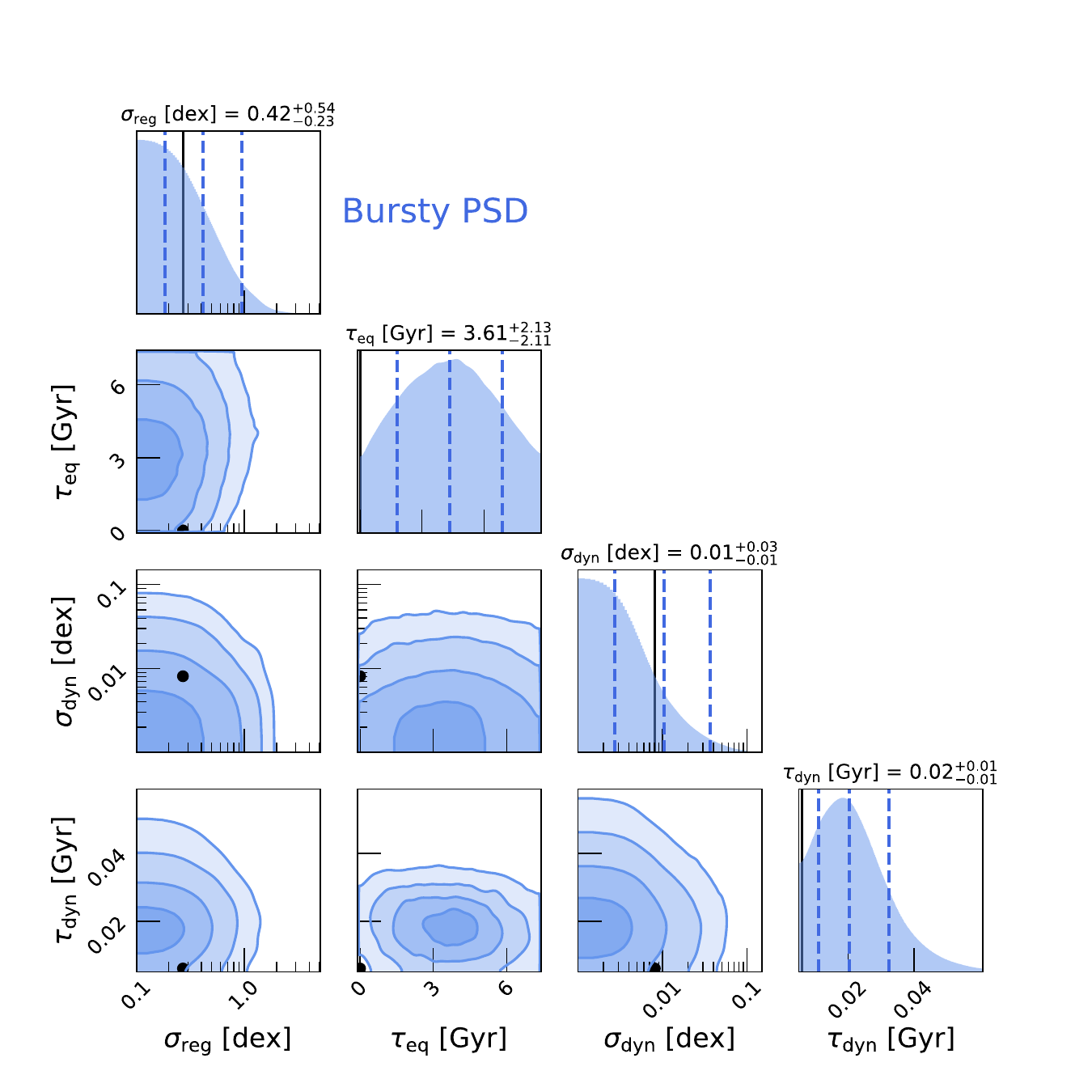}
         \label{fig:bursty PSD}
     \end{subfigure}
     \hfill
    \begin{subfigure}[b]{0.49\textwidth}
         \centering
         \includegraphics[width=\textwidth]{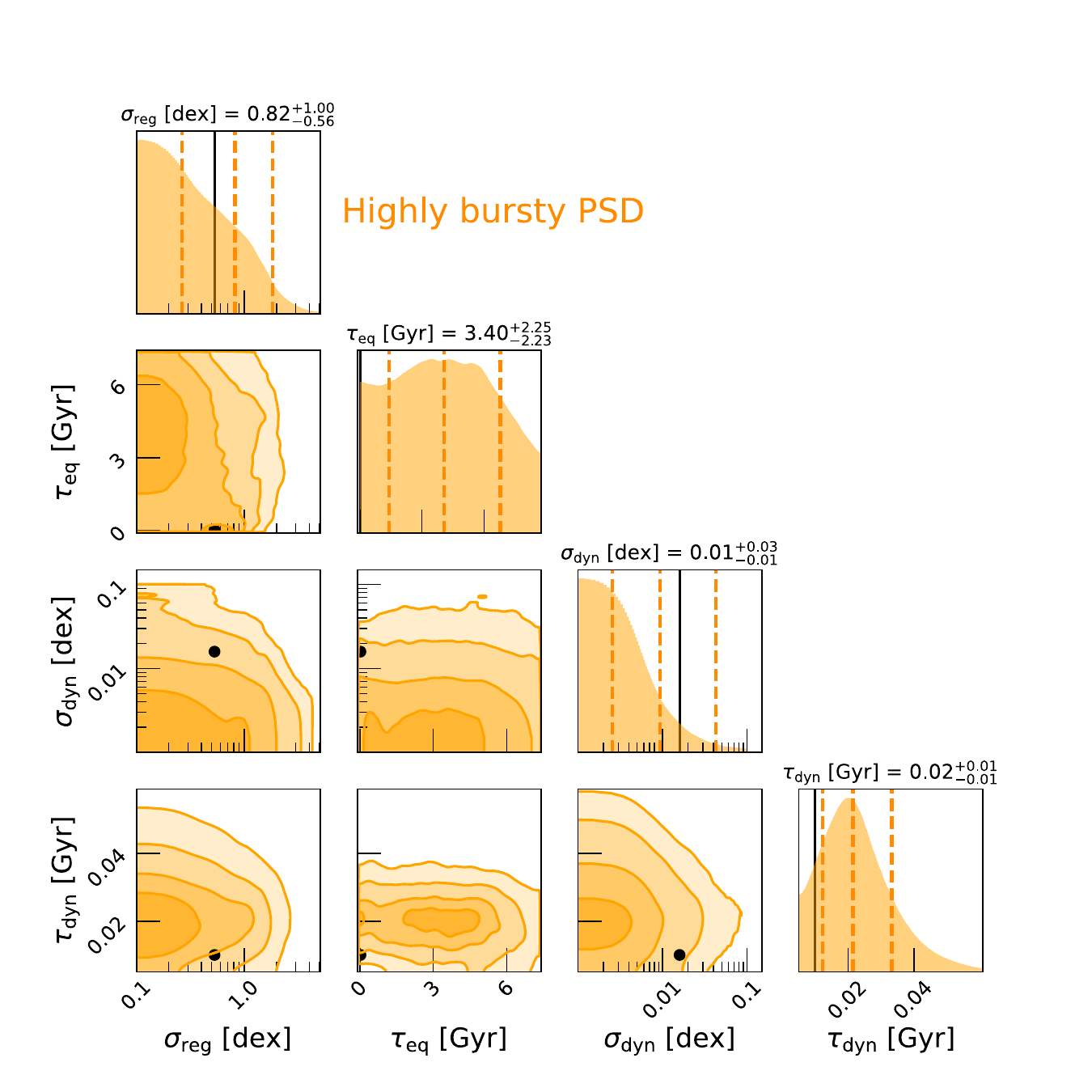}
         \label{fig:highly bursty PSD}
     \end{subfigure}
        \caption{Combined PSD parameters' posteriors for our samples of smooth, varying, bursty, and highly bursty PSD regime mock galaxies. The contours indicate the [0.5, 1, 1.5, 2]-$\sigma$ levels for each pair of parameters. The [16, 50, 84]-th percentiles of the posterior distributions of each parameter are marked with dashed lines. The true parameter values are indicated by a solid black line on the posteriors, and by a black dot on the contours. $\sigreg$ is reliably recovered in all instances, while the equilibrium timescale ($\taueq$) tends towards 0.5$t_H$, and the dynamical PSD parameters ($\sigdyn$, $\taudyn$) are prior-dominated.}
        \label{fig:psd params}
\end{figure*}

In Figure \ref{fig:psd params}, we show how well {\typewriter Prospector} is able to recover the PSD parameters of the galaxies in our four PSD regimes when configured with the stochastic SFH prior. While we expect the constraining power of a single galaxy on the PSD parameters to be relatively weak, we can combine constraints for large samples of galaxies to understand how the variability in SFHs changes across the galaxy population. The implementation of a true Bayesian hierarchical model is outside the scope of this work; however, given that the PSD parameters of each group of galaxies come from the same distribution, each individual galaxy's fitting results can be thought of as an independent estimate of the PSD parameters' posteriors. These results can then be merged into a single run to provide the posterior PSD parameter estimates for the entire sample. 

Thus, we combine the individual {\typewriter dynesty} runs of the 100 galaxies in each regime to obtain the combined, population-level posterior PSD parameter estimates. (We note that our current method of combining posteriors simply merges the individual {\typewriter dynesty} runs for each galaxy in a regime, rather than taking the prior-weighted product of the posteriors. We discuss this choice further in Appendix \ref{sec:combining_posteriors}.) The key takeaways are as follows. In all four cases, {\typewriter Prospector} is able to reasonably estimate $\sigma_{\rm{reg}}$ (which encapsulates the overall variability in the gas inflow and cycling processes in galaxies) from the mock data. The remaining three free PSD parameters -- $\taueq$ (the equilibrium gas cycling timescale), $\sigdyn$ (the variability in short-timescale, dynamical processes), and $\taudyn$ (the timescale relating to short-term dynamical processes) -- are unable to be constrained. The posteriors of all three parameters are prior-dominated.

This result is unsurprising, given that $\sigreg$ is the parameter which has the largest effect on the SFH prior distribution (refer to Section \ref{sec:fixed psd params} for details). The broadness of the SFR range that the model is able to reach is primarily controlled by $\sigreg$, with only a minor influence from $\taueq$ and negligible impact from $\sigdyn$ and $\taudyn$. As such, as long as $\sigreg$ is well-fit, the values of the remaining PSD parameters will not significantly impact the goodness-of-fit of the model. We find that when $\sigreg$ is fixed in the SED-fitting model, the recovery of $\taueq$ improves.

Additionally, it's important to note that the SFHs of the mock galaxies in the four different PSD regimes were generated using very different sets of PSD parameters. In particular, $\taueq$ ranges over three orders of magnitude across the regimes. We use the same $\taueq$ prior in every case -- a uniform distribution between 0.01 Gyr and the age of the universe. This means that longer timescales (i.e. $\taueq > 1$ Gyr) will be preferentially sampled at the expense of shorter ones, and will result in a posterior distribution of $\taueq$ that tends towards these Gyr-length timescales. 
(On the other hand, if we assume a log-uniform distribution on $\taueq$ between 0.01 Gyr and the age of the universe, timescales $< 1$ Gyr are more thoroughly sampled, resulting in posterior distributions that tend toward $\sim 500$ Myr-length timescales.) Because our mock tests are geared towards the massive, $z \sim 0.6 - 1.0$ galaxies in the LEGA-C survey, which we expect to have Gyr-length equilibrium timescales, we opt to use a uniform prior. Such galaxies are most similar those in the smooth PSD regime, which is where the recovery is the most accurate. The model's performance in the other regimes provides a demonstration of the parameter space limits that the stochastic SFH prior can reach.

When fitting real data, the priors on the PSD parameters should, naturally, be adjusted to better match what is expected from the observations. For example, when fitting $z \sim 6$ galaxies, instead of having an extremely broad prior distribution for $\taueq$, you might only let it vary between $1 - 50$ Myr, since you would expect such systems to be low-mass and cycle gas very rapidly. (See Section \ref{sec:prior flexibility} for more detailed examples of how the stochastic SFH prior can be modified.)

\section{Recovery tests II: photometry at \texorpdfstring{\MakeLowercase{\textit{z} = 8}}{z = 8}}
\label{sec:highz recovery}

In this section, we validate the performance of the stochastic prior when used to model high-redshift galaxies. The mock data used in these tests are designed for direct comparison to the photometry able to be obtained with NIRCam on JWST. In Section \ref{sec:highz model}, we describe the physical model used to generate and model our mock systems in {\typewriter Prospector}. We detail the process of generating mock observations in Section \ref{sec:highz mock data}. The results of our high-$z$ validation tests are presented in Section \ref{sec:highz recovery tests}.

\subsection{Physical SED model and priors}
\label{sec:highz model}

We use {\typewriter Prospector} to constrain the stellar populations of the mock high-$z$ galaxy sample. The adopted model is similar to that described in Section \ref{sec:model_and_priors}, with a few adjustments. Because we are working at a much higher redshift where galaxies have had less time to build up their stellar populations, we decrease the minimum of the stellar mass ($\log \Mstar$) prior to $10^6 \Msun$. In the same vein, we also expect galaxies to be less metal-enriched. Therefore, we use a clipped-normal prior for stellar metallicity ($\log \mathrm{Z}_*)$ that has a mean of $-1.5$ and a standard deviation of $0.5$ between $-2.0$ and $0.0$, which is the usual prior assumed for high-$z$ SED-modelling \citep{Whitler2023b, Tacchella2023}.

Furthermore, the high$-z$ mock galaxy sample contains only photometric data. Thus, we do not include stellar velocity dispersion ($\sigma_*$), gas velocity dispersion ($\sigma_{\rm{gas}}$), and spectral outlier fraction ($f_{\rm{out}}$) as model parameters. We also do not fit for the redshifts of the systems and instead, fix the redshifts to the true value, $z = 8$.

We model the galaxies' SFHs with nine time bins -- the first two bins cover lookback times of $0-5$ and $5-10$ Myr, and the remaining bins are log-spaced out to $z = 15$. We place the stochastic SFH prior on the ratios between the SFRs in adjacent time bins. The priors on the PSD parameters are adjusted to reflect the increased star-formation variability and shorter timescales characteristic of the high-redshift universe.

Additionally, because the photometry probes shortward of $1216 \Angstrom$ in the rest-frame at $z = 8$, we include an IGM attenuation model following \citet{Madau1995}. This IGM model includes a free parameter that scales the IGM optical depth ($f_{\rm{IGM}}$), which accounts for variations in the total opacity along the line of sight. We place a clipped Gaussian prior on $f_{\rm{IGM}}$, which is centered at 1.0 with a dispersion of 0.3 and clipped at 0.0 and 2.0.

The parameters and their associated priors are summarized in Table \ref{tab:highz priors}.

\begin{table*}
    \centering
    \caption{Description of the free parameters and corresponding priors used to fit the photometry of the high-redshift mock galaxy sample in {\typewriter Prospector}.}
    \label{tab:highz priors}
    \begin{tabular}{lll}
        \hline
        \hline
        Parameter & Description & Prior \\
        \hline
        log(M$_*$ / M$_\odot$) & Stellar mass & Uniform~(6.0, 12.0) \\

log(Z$_*$ / Z$_\odot$) & Stellar metallicity & Clipped-Normal~(min $=-2.0$, max $=0.0$, $\mu=-1.5$, $\sigma=0.5$) \\

log SFR ratios & Ratio of SFRs in adjacent time bins & Multivariate-Normal~($\bm{\mu} = 0$, $\Sigma = \mathrm{ACF}_{\mathrm{SFH}}(\sigma_{\mathrm{reg}}, \taueq, \tauin, \sigdyn, \taudyn)$) \\

$\sigma_{\rm{reg}} / \rm{dex}$ & Overall variability in gas inflow and cycling processes & log-Uniform~(0.1, 5.0) \\

$\tau_{\rm{eq}}$ / Gyr & Equilibrium timescale & log-Uniform~(0.05, 0.5)\\

$\sigma_{\rm{dyn}} / \rm{dex}$ & Overall variability in short-timescale dynamical processes & log-Uniform~(0.01, 1.0) \\

$\tau_{\rm{dyn}}$ / Gyr & Dynamical timescale & log-Uniform~(0.001, 0.02) \\

$n$ & Power-law multiplicative modifier to Calzetti law & Uniform~($-1.0$, 0.4) \\

$\hat{\tau}_{\rm{dust, 2}}$ & Diffuse dust optical depth & Clipped-Normal~(min $=0.0$, max $=4.0$, $\mu=0.0$, $\sigma=0.5$) \\

$\hat{\tau}_{\rm{dust, 1}}$ & Birth cloud optical depth & Clipped-Normal in ($\hat{\tau}_{\rm{dust, 1}}$ / $\hat{\tau}_{\rm{dust, 2}}$) (min $=0.0$, max =$2.0$, $\mu=1.0$, $\sigma=0.3$) \\

$U_{\rm{min}}$ & Minimum radiation field strength & Clipped-Normal~(min $=0.1$, max $=15.0$, $\mu=2.0$, $\sigma=1.0$) \\

$\gamma_e$ & Fraction of dust heated at radiation intensity $U_{\rm{min}}$ & log-Uniform~($10^{-4}$, $0.1$) \\

$q_{\rm{PAH}}$ & Fraction of dust grain mass in PAHs & Uniform~(0.5, 7.0) \\

$\log(\mathrm{Z} / \mathrm{Z_\odot})$ & Gas-phase metallicity & Uniform~($-2.0$, $0.5$) \\

$\log(U)$ & Gas ionization parameter for nebular emission & Uniform~($-4.0$, $-1.0$) \\

$f_{\rm{IGM}}$ & Scaling of the IGM attenuation curve & Clipped-Normal~(min $=0.0$, max $=2.0$, $\mu=1.0$, $\sigma=0.3$) \\
        \hline
    \end{tabular}
\end{table*}

\subsection{Mock observations}
\label{sec:highz mock data}

We create four sets of 30 mock high-redshift galaxies, each with a different signal-to-noise ratio (SNR), to test the performance of the stochastic SFH prior when applied to photometric observations of high-$z$ systems. In order to incorporate realistic high-redshift star-formation information in our mock galaxies, we use the SFHs of the $z = 8$ galaxies in the SPHINX suite of simulations \citep{Rosdahl2018, Rosdahl2022, Katz2023}. We select the galaxies with stellar masses above $10^8~\Msun$ and degrade their individual high-resolution SFHs to the nine bins described in Section \ref{sec:highz model}. \textcolor{black}{This mass cut is implemented because at stellar masses of $\sim 10^7 \Msun$ and lower, the formation of individual star clusters begins to be important, which would require separate consideration that is outside the scope of this work.} Here, we focus on the SFHs of the 30 galaxies at $z = 8$ in the SPHINX simulations with stellar masses exceeding $10^8~\Msun$.

All mock galaxies assume a \citet{Kroupa2001} IMF, a total stellar mass of $10^9~\Msun$, and a redshift of \textcolor{black}{$z = 8$}. Additionally, because the mock galaxies' SFHs are derived from the SPHINX simulations (rather than drawn from a power spectrum, like in Section \ref{sec:mock_data}), they do not have ``true'' PSD parameter values. We vary the stellar metallicities (log $Z_*$), dust attenuation and emission parameters ($n$, $\hat{\tau}_{\rm{dust,1}}$, $\hat{\tau}_{\rm{dust,2}}$, $U_{\rm{min}}$, $\gamma_e$, $q_{\rm{PAH}}$), nebular emission parameters ($\log \rm{Z_{gas}}$, $\log U$), and the IGM factor ($f_{\rm{IGM}})$ for each galaxy by drawing from their associated priors (Table \ref{tab:highz priors}). These parameter values remain the same for the galaxies in each SNR sample. 

The stellar population parameters and SFHs of the mock galaxies are passed through the {\typewriter predict()} function in {\typewriter Prospector} to generate the photometric data. We model the photometry in the JWST NIRCam F090W, F115W, F150W, F182M, F200W, F210M, F277W, F335M, F356W, F410M, F430M, F444W, F460M, F480M filters, as well as the HST ACS/WFC filters F435W, F606W, F775W, F814W, F850LP. The motivation for the chosen filterset stems from the data available in the Hubble Ultra Deep Field that is covered by the JADES survey \citep{JADES2023} and the JEMS survey \citep{JEMS2023}.

We generate observations with SNR = 5, 10, 20, and 100. We define the SNR as the ratio between the flux in the F444W filter ($f_{444}$) and the uncertainty ($\sigma_{444}$). In each SNR regime, we calculate $\sigma_{444}$ of each galaxy by dividing $f_{444}$ by the chosen SNR. The noise of every photometric point is then randomly drawn from a normal distribution with $\mu = 0$ and $\sigma = \sigma_{444}$, and these values are added to the intrinsic fluxes from {\typewriter Prospector}.

\subsection{Results}
\label{sec:highz recovery tests}

In this section, we explore how well the stochastic model is able to constrain the basic properties and SFHs of high-$z$ galaxies. We investigate the recovery of the stellar population parameters (Section \ref{sec:highz recovery sps}), SFHs (Section \ref{sec:highz recovery sfh}), and recent SFRs and ages (Section \ref{sec:highz recovery higher order}) of these systems, and compare the results against the standard continuity prior.

\begin{figure}
    \centering
        \begin{subfigure}[b]{0.48\textwidth}
        \centering
        \includegraphics[width=\textwidth]{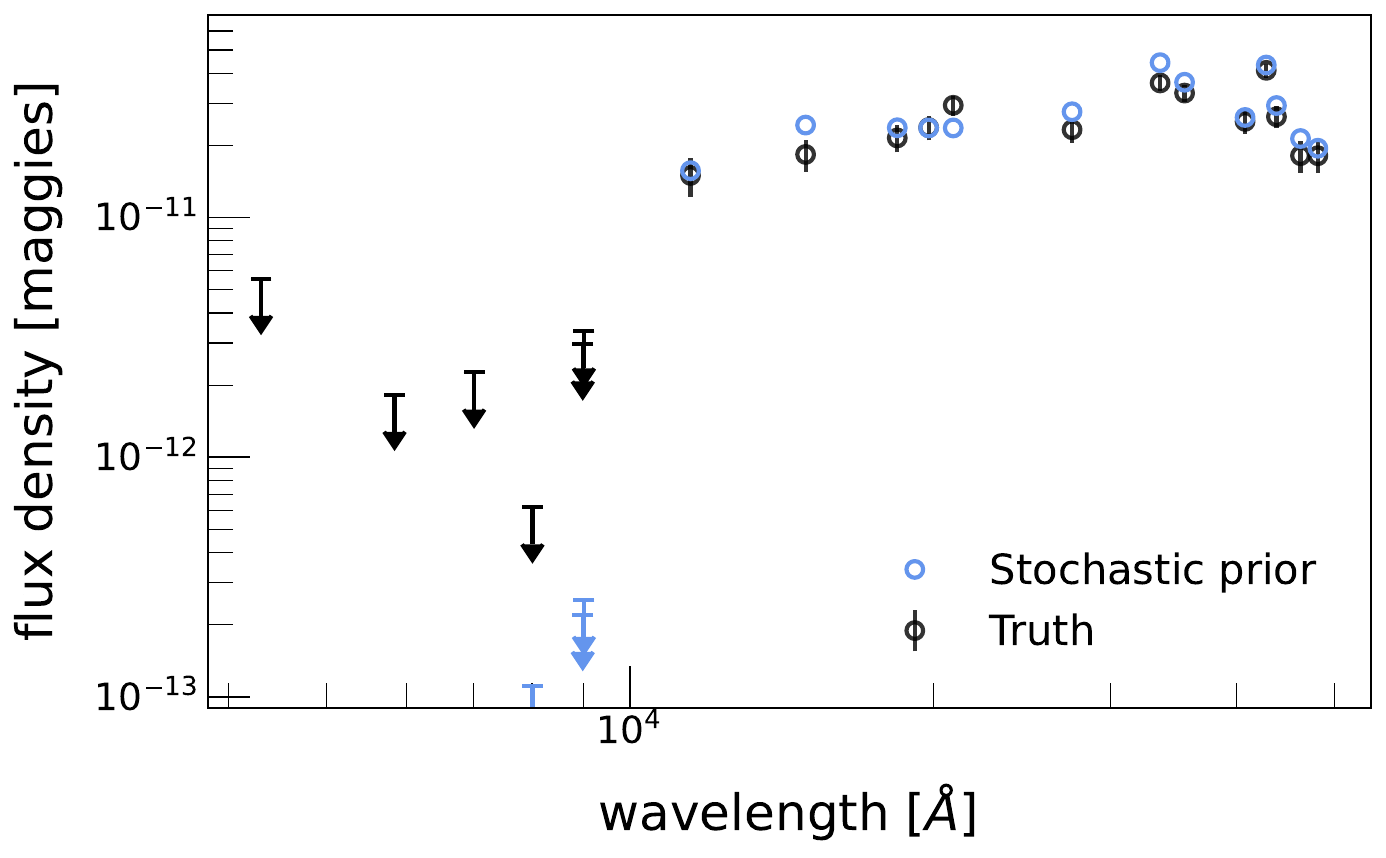}
    \end{subfigure}
    \hfill
    \begin{subfigure}[b]{0.48\textwidth}
        \centering
        \includegraphics[width=\textwidth]{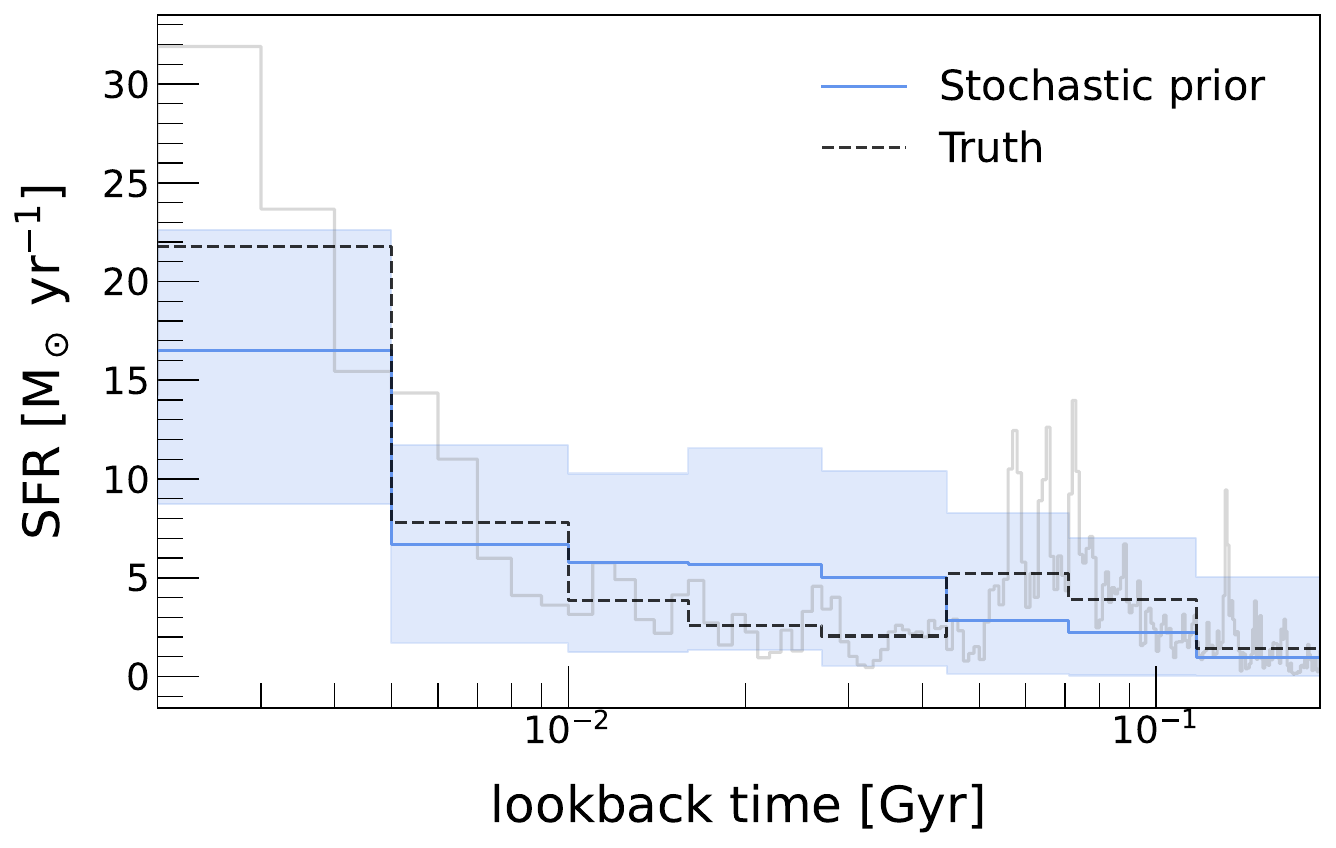}
    \end{subfigure}
    \caption{Example {\typewriter Prospector} fit for one mock high-redshift galaxy. The top panel shows the photometric data (black), along with the median model fit from the stochastic prior (blue), for this example galaxy. The bottom panel shows the posterior SFHs obtained from the stochastic prior in blue, with the truth over-plotted with a dashed black line. The shaded regions show the 16th$-$84th percentile. The stochastic prior provides a good fit to the photometric data and is able to recover the true SFH reasonably well.}
    \label{fig:example highz fit}
\end{figure}

We show an example of a mock high-$z$ galaxy and its associated {\typewriter Prospector} fits using both the stochastic and continuity priors in Figure \ref{fig:example highz fit}. The stochastic model fits the photometric data well and is able to recover the input SFH to within $1\sigma$. We verify that this is the case across the mock sample galaxies. The subsequent sections delve further into the recovery of high-redshift galaxy properties.

\subsubsection{Stellar population parameters}
\label{sec:highz recovery sps}

\begin{figure}
    \centering
    \includegraphics[width=0.48\textwidth]{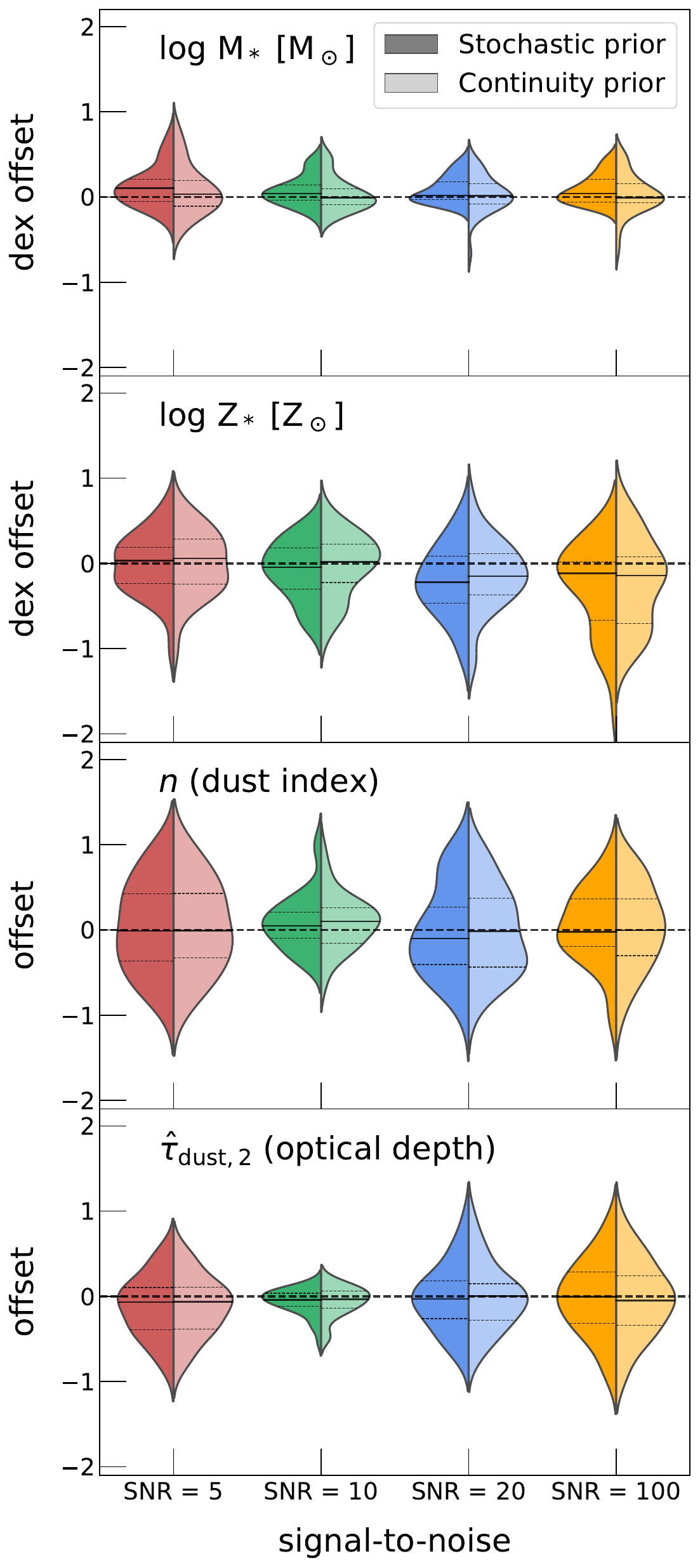}
    \caption{Histograms showing the dex offset between the median posterior recovered stellar mass, metallicity, dust index, and diffuse dust optical depth values and the true values for 31 mock galaxies in each of our four SNR regimes. For each regime, the dark shaded histogram (left half of each pair) corresponds to the stochastic SFH prior, and the light shaded histogram (right half) corresponds to the continuity prior. In all cases, these quantities are well-recovered with the stochastic prior. The bias and scatter in the recovery performance of the stochastic prior is similar to that of the continuity prior for all SNRs.}
    \label{fig:highz compare sps params}
\end{figure}

Figure \ref{fig:highz compare sps params} shows histograms of the offset between the median posterior stellar mass (log $\Mstar$), metallicity ($\log \mathrm{Z_*}$), dust index ($n$), and diffuse dust optical depth ($\hat{\tau}_{\rm{dust,2}}$) values and the input values for the mock high-$z$ galaxies in each of our four SNR regimes. The left (dark shaded) half of each histogram represents the results from fits using the stochastic SFH prior, and the right (light shaded) half represents those from the continuity prior.

The offset between the recovered and input stellar masses of the galaxies is greatest when the SNR of the photometry is lowest (SNR = 5), at 0.10 dex. In the remaining three SNR regimes, this bias is $\sim 0.03$ dex. The scatter is the same across all SNRs, at 0.17 dex. The bias comparable, albiet slightly smaller, when the continuity prior is used, averaging to 0.02 dex across all SNR samples. The scatter is also comparable to that of the stochastic prior fits, at $\sim 0.19$ dex for all four SNR samples.

The stellar metallicities are recovered to within $\lesssim 0.04$ dex of their true values in the SNR = 5 and 10 regimes with the stochastic prior. However, this bias is larger in the higher SNR regimes. When the SNR of the data is 20, the estimated metallicities are biased low by 0.22 dex, and when SNR = 100, they are biased low by 0.11 dex. The scatters in the SNR = 5, 10, and 20 samples are $\sim 0.35$ dex, and the scatter in the SNR = 100 sample is 0.58 dex. The continuity prior performs similarly in measuring the stellar metallicities of our mock high-$z$ galaxies. The systematic bias is minimal in the SNR = 5 and 10 samples, at 0.06 dex and 0.02 dex, respectively. This offset is larger in the SNR = 20 and 100 samples, where the recovered metallicites are biased low by 0.15 dex in both cases. The scatters are nearly identical to those obtained with the stochastic prior -- $\sim 0.37$ dex in the SNR = 5, 10, and 20 regimes, and 0.61 dex when SNR = 100.

Finally, the dust indices ($n$) and the diffuse dust optical depths ($\hat{\tau}_{\rm{dust,2}}$) are well-recovered across all SNRs with both the stochastic and continuity priors. The average bias in the estimated dust indices is 0.04 dex with the stochastic model, compared to 0.03 dex with the continuity model. The average scatter obtained from both models is 0.47 dex. Similarly, the average offset between the measured and input $\hat{\tau_{\rm{dust,2}}}$ values is 0.03 dex for the fits using both the stochastic and continuity priors, and the scatter is 0.31 dex. There is no significant difference in recovery between the different SNR regimes.

Overall, we find that we are able to recover the basic properties of these mock high-redshift galaxies with minimal bias and reasonable scatter when {\typewriter Prospector} is configured with the stochastic SFH prior. Furthermore, there are no major differences between the stochastic and continuity priors in how well these parameters are recovered, regardless of SNR.

\subsubsection{Star formation histories}
\label{sec:highz recovery sfh}

\begin{figure*}
    \centering
    \includegraphics[width=0.9\textwidth]{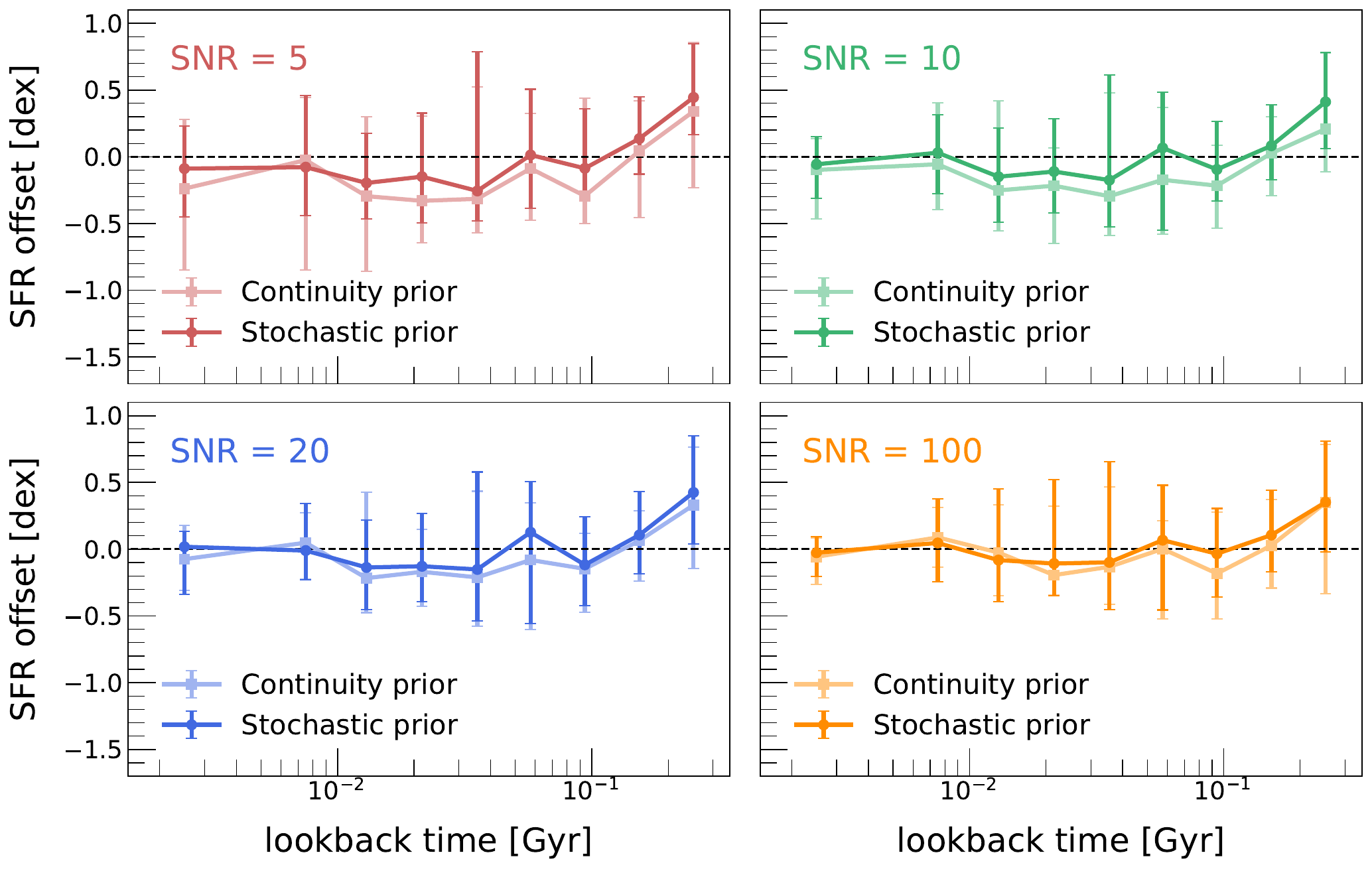}
    \caption{The dex offsets between the median posterior recovered SFRs and their true values in each time bin for the mock high-redshift galaxies with varying SNRs. The dark-coloured points correspond to the stochastic SFH prior, and the light-coloured points correspond to the continuity prior. Error bars show the 16th$-$84th percentiles. The SFRs in each time are tend to be biased low at early lookback times in both the stochastic and continuity models, but this bias is stronger with the continuity prior. Additionally, the SFHs are recovered more precisely with the stochastic prior than the continuity prior.}
    \label{fig:highz SFH recovery}
\end{figure*}

The differences between the median posterior SFRs and their input values in each SFH time bin is shown Figure \ref{fig:highz SFH recovery}. We compare the recovery performance between varying photometric SNRs, as well as between the stochastic prior (marked by the dark-coloured points) and the continuity prior (light-coloured points).

We find that the recovered SFHs obtained from the stochastic prior fits are generally consistent with the true SFHs of our mock high-$z$ galaxies. However, the SFRs in the first five bins tends to be biased low by $0.12$ dex on average in the SNR = 5 sample, $0.06$ dex in the SNR = 10 sample, $0.05$ dex in the SNR = 20 sample, and $0.03$ dex in the SNR = 100 sample. Unsurprisingly, the SFHs become more accurately estimated as the SNR of the photometric data increases. 

This tendency to underestimate the SFRs in the most recent time bins is also present in the fits with the continuity SFH prior. In fact, the biases are more significant when the continuity prior is used. We find that the average offset across the first five bins is $-0.22$ dex in the SNR = 5 sample, $-0.18$ dex in the SNR = 10 sample, $-0.12$ dex in the SNR = 20 sample, and $-0.05$ dex in the SNR = 100 sample. The stochastic prior, then, is $\sim 2\times$ more accurate than the continuity prior in these time bins, especially when the SNR of the data is lower.

Another key feature seen in Figure \ref{fig:highz SFH recovery} is the offset between the estimated and true SFRs in the last time bin. The modelled SFRs obtained from the stochastic prior fits are biased high by $\sim 0.4$ dex in the final bin across all four SNRs. This overestimation of the SFR at late times is not unique to the stochastic prior. When the continuity prior is used, the recovered SFRs are overestimated by $\sim 0.3$ dex in the final bin. This behavior, however, is not unexpected. First of all, it is inherently more difficult to estimate the SFHs of galaxies at large lookback times because of outshining. Secondly, most of the mock galaxies in our high-$z$ sample experience very low, near-zero SFRs in this time bin. The combination of these two factors makes accurately reproducing the SFRs in the final time bin a challenge. Indeed, both priors struggle to reproduce very low ongoing star-formation activity occurring at these large lookback times.

\textcolor{black}{Additionally, we investigate how the scatter in the recovered SFRs depends on the SNR of the data. When the stochastic prior is used, the scatter in the most recent SFH bin decreases from $\sim0.3$ dex to $\sim0.1$ dex when the SNR of the photometric data is increased from 5 to 100. The scatter also decreases from $\sim0.5$ dex to $\sim0.4$ dex in the second-most recent bin going from an SNR of 5 to 100. In the remaining bins however, there is no significant evolution in the recovery scatter with SNR. This is likely because (i) the further back in time one goes in a galaxy’s SFH, the more difficult it becomes to constrain the SFR due to outshining, and (ii) detailed constraints on SFHs are really only possible when fitting both photometry and spectroscopy -- spectroscopy is needed to break the degeneracy between the SFH, dust attenuation, and metallicity (e.g. \citealt{Leja2017, Tacchella2022Halo7D}). Thus, increasing the SNR of the photometry alone will not significantly improve the SED-fitting constraints in this regime.} The average scatter across these time bins is comparable between the stochastic and continuity priors, at $\sim 0.45$ dex and $\sim 0.47$ dex, respectively.

In general, we find that the stochastic prior performs similarly to the continuity prior in modelling the SFHs of the high-$z$ mock galaxy sample. Both priors are able to approximate the SFHs to within an average of $\sim 0.1$ dex in all four SNR regimes. The difference between the two priors is most evident in the lower SNR samples (SNR = 5 and 10), where the stochastic prior is twice as accurate than the continuity prior in the time bins at early lookback times.

\subsubsection{Recent SFRs and ages}
\label{sec:highz recovery higher order}

\begin{figure*}
    \centering
    \includegraphics[width=0.9\textwidth]{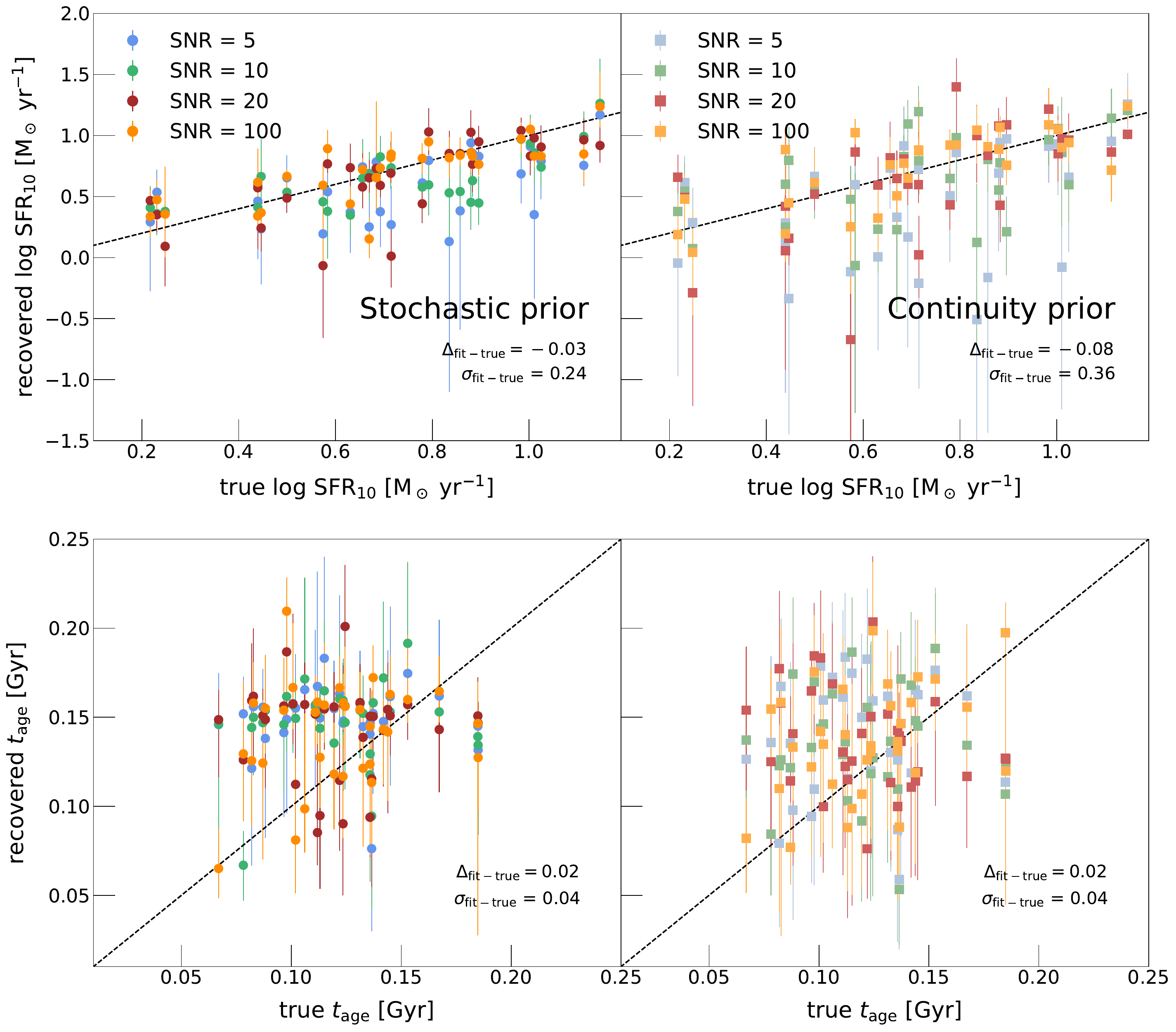}
    \caption{Recovered vs. true log SFR$_{10}$ (top) and mass-weighted age ($t_{\rm{age}}$; bottom) of the high-redshift mock galaxy sample, fit with the stochastic and continuity SFH models. The perfect, one-to-one recovery scenario is denoted with a black dashed line. The stochastic SFH model recovers both the recent SFRs with reasonable scatter and bias, and provides a more precise estimate of SFR$_{10}$ than the continuity prior. Both the stochastic and continuity priors perform similarly in recovering the mass-weighted ages of the mock galaxies. The estimated values are typically consistent with the truth to within $1\sigma$. However, the errors are quite significant in both cases.}
    \label{fig:highz sfr10 mwa recovery}
\end{figure*}

Lastly, we explore the recovery of the recent SFRs and mass-weighted ages by the stochastic SFH prior. The top panel of Figure \ref{fig:highz sfr10 mwa recovery} shows the recovered vs. true log SFRs averaged over the most recent 10 Myr (log SFR$_{10}$) for the mock galaxies, with points colour-coded by SNR regime; the bottom panel shows the recovered vs. true mass-weighted age ($t_{\rm{age}}$). The left column shows the results from the stochastic prior, and the right column shows the continuity prior for comparison.

The recent SFRs are recovered accurately for the high-$z$ mock galaxies with the stochastic model, with 0.24 dex of scatter and an offset of $-0.03$ dex, on average. The stochastic prior is both more accurate and more precise in estimating the SFR$_{10}$ than the continuity prior, which recovers SFR$_{10}$ with an average offset of $-0.08$ dex and scatter of 0.36 dex. This improvement is most evident in the SNR = 5 sample. When the SNR = 5 galaxies are fit using the continuity prior, the posterior SFR$_{10}$ values are biased low by 0.24 dex with a scatter of 0.44 dex. On the other hand, when the stochastic prior is used, the recovered SFR$_{10}$ values are only offset by $-0.09$ dex with a scatter of 0.29.

The mass-weighted ages are recovered similarly well with both the stochastic and continuity priors. Both models measure $t_{\rm{age}}$ values which are biased high by $\sim 20$ Myr with $\sim 40$ Myr of scatter. The tendency to overestimate the ages indicates that both models infer too much mass formed at large lookback times. This indeed is the case, given that we find the posterior SFRs in the final time bin are systematically biased high, regardless of which prior is used (see Figure \ref{fig:highz SFH recovery}).

Additionally, the error bars obtained from both priors are fairly large, reaching $\sim 50 - 100$ Myr in some cases. This primarily shows that the current observational data cannot give us tight constraints on the ages of galaxies at these high redshifts. Additionally, it is difficult to break the dust-age degeneracy with photometry alone. We find that the attenuation law is not well-constrained with our mock data. Being able to accurately model the UV attenuation is key to measuring accurate ages, which highlights the important role that joint spectroscopic and photometric information plays for galaxies in the high-redshift universe.

\section{Discussion}
\label{sec:discussion}

We discuss here the implications and limitations of the stochastic SFH model developed in this paper. We highlight ways in which the model can be improved, as well as avenues for future work.

\subsection{Model limitations \& caveats}

Our mock tests show that the stochastic SFH model is well-suited to recover the basic properties of star-forming galaxies at a redshift range of $z \sim 1 - 8$, as well as higher-order quantities such as recent SFR (SFR$_{100}$ or SFR$_{10}$) and mass-weighted age ($t_{\rm{age}}$). It can also provide reasonable constraints on the long-timescale PSD parameters of such galaxies. On the other hand, the short-term, dynamical component of the PSD is much more difficult to constrain, since extremely high-resolution data is necessary to tease out information about galaxy processes that occur on very short timescales.

It is important to note that the mock observations used in this work consist of 1) high-S/N ($\sim 27 \Angstrom^{-1}$) spectra + photometry of massive galaxies at redshift $z = 0.7$ intended to be analogous to the galaxies of the LEGA-C survey, and 2) photometry spanning a range of SNRs for galaxies at $z = 8$ analogous to observations from NIRCam on JWST. The priors placed on the PSD parameters in each case were also specifically chosen to fit massive galaxies at intermediate redshift and bursty galaxies at high redshift, respectively. However, the stochastic prior can be easily adjusted to accommodate wide range of galaxy types (see Section \ref{sec:prior flexibility}). Thus, in principle, the performance of the stochastic SFH model seen in this work should extend to data obtained from a wide range of surveys and instruments.

Additionally, the generation and fitting of our mock galaxies were done with FSPS as a backbone, configured with the MIST isochrones and MILES spectral library. A different choice of isochrones and/or spectral library may alter the stochastic model's performance.

\subsection{Future work}

The stochastic SFH prior presented in this work represents a first step towards a SFH model with the ability to constrain the power spectra of galaxy SFHs and extract key information about the processes and timescales over which galaxies built up their stellar masses. However, many avenues for future work remain.

When the PSD parameters remain free, the SED-fitting model adopts a hierarchical structure. To explore the full parameter space, we must first sample the space of the PSD parameters. Each iteration of the PSD parameters in turn varies the prior distribution of the log SFR ratios. Thus, the top level of the hierarchical model constrains the PSD parameters. The base level delivers the usual constraints on the SFH of the galaxy, along with its various stellar population parameters. In this model, the stellar population parameters and PSD parameters of a galaxy are constrained simultaneously. However, the model implemented in this work is not a true Bayesian hierarchical model. Rather than fitting for the PSD parameters of a sample of galaxies collectively, we fit each galaxy individually and merge their posteriors, as hierarchical modelling is currently not possible with {\typewriter Prospector}. 

The next step towards improving the stochastic SFH model is to use Bayesian hierarchical modelling to obtain PSD constraints for \textit{populations} of galaxies. In a hierarchical model, a sample of $n$ galaxies would be fit at once, each with their own physical parameters (e.g. redshift, stellar mass, metallicity), but also a set of shared, population-level parameters -- namely, the PSD parameters. This will allow us to take into account the differences and similarities among a galaxy sample, and pool star-formation information across the sample to learn about the PSD of the entire group of galaxies. While this would require some restructuring in an SED-fitting code like {\typewriter Prospector}, building a hierarchical model could be accomplished more easily in another code, such as {\typewriter Dense Basis} \citep{Iyer2017, Iyer2021}.

Additionally, while a flat baseline SFH is a reasonable null expectation, it is inconsistent with what has been observed in real galaxies. Instead, on average, galaxies experience a mass-dependent rise and fall in their SFHs \citep[e.g.][]{Leitner2012, Pacifici2016}. Galaxies tend to have rising SFHs at early times and falling SFHs at late times, as well as as a mass dependence, and more massive systems tend to have formed earlier and over a shorter period of time (``downsizing''). In the future, we can take these effects into account by adjusting the baseline SFH assumption to match the cosmic star-formation rate densities expected at the lookback times associated with each SFH bin, and shifting the time at which star formation begins, similar to what has been done in \citet{Wang2023}.

\section{Conclusions}
\label{sec:conclusion}

In this paper, we construct a physically-motivated prior. which is derived from the power spectral density (PSD) formalism developed in \cite{CT2019}, \cite{TFC2020}, \cite{Iyer2020}, and \citet{Iyer2022}, to fit non-parametric SFHs in the SED-modelling code {\typewriter Prospector} \citep{JohnsonLeja2017, Johnson2021}. We model a galaxy's SFH as a stochastic time series, which is defined via the PSD and its Fourier pair, the auto-covariance function (ACF). The PSD and ACF are determined by parameters relating to large-scale gas processes in the galaxy ($\sigreg$, the overall variability in gas inflow into the galaxy; $\taudyn$, the characteristic timescale for gas cycling in equilibrium; and $\tauin$, the characterstic timescale for gas), as well as the overall variability and timescale associated with short-term dynamical processes ($\sigdyn$ and $\taudyn$, respectively). 

This stochastic SFH prior parametrizes the log of the SFR ratio between adjacent bins in the SFH with a multivariate-normal distribution that has a zero mean vector and a covariance matrix given by the ACF (as a function of $\sigreg$, $\taueq$, $\tauin$, $\sigdyn$, and $\taudyn$, i.e. the ``PSD parameters''). With the stochastic model, we are able to simultaneously constrain the SFH of a galaxy along with the PSD parameters, which contain key information about the timescales important to the growth history of the galaxy.

We create mock galaxy observations with SFHs illustrative of four different burstiness regimes at intermediate redshift (modelled after the LEGA-C survey), as well as a sample of high-$z$ galaxies (modelled after NIRCam observations). We fit these mock observations using {\typewriter Prospector} and test the ability of the stochastic SFH prior to recover the input properties of each mock galaxy, including the SFH and PSD parameters. The performance of the stochastic prior is then compared against that of the commonly-used continuity prior.

We find that in both the $z = 0.7$ and $z = 8$ mock samples, the stochastic prior is able to recover basic stellar population parameters (stellar mass, stellar metallicity, dust index, etc.) to a high level of accuracy and with very minimal bias, comparable to the performance of the continuity prior. Higher-order quantities (i.e. the recent SFRs and the mass-weighted ages) are also well-recovered, and with significantly less uncertainty than the continuity prior. \textcolor{black}{The largest impact of the stochastic prior can be seen in the recovery of the SFHs of galaxies. The physically-motivated information about star-formation burstiness that is baked into the stochastic prior leads to an improvement in accuracy, and a larger improvement in precision, when estimating galaxies' SFHs at all redshifts. This makes the stochastic model an especially powerful tool in estimating the SFHs of high-$z$ systems, where the inferred SFHs are heavily dependent on the choice of prior \citep[e.g.][]{Tacchella2022-bursty-sfh-prior, Whitler2023}.}

Furthermore, the model reliably estimates $\sigreg$, the long-term variability PSD term, for all four burstiness regimes. It is difficult to obtain constraints the equilibrium timescale ($\taueq$) and the dynamical component PSD parameters, $\sigdyn$ and $\taudyn$, which describe processes occurring on $\sim10$ Myr timescales. We can improve the model's ability to recover $\taueq$ by a) fixing $\sigreg$, or b) adjusting its prior towards shorter timescales. The dynamical component PSD parameters do not affect the SFH prior distribution significantly, and as such, will always be difficult to measure. 

Overall, the tests performed in this work demonstrate that the stochastic SFH prior provides a powerful avenue to probe fluctuations in the star-formation activity of galaxies and learn about how this variability evolves over cosmic time. The stochastic prior can accurately recover the SFHs and basic stellar population parameters of a wide range galaxies, ensuring that we can trust the results obtained from fitting real observational data with the stochastic model. Tests of this nature, which should be tailored to target the science questions, parameters, and instruments of interest, are highly recommended for future studies as well. 

We publicly provide the stochastic SFH prior developed in this work as a part of {\typewriter Prospector} for the community to use.

\section*{Acknowledgements}

We thank the anonymous referee for their helpful comments that strengthened this work. JTW was supported by the Churchill Scholarship through the Winston Churchill Foundation. JSS acknowledges the support of the Natural Sciences and Engineering Research Council of Canada (NSERC), Discovery Grant RGPIN-2023-04849. This work was performed using resources provided by the Cambridge Service for Data Driven Discovery (CSD3) operated by the University of Cambridge Research Computing Service (\url{www.csd3.cam.ac.uk}), provided by Dell EMC and Intel using Tier-2 funding from the Engineering and Physical Sciences Research Council (capital grant EP/T022159/1), and DiRAC funding from the Science and Technology Facilities Council (\url{www.dirac.ac.uk}).

\textit{Software}: {\typewriter astropy} \citep{Astropy2022}, {\typewriter dynesty} \citep{Speagle2020}, {\typewriter GP-SFH} \citep{Iyer2022}, {\typewriter matplotlib} \citep{Matplotlib2007}, {\typewriter numpy} \citep{Numpy2020}, {\typewriter Prospector} \citep{Johnson2021}, {\typewriter seaborn} \citep{Seaborn2021}

\section*{Data Availability}

The LEGA-C DR3 data used to generate spectra and photometry for our sample of mock $z = 0.7$ galaxies is publicly available at \url{https://users.ugent.be/~avdrwel/research.html#legac}.
The SPHINX data used to generate photometry for our sample of mock high-redshift galaxies can be found at \url{https://github.com/HarleyKatz/SPHINX-20-data}.



\bibliographystyle{mnras}
\bibliography{stochastic_prior} 

\begin{thebibliography}{}
\makeatletter
\relax
\def\mn@urlcharsother{\let\do\@makeother \do\$\do\&\do\#\do\^\do\_\do\%\do\~}
\def\mn@doi{\begingroup\mn@urlcharsother \@ifnextchar [ {\mn@doi@}
  {\mn@doi@[]}}
\def\mn@doi@[#1]#2{\def\@tempa{#1}\ifx\@tempa\@empty \href
  {http://dx.doi.org/#2} {doi:#2}\else \href {http://dx.doi.org/#2} {#1}\fi
  \endgroup}
\def\mn@eprint#1#2{\mn@eprint@#1:#2::\@nil}
\def\mn@eprint@arXiv#1{\href {http://arxiv.org/abs/#1} {{\tt arXiv:#1}}}
\def\mn@eprint@dblp#1{\href {http://dblp.uni-trier.de/rec/bibtex/#1.xml}
  {dblp:#1}}
\def\mn@eprint@#1:#2:#3:#4\@nil{\def\@tempa {#1}\def\@tempb {#2}\def\@tempc
  {#3}\ifx \@tempc \@empty \let \@tempc \@tempb \let \@tempb \@tempa \fi \ifx
  \@tempb \@empty \def\@tempb {arXiv}\fi \@ifundefined
  {mn@eprint@\@tempb}{\@tempb:\@tempc}{\expandafter \expandafter \csname
  mn@eprint@\@tempb\endcsname \expandafter{\@tempc}}}

\bibitem[\protect\citeauthoryear{{Aigrain} \& {Foreman-Mackey}}{{Aigrain} \&
  {Foreman-Mackey}}{2022}]{Aigrain2022}
{Aigrain} S.,  {Foreman-Mackey} D.,  2022, \mn@doi [arXiv e-prints]
  {10.48550/arXiv.2209.08940}, \href
  {https://ui.adsabs.harvard.edu/abs/2022arXiv220908940A} {p. arXiv:2209.08940}

\bibitem[\protect\citeauthoryear{{Astropy Collaboration} et~al.,}{{Astropy
  Collaboration} et~al.}{2022}]{Astropy2022}
{Astropy Collaboration} et~al., 2022, \mn@doi [\apj]
  {10.3847/1538-4357/ac7c74}, \href
  {https://ui.adsabs.harvard.edu/abs/2022ApJ...935..167A} {935, 167}

\bibitem[\protect\citeauthoryear{{Barbaro} \& {Poggianti}}{{Barbaro} \&
  {Poggianti}}{1997}]{Barbaro1997}
{Barbaro} G.,  {Poggianti} B.~M.,  1997, \mn@doi [\aap]
  {10.48550/arXiv.astro-ph/9702129}, \href
  {https://ui.adsabs.harvard.edu/abs/1997A&A...324..490B} {324, 490}

\bibitem[\protect\citeauthoryear{{Behroozi}, {Wechsler}, {Hearin}  \&
  {Conroy}}{{Behroozi} et~al.}{2019}]{Behroozi2019}
{Behroozi} P.,  {Wechsler} R.~H.,  {Hearin} A.~P.,   {Conroy} C.,  2019,
  \mn@doi [\mnras] {10.1093/mnras/stz1182}, \href
  {https://ui.adsabs.harvard.edu/abs/2019MNRAS.488.3143B} {488, 3143}

\bibitem[\protect\citeauthoryear{{Boogaard} et~al.,}{{Boogaard}
  et~al.}{2018}]{Boogaard2018}
{Boogaard} L.~A.,  et~al., 2018, \mn@doi [\aap] {10.1051/0004-6361/201833136},
  \href {https://ui.adsabs.harvard.edu/abs/2018A&A...619A..27B} {619, A27}

\bibitem[\protect\citeauthoryear{{Bouch{\'e}} et~al.,}{{Bouch{\'e}}
  et~al.}{2010}]{Bouche2010}
{Bouch{\'e}} N.,  et~al., 2010, \mn@doi [\apj] {10.1088/0004-637X/718/2/1001},
  \href {https://ui.adsabs.harvard.edu/abs/2010ApJ...718.1001B} {718, 1001}

\bibitem[\protect\citeauthoryear{{Brinchmann}, {Charlot}, {White}, {Tremonti},
  {Kauffmann}, {Heckman}  \& {Brinkmann}}{{Brinchmann}
  et~al.}{2004}]{Brinchmann2004}
{Brinchmann} J.,  {Charlot} S.,  {White} S.~D.~M.,  {Tremonti} C.,  {Kauffmann}
  G.,  {Heckman} T.,   {Brinkmann} J.,  2004, \mn@doi [\mnras]
  {10.1111/j.1365-2966.2004.07881.x}, \href
  {https://ui.adsabs.harvard.edu/abs/2004MNRAS.351.1151B} {351, 1151}

\bibitem[\protect\citeauthoryear{{Caplar} \& {Tacchella}}{{Caplar} \&
  {Tacchella}}{2019}]{CT2019}
{Caplar} N.,  {Tacchella} S.,  2019, \mn@doi [\mnras] {10.1093/mnras/stz1449},
  \href {https://ui.adsabs.harvard.edu/abs/2019MNRAS.487.3845C} {487, 3845}

\bibitem[\protect\citeauthoryear{{Charlot} \& {Fall}}{{Charlot} \&
  {Fall}}{2000}]{CharlotFall2000}
{Charlot} S.,  {Fall} S.~M.,  2000, \mn@doi [\apj] {10.1086/309250}, \href
  {https://ui.adsabs.harvard.edu/abs/2000ApJ...539..718C} {539, 718}

\bibitem[\protect\citeauthoryear{{Choi}, {Dotter}, {Conroy}, {Cantiello},
  {Paxton}  \& {Johnson}}{{Choi} et~al.}{2016}]{Choi2016}
{Choi} J.,  {Dotter} A.,  {Conroy} C.,  {Cantiello} M.,  {Paxton} B.,
  {Johnson} B.~D.,  2016, \mn@doi [\apj] {10.3847/0004-637X/823/2/102}, \href
  {https://ui.adsabs.harvard.edu/abs/2016ApJ...823..102C} {823, 102}

\bibitem[\protect\citeauthoryear{{Conroy} \& {Gunn}}{{Conroy} \&
  {Gunn}}{2010}]{ConroyGunn2010}
{Conroy} C.,  {Gunn} J.~E.,  2010, \mn@doi [\apj]
  {10.1088/0004-637X/712/2/833}, \href
  {https://ui.adsabs.harvard.edu/abs/2010ApJ...712..833C} {712, 833}

\bibitem[\protect\citeauthoryear{{Conroy}, {Gunn}  \& {White}}{{Conroy}
  et~al.}{2009}]{Conroy2009}
{Conroy} C.,  {Gunn} J.~E.,   {White} M.,  2009, \mn@doi [\apj]
  {10.1088/0004-637X/699/1/486}, \href
  {https://ui.adsabs.harvard.edu/abs/2009ApJ...699..486C} {699, 486}

\bibitem[\protect\citeauthoryear{{Daddi} et~al.,}{{Daddi}
  et~al.}{2007}]{Daddi2007}
{Daddi} E.,  et~al., 2007, \mn@doi [\apj] {10.1086/521818}, \href
  {https://ui.adsabs.harvard.edu/abs/2007ApJ...670..156D} {670, 156}

\bibitem[\protect\citeauthoryear{{Dotter}}{{Dotter}}{2016}]{Dotter2016}
{Dotter} A.,  2016, \mn@doi [\apjs] {10.3847/0067-0049/222/1/8}, \href
  {https://ui.adsabs.harvard.edu/abs/2016ApJS..222....8D} {222, 8}

\bibitem[\protect\citeauthoryear{{Draine} \& {Li}}{{Draine} \&
  {Li}}{2007}]{DraineLi2007}
{Draine} B.~T.,  {Li} A.,  2007, \mn@doi [\apj] {10.1086/511055}, \href
  {https://ui.adsabs.harvard.edu/abs/2007ApJ...657..810D} {657, 810}

\bibitem[\protect\citeauthoryear{{Eisenstein} et~al.,}{{Eisenstein}
  et~al.}{2023}]{JADES2023}
{Eisenstein} D.~J.,  et~al., 2023, \mn@doi [arXiv e-prints]
  {10.48550/arXiv.2306.02465}, \href
  {https://ui.adsabs.harvard.edu/abs/2023arXiv230602465E} {p. arXiv:2306.02465}

\bibitem[\protect\citeauthoryear{{Elbaz} et~al.,}{{Elbaz}
  et~al.}{2007}]{Elbaz2007}
{Elbaz} D.,  et~al., 2007, \mn@doi [\aap] {10.1051/0004-6361:20077525}, \href
  {https://ui.adsabs.harvard.edu/abs/2007A&A...468...33E} {468, 33}

\bibitem[\protect\citeauthoryear{{Falc{\'o}n-Barroso},
  {S{\'a}nchez-Bl{\'a}zquez}, {Vazdekis}, {Ricciardelli}, {Cardiel}, {Cenarro},
  {Gorgas}  \& {Peletier}}{{Falc{\'o}n-Barroso} et~al.}{2011}]{MILES2011}
{Falc{\'o}n-Barroso} J.,  {S{\'a}nchez-Bl{\'a}zquez} P.,  {Vazdekis} A.,
  {Ricciardelli} E.,  {Cardiel} N.,  {Cenarro} A.~J.,  {Gorgas} J.,
  {Peletier} R.~F.,  2011, \mn@doi [\aap] {10.1051/0004-6361/201116842}, \href
  {https://ui.adsabs.harvard.edu/abs/2011A&A...532A..95F} {532, A95}

\bibitem[\protect\citeauthoryear{{Flores Vel{\'a}zquez} et~al.,}{{Flores
  Vel{\'a}zquez} et~al.}{2021}]{Flores2021}
{Flores Vel{\'a}zquez} J.~A.,  et~al., 2021, \mn@doi [\mnras]
  {10.1093/mnras/staa3893}, \href
  {https://ui.adsabs.harvard.edu/abs/2021MNRAS.501.4812F} {501, 4812}

\bibitem[\protect\citeauthoryear{{Fumagalli} et~al.,}{{Fumagalli}
  et~al.}{2014}]{Fumagalli2014}
{Fumagalli} M.,  et~al., 2014, \mn@doi [\apj] {10.1088/0004-637X/796/1/35},
  \href {https://ui.adsabs.harvard.edu/abs/2014ApJ...796...35F} {796, 35}

\bibitem[\protect\citeauthoryear{{Gunn} \& {Gott}}{{Gunn} \&
  {Gott}}{1972}]{GunnGott1972}
{Gunn} J.~E.,  {Gott} J.~Richard I.,  1972, \mn@doi [\apj] {10.1086/151605},
  \href {https://ui.adsabs.harvard.edu/abs/1972ApJ...176....1G} {176, 1}

\bibitem[\protect\citeauthoryear{Harris et~al.,}{Harris
  et~al.}{2020}]{Numpy2020}
Harris C.~R.,  et~al., 2020, \mn@doi [Nature] {10.1038/s41586-020-2649-2}, 585,
  357

\bibitem[\protect\citeauthoryear{{Hernquist}}{{Hernquist}}{1989}]{Hernquist1989}
{Hernquist} L.,  1989, \mn@doi [\nat] {10.1038/340687a0}, \href
  {https://ui.adsabs.harvard.edu/abs/1989Natur.340..687H} {340, 687}

\bibitem[\protect\citeauthoryear{Hunter}{Hunter}{2007}]{Matplotlib2007}
Hunter J.~D.,  2007, \mn@doi [Computing in Science \& Engineering]
  {10.1109/MCSE.2007.55}, 9, 90

\bibitem[\protect\citeauthoryear{{Iyer} \& {Gawiser}}{{Iyer} \&
  {Gawiser}}{2017}]{Iyer2017}
{Iyer} K.,  {Gawiser} E.,  2017, \mn@doi [\apj] {10.3847/1538-4357/aa63f0},
  \href {https://ui.adsabs.harvard.edu/abs/2017ApJ...838..127I} {838, 127}

\bibitem[\protect\citeauthoryear{{Iyer} et~al.,}{{Iyer}
  et~al.}{2020}]{Iyer2020}
{Iyer} K.~G.,  et~al., 2020, \mn@doi [\mnras] {10.1093/mnras/staa2150}, \href
  {https://ui.adsabs.harvard.edu/abs/2020MNRAS.498..430I} {498, 430}

\bibitem[\protect\citeauthoryear{{Iyer}, {Gawiser}, {Faber}, {Ferguson},
  {Kartaltepe}, {Koekemoer}, {Pacifici}  \& {Somerville}}{{Iyer}
  et~al.}{2021}]{Iyer2021}
{Iyer} K.~G.,  {Gawiser} E.,  {Faber} S.~M.,  {Ferguson} H.~C.,  {Kartaltepe}
  J.,  {Koekemoer} A.~M.,  {Pacifici} C.,   {Somerville} R.~S.,  2021,
  {dense\_basis: Dense Basis SED fitting}, Astrophysics Source Code Library,
  record ascl:2104.015 (\mn@eprint {ascl} {2104.015})

\bibitem[\protect\citeauthoryear{{Iyer}, {Speagle}, {Caplar}, {Forbes},
  {Gawiser}, {Leja}  \& {Tacchella}}{{Iyer} et~al.}{2024}]{Iyer2022}
{Iyer} K.~G.,  {Speagle} J.~S.,  {Caplar} N.,  {Forbes} J.~C.,  {Gawiser} E.,
  {Leja} J.,   {Tacchella} S.,  2024, \mn@doi [\apj]
  {10.3847/1538-4357/acff64}, \href
  {https://ui.adsabs.harvard.edu/abs/2024ApJ...961...53I} {961, 53}

\bibitem[\protect\citeauthoryear{{Jin} et~al.,}{{Jin}
  et~al.}{2018}]{Superdeblended}
{Jin} S.,  et~al., 2018, \mn@doi [\apj] {10.3847/1538-4357/aad4af}, \href
  {https://ui.adsabs.harvard.edu/abs/2018ApJ...864...56J} {864, 56}

\bibitem[\protect\citeauthoryear{{Johnson} \& {Leja}}{{Johnson} \&
  {Leja}}{2017}]{JohnsonLeja2017}
{Johnson} B.,  {Leja} J.,  2017, {Bd-J/Prospector: Initial Release}, Zenodo,
  \mn@doi{10.5281/zenodo.1116491}

\bibitem[\protect\citeauthoryear{{Johnson}, {Leja}, {Conroy}  \&
  {Speagle}}{{Johnson} et~al.}{2021}]{Johnson2021}
{Johnson} B.~D.,  {Leja} J.,  {Conroy} C.,   {Speagle} J.~S.,  2021, \mn@doi
  [\apjs] {10.3847/1538-4365/abef67}, \href
  {https://ui.adsabs.harvard.edu/abs/2021ApJS..254...22J} {254, 22}

\bibitem[\protect\citeauthoryear{{Katz} et~al.,}{{Katz}
  et~al.}{2023}]{Katz2023}
{Katz} H.,  et~al., 2023, \mn@doi [arXiv e-prints] {10.48550/arXiv.2309.03269},
  \href {https://ui.adsabs.harvard.edu/abs/2023arXiv230903269K} {p.
  arXiv:2309.03269}

\bibitem[\protect\citeauthoryear{{Kauffmann} et~al.,}{{Kauffmann}
  et~al.}{2003}]{Kauffmann2003}
{Kauffmann} G.,  et~al., 2003, \mn@doi [\mnras]
  {10.1046/j.1365-8711.2003.06291.x}, \href
  {https://ui.adsabs.harvard.edu/abs/2003MNRAS.341...33K} {341, 33}

\bibitem[\protect\citeauthoryear{{Kroupa}}{{Kroupa}}{2001}]{Kroupa2001}
{Kroupa} P.,  2001, \mn@doi [\mnras] {10.1046/j.1365-8711.2001.04022.x}, \href
  {https://ui.adsabs.harvard.edu/abs/2001MNRAS.322..231K} {322, 231}

\bibitem[\protect\citeauthoryear{{Krumholz} \& {Kruijssen}}{{Krumholz} \&
  {Kruijssen}}{2015}]{Krumholz2015}
{Krumholz} M.~R.,  {Kruijssen} J.~M.~D.,  2015, \mn@doi [\mnras]
  {10.1093/mnras/stv1670}, \href
  {https://ui.adsabs.harvard.edu/abs/2015MNRAS.453..739K} {453, 739}

\bibitem[\protect\citeauthoryear{{Leitner}}{{Leitner}}{2012}]{Leitner2012}
{Leitner} S.~N.,  2012, \mn@doi [\apj] {10.1088/0004-637X/745/2/149}, \href
  {https://ui.adsabs.harvard.edu/abs/2012ApJ...745..149L} {745, 149}

\bibitem[\protect\citeauthoryear{{Leja}, {Johnson}, {Conroy}, {van Dokkum}  \&
  {Byler}}{{Leja} et~al.}{2017}]{Leja2017}
{Leja} J.,  {Johnson} B.~D.,  {Conroy} C.,  {van Dokkum} P.~G.,   {Byler} N.,
  2017, \mn@doi [\apj] {10.3847/1538-4357/aa5ffe}, \href
  {https://ui.adsabs.harvard.edu/abs/2017ApJ...837..170L} {837, 170}

\bibitem[\protect\citeauthoryear{{Leja}, {Carnall}, {Johnson}, {Conroy}  \&
  {Speagle}}{{Leja} et~al.}{2019}]{Leja2019}
{Leja} J.,  {Carnall} A.~C.,  {Johnson} B.~D.,  {Conroy} C.,   {Speagle} J.~S.,
   2019, \mn@doi [\apj] {10.3847/1538-4357/ab133c}, \href
  {https://ui.adsabs.harvard.edu/abs/2019ApJ...876....3L} {876, 3}

\bibitem[\protect\citeauthoryear{{Leja} et~al.,}{{Leja}
  et~al.}{2022}]{Leja2022}
{Leja} J.,  et~al., 2022, \mn@doi [\apj] {10.3847/1538-4357/ac887d}, \href
  {https://ui.adsabs.harvard.edu/abs/2022ApJ...936..165L} {936, 165}

\bibitem[\protect\citeauthoryear{{Lilly}, {Carollo}, {Pipino}, {Renzini}  \&
  {Peng}}{{Lilly} et~al.}{2013}]{Lilly2013}
{Lilly} S.~J.,  {Carollo} C.~M.,  {Pipino} A.,  {Renzini} A.,   {Peng} Y.,
  2013, \mn@doi [\apj] {10.1088/0004-637X/772/2/119}, \href
  {https://ui.adsabs.harvard.edu/abs/2013ApJ...772..119L} {772, 119}

\bibitem[\protect\citeauthoryear{{Madau}}{{Madau}}{1995}]{Madau1995}
{Madau} P.,  1995, \mn@doi [\apj] {10.1086/175332}, \href
  {https://ui.adsabs.harvard.edu/abs/1995ApJ...441...18M} {441, 18}

\bibitem[\protect\citeauthoryear{{Matthee} \& {Schaye}}{{Matthee} \&
  {Schaye}}{2019}]{Matthee2019}
{Matthee} J.,  {Schaye} J.,  2019, \mn@doi [\mnras] {10.1093/mnras/stz030},
  \href {https://ui.adsabs.harvard.edu/abs/2019MNRAS.484..915M} {484, 915}

\bibitem[\protect\citeauthoryear{{Myers}, {Dame}, {Thaddeus}, {Cohen},
  {Silverberg}, {Dwek}  \& {Hauser}}{{Myers} et~al.}{1986}]{Myers1986}
{Myers} P.~C.,  {Dame} T.~M.,  {Thaddeus} P.,  {Cohen} R.~S.,  {Silverberg}
  R.~F.,  {Dwek} E.,   {Hauser} M.~G.,  1986, \mn@doi [\apj] {10.1086/163909},
  \href {https://ui.adsabs.harvard.edu/abs/1986ApJ...301..398M} {301, 398}

\bibitem[\protect\citeauthoryear{{Naab} \& {Ostriker}}{{Naab} \&
  {Ostriker}}{2017}]{NaabOstriker2017}
{Naab} T.,  {Ostriker} J.~P.,  2017, \mn@doi [\araa]
  {10.1146/annurev-astro-081913-040019}, \href
  {https://ui.adsabs.harvard.edu/abs/2017ARA&A..55...59N} {55, 59}

\bibitem[\protect\citeauthoryear{{Orr}, {Hayward}  \& {Hopkins}}{{Orr}
  et~al.}{2019}]{Orr2019}
{Orr} M.~E.,  {Hayward} C.~C.,   {Hopkins} P.~F.,  2019, \mn@doi [\mnras]
  {10.1093/mnras/stz1156}, \href
  {https://ui.adsabs.harvard.edu/abs/2019MNRAS.486.4724O} {486, 4724}

\bibitem[\protect\citeauthoryear{{Pacifici} et~al.,}{{Pacifici}
  et~al.}{2016}]{Pacifici2016}
{Pacifici} C.,  et~al., 2016, \mn@doi [\apj] {10.3847/0004-637X/832/1/79},
  \href {https://ui.adsabs.harvard.edu/abs/2016ApJ...832...79P} {832, 79}

\bibitem[\protect\citeauthoryear{{Paxton}, {Bildsten}, {Dotter}, {Herwig},
  {Lesaffre}  \& {Timmes}}{{Paxton} et~al.}{2011}]{Paxton2011}
{Paxton} B.,  {Bildsten} L.,  {Dotter} A.,  {Herwig} F.,  {Lesaffre} P.,
  {Timmes} F.,  2011, \mn@doi [\apjs] {10.1088/0067-0049/192/1/3}, \href
  {https://ui.adsabs.harvard.edu/abs/2011ApJS..192....3P} {192, 3}

\bibitem[\protect\citeauthoryear{{Paxton} et~al.,}{{Paxton}
  et~al.}{2013}]{Paxton2013}
{Paxton} B.,  et~al., 2013, \mn@doi [\apjs] {10.1088/0067-0049/208/1/4}, \href
  {https://ui.adsabs.harvard.edu/abs/2013ApJS..208....4P} {208, 4}

\bibitem[\protect\citeauthoryear{{Paxton} et~al.,}{{Paxton}
  et~al.}{2015}]{Paxton2015}
{Paxton} B.,  et~al., 2015, \mn@doi [\apjs] {10.1088/0067-0049/220/1/15}, \href
  {https://ui.adsabs.harvard.edu/abs/2015ApJS..220...15P} {220, 15}

\bibitem[\protect\citeauthoryear{{Rodr{\'\i}guez-Puebla}, {Primack}, {Behroozi}
   \& {Faber}}{{Rodr{\'\i}guez-Puebla} et~al.}{2016}]{Rodriguez2016}
{Rodr{\'\i}guez-Puebla} A.,  {Primack} J.~R.,  {Behroozi} P.,   {Faber} S.~M.,
  2016, \mn@doi [\mnras] {10.1093/mnras/stv2513}, \href
  {https://ui.adsabs.harvard.edu/abs/2016MNRAS.455.2592R} {455, 2592}

\bibitem[\protect\citeauthoryear{{Rosdahl} et~al.,}{{Rosdahl}
  et~al.}{2018}]{Rosdahl2018}
{Rosdahl} J.,  et~al., 2018, \mn@doi [\mnras] {10.1093/mnras/sty1655}, \href
  {https://ui.adsabs.harvard.edu/abs/2018MNRAS.479..994R} {479, 994}

\bibitem[\protect\citeauthoryear{{Rosdahl} et~al.,}{{Rosdahl}
  et~al.}{2022}]{Rosdahl2022}
{Rosdahl} J.,  et~al., 2022, \mn@doi [\mnras] {10.1093/mnras/stac1942}, \href
  {https://ui.adsabs.harvard.edu/abs/2022MNRAS.515.2386R} {515, 2386}

\bibitem[\protect\citeauthoryear{{Scalo} \& {Struck-Marcell}}{{Scalo} \&
  {Struck-Marcell}}{1984}]{Scalo1984}
{Scalo} J.~M.,  {Struck-Marcell} C.,  1984, \mn@doi [\apj] {10.1086/161593},
  \href {https://ui.adsabs.harvard.edu/abs/1984ApJ...276...60S} {276, 60}

\bibitem[\protect\citeauthoryear{{Schreiber} et~al.,}{{Schreiber}
  et~al.}{2015}]{Schreiber2015}
{Schreiber} C.,  et~al., 2015, \mn@doi [\aap] {10.1051/0004-6361/201425017},
  \href {https://ui.adsabs.harvard.edu/abs/2015A&A...575A..74S} {575, A74}

\bibitem[\protect\citeauthoryear{{Scoville} \& {Good}}{{Scoville} \&
  {Good}}{1989}]{ScovilleGood1989}
{Scoville} N.~Z.,  {Good} J.~C.,  1989, \mn@doi [\apj] {10.1086/167283}, \href
  {https://ui.adsabs.harvard.edu/abs/1989ApJ...339..149S} {339, 149}

\bibitem[\protect\citeauthoryear{{Shin}, {Tacchella}, {Kim}, {Iyer}  \&
  {Semenov}}{{Shin} et~al.}{2023}]{Shin2023}
{Shin} E.-j.,  {Tacchella} S.,  {Kim} J.-h.,  {Iyer} K.~G.,   {Semenov} V.~A.,
  2023, \mn@doi [\apj] {10.3847/1538-4357/acc251}, \href
  {https://ui.adsabs.harvard.edu/abs/2023ApJ...947...61S} {947, 61}

\bibitem[\protect\citeauthoryear{{Shivaei} et~al.,}{{Shivaei}
  et~al.}{2018}]{Shivaei2018}
{Shivaei} I.,  et~al., 2018, \mn@doi [\apj] {10.3847/1538-4357/aaad62}, \href
  {https://ui.adsabs.harvard.edu/abs/2018ApJ...855...42S} {855, 42}

\bibitem[\protect\citeauthoryear{{Speagle}}{{Speagle}}{2020}]{Speagle2020}
{Speagle} J.~S.,  2020, \mn@doi [\mnras] {10.1093/mnras/staa278}, \href
  {https://ui.adsabs.harvard.edu/abs/2020MNRAS.493.3132S} {493, 3132}

\bibitem[\protect\citeauthoryear{{Speagle}, {Steinhardt}, {Capak}  \&
  {Silverman}}{{Speagle} et~al.}{2014}]{Speagle2014}
{Speagle} J.~S.,  {Steinhardt} C.~L.,  {Capak} P.~L.,   {Silverman} J.~D.,
  2014, \mn@doi [\apjs] {10.1088/0067-0049/214/2/15}, \href
  {https://ui.adsabs.harvard.edu/abs/2014ApJS..214...15S} {214, 15}

\bibitem[\protect\citeauthoryear{{Tacchella}, {Dekel}, {Carollo}, {Ceverino},
  {DeGraf}, {Lapiner}, {Mandelker}  \& {Primack Joel}}{{Tacchella}
  et~al.}{2016}]{Tacchella2016}
{Tacchella} S.,  {Dekel} A.,  {Carollo} C.~M.,  {Ceverino} D.,  {DeGraf} C.,
  {Lapiner} S.,  {Mandelker} N.,   {Primack Joel} R.,  2016, \mn@doi [\mnras]
  {10.1093/mnras/stw131}, \href
  {https://ui.adsabs.harvard.edu/abs/2016MNRAS.457.2790T} {457, 2790}

\bibitem[\protect\citeauthoryear{{Tacchella}, {Bose}, {Conroy}, {Eisenstein}
  \& {Johnson}}{{Tacchella} et~al.}{2018}]{Tacchella2018}
{Tacchella} S.,  {Bose} S.,  {Conroy} C.,  {Eisenstein} D.~J.,   {Johnson}
  B.~D.,  2018, \mn@doi [\apj] {10.3847/1538-4357/aae8e0}, \href
  {https://ui.adsabs.harvard.edu/abs/2018ApJ...868...92T} {868, 92}

\bibitem[\protect\citeauthoryear{{Tacchella}, {Forbes}  \&
  {Caplar}}{{Tacchella} et~al.}{2020}]{TFC2020}
{Tacchella} S.,  {Forbes} J.~C.,   {Caplar} N.,  2020, \mn@doi [\mnras]
  {10.1093/mnras/staa1838}, \href
  {https://ui.adsabs.harvard.edu/abs/2020MNRAS.497..698T} {497, 698}

\bibitem[\protect\citeauthoryear{{Tacchella} et~al.,}{{Tacchella}
  et~al.}{2022a}]{Tacchella2022}
{Tacchella} S.,  et~al., 2022a, \mn@doi [\mnras] {10.1093/mnras/stac818}, \href
  {https://ui.adsabs.harvard.edu/abs/2022MNRAS.513.2904T} {513, 2904}

\bibitem[\protect\citeauthoryear{{Tacchella} et~al.,}{{Tacchella}
  et~al.}{2022b}]{Tacchella2022Halo7D}
{Tacchella} S.,  et~al., 2022b, \mn@doi [\apj] {10.3847/1538-4357/ac449b},
  \href {https://ui.adsabs.harvard.edu/abs/2022ApJ...926..134T} {926, 134}

\bibitem[\protect\citeauthoryear{{Tacchella} et~al.,}{{Tacchella}
  et~al.}{2022c}]{Tacchella2022-bursty-sfh-prior}
{Tacchella} S.,  et~al., 2022c, \mn@doi [\apj] {10.3847/1538-4357/ac4cad},
  \href {https://ui.adsabs.harvard.edu/abs/2022ApJ...927..170T} {927, 170}

\bibitem[\protect\citeauthoryear{{Tacchella} et~al.,}{{Tacchella}
  et~al.}{2023}]{Tacchella2023}
{Tacchella} S.,  et~al., 2023, \mn@doi [\apj] {10.3847/1538-4357/acdbc6}, \href
  {https://ui.adsabs.harvard.edu/abs/2023ApJ...952...74T} {952, 74}

\bibitem[\protect\citeauthoryear{{Utomo}, {Kriek}, {Labb{\'e}}, {Conroy}  \&
  {Fumagalli}}{{Utomo} et~al.}{2014}]{Utomo2014}
{Utomo} D.,  {Kriek} M.,  {Labb{\'e}} I.,  {Conroy} C.,   {Fumagalli} M.,
  2014, \mn@doi [\apjl] {10.1088/2041-8205/783/2/L30}, \href
  {https://ui.adsabs.harvard.edu/abs/2014ApJ...783L..30U} {783, L30}

\bibitem[\protect\citeauthoryear{{Wang} \& {Lilly}}{{Wang} \&
  {Lilly}}{2020a}]{WangLilly2020a}
{Wang} E.,  {Lilly} S.~J.,  2020a, \mn@doi [\apj] {10.3847/1538-4357/ab7b7d},
  \href {https://ui.adsabs.harvard.edu/abs/2020ApJ...892...87W} {892, 87}

\bibitem[\protect\citeauthoryear{{Wang} \& {Lilly}}{{Wang} \&
  {Lilly}}{2020b}]{WangLilly2020}
{Wang} E.,  {Lilly} S.~J.,  2020b, \mn@doi [\apj] {10.3847/1538-4357/ab8b5e},
  \href {https://ui.adsabs.harvard.edu/abs/2020ApJ...895...25W} {895, 25}

\bibitem[\protect\citeauthoryear{{Wang} et~al.,}{{Wang}
  et~al.}{2023}]{Wang2023}
{Wang} B.,  et~al., 2023, \mn@doi [\apjl] {10.3847/2041-8213/acba99}, \href
  {https://ui.adsabs.harvard.edu/abs/2023ApJ...944L..58W} {944, L58}

\bibitem[\protect\citeauthoryear{Waskom}{Waskom}{2021}]{Seaborn2021}
Waskom M.~L.,  2021, \mn@doi [Journal of Open Source Software]
  {10.21105/joss.03021}, 6, 3021

\bibitem[\protect\citeauthoryear{{Weaver} et~al.,}{{Weaver}
  et~al.}{2022}]{COSMOS2020}
{Weaver} J.~R.,  et~al., 2022, \mn@doi [\apjs] {10.3847/1538-4365/ac3078},
  \href {https://ui.adsabs.harvard.edu/abs/2022ApJS..258...11W} {258, 11}

\bibitem[\protect\citeauthoryear{{Whitaker}, {van Dokkum}, {Brammer}  \&
  {Franx}}{{Whitaker} et~al.}{2012}]{Whitaker2012}
{Whitaker} K.~E.,  {van Dokkum} P.~G.,  {Brammer} G.,   {Franx} M.,  2012,
  \mn@doi [\apjl] {10.1088/2041-8205/754/2/L29}, \href
  {https://ui.adsabs.harvard.edu/abs/2012ApJ...754L..29W} {754, L29}

\bibitem[\protect\citeauthoryear{{Whitler}, {Endsley}, {Stark}, {Topping},
  {Chen}  \& {Charlot}}{{Whitler} et~al.}{2023a}]{Whitler2023b}
{Whitler} L.,  {Endsley} R.,  {Stark} D.~P.,  {Topping} M.,  {Chen} Z.,
  {Charlot} S.,  2023a, \mn@doi [\mnras] {10.1093/mnras/stac3535}, \href
  {https://ui.adsabs.harvard.edu/abs/2023MNRAS.519..157W} {519, 157}

\bibitem[\protect\citeauthoryear{{Whitler}, {Stark}, {Endsley}, {Leja},
  {Charlot}  \& {Chevallard}}{{Whitler} et~al.}{2023b}]{Whitler2023}
{Whitler} L.,  {Stark} D.~P.,  {Endsley} R.,  {Leja} J.,  {Charlot} S.,
  {Chevallard} J.,  2023b, \mn@doi [\mnras] {10.1093/mnras/stad004}, \href
  {https://ui.adsabs.harvard.edu/abs/2023MNRAS.519.5859W} {519, 5859}

\bibitem[\protect\citeauthoryear{{Williams} et~al.,}{{Williams}
  et~al.}{2023}]{JEMS2023}
{Williams} C.~C.,  et~al., 2023, \mn@doi [arXiv e-prints]
  {10.48550/arXiv.2301.09780}, \href
  {https://ui.adsabs.harvard.edu/abs/2023arXiv230109780W} {p. arXiv:2301.09780}

\bibitem[\protect\citeauthoryear{{Worthey} \& {Ottaviani}}{{Worthey} \&
  {Ottaviani}}{1997}]{Worthey1997}
{Worthey} G.,  {Ottaviani} D.~L.,  1997, \mn@doi [\apjs] {10.1086/313021},
  \href {https://ui.adsabs.harvard.edu/abs/1997ApJS..111..377W} {111, 377}

\bibitem[\protect\citeauthoryear{{Yu}, {Ho}  \& {Wang}}{{Yu}
  et~al.}{2021}]{Yu2021}
{Yu} S.-Y.,  {Ho} L.~C.,   {Wang} J.,  2021, \mn@doi [\apj]
  {10.3847/1538-4357/ac0c77}, \href
  {https://ui.adsabs.harvard.edu/abs/2021ApJ...917...88Y} {917, 88}

\bibitem[\protect\citeauthoryear{{van der Wel} et~al.,}{{van der Wel}
  et~al.}{2021}]{LEGAC2021}
{van der Wel} A.,  et~al., 2021, \mn@doi [\apjs] {10.3847/1538-4365/ac1356},
  \href {https://ui.adsabs.harvard.edu/abs/2021ApJS..256...44V} {256, 44}

\makeatother
\end{thebibliography}



\appendix

\section{Testing the stochastic SFH prior on quiescent galaxies}
\label{sec:quiescent recovery}

The systems in both the intermediate- and high-redshift mock galaxy samples are star-forming. As such, we also create a sample of 100 mock quiescent galaxies in order to test the performance of the stochastic SFH model on a broad range of SFHs. 

\subsection{Mock observations}
\label{sec:quiescent sample}

We take a relatively simple approach to generating the quiescent galaxy sample. We select a representative quiescent galaxy from the LEGA-C catalog (ID 28041 \footnote{LEGA-C galaxy 28041 has a spectroscopic redshift of $z = 0.8088$, and its measured spectrum has an average S/N of $27.5 \Angstrom^{-1}$.}) to serve as the template. We fit the galaxy in {\typewriter Prospector} using the \textit{continuity} prior and the various other model components described in Section \ref{sec:model_and_priors}. The posterior distributions of the log SFR ratios from the resulting fit are used to generate 100 SFHs representative of quiescent galaxies. Because we expect quiescent systems to, on average, be more massive than star-forming galaxies, we set the stellar masses for our mock quiescent galaxies by sampled from a uniform distribution in log $\Mstar/\Msun$ between $11 - 12$. The remaining galaxy parameters (e.g. metallicity, dust parameters, stellar velocity dispersion) are drawn from the priors listed in Table \ref{tab:priors}. 

As with the mock samples in Sections \ref{sec:legac recovery} and \ref{sec:highz recovery}, the mock quiescent SFHs and stellar population parameters are passed through the {\typewriter predict()} function in {\typewriter Prospector} to generate the associated photometry and spectra. The noise of every spectral and photometric point for each mock galaxy is randomly drawn from a normal distribution with $\mu = 0$ and $\sigma = $ the uncertainty at each corresponding point in the template data. We then add the noise values to the intrinsic fluxes from {\typewriter Prospector}.

\subsection{Results}

We investigate how well our stochastic SFH model is able to describe quiescent galaxies, in addition to the (primarily) star-forming galaxies of the intermediate and high-redshift mock samples. Our goal is to demonstrate that the stochastic SFH prior is suitable for general use, independent of whether the systems of interest are star-forming or not.

\begin{figure}
    \centering
    \includegraphics[width=0.48\textwidth]{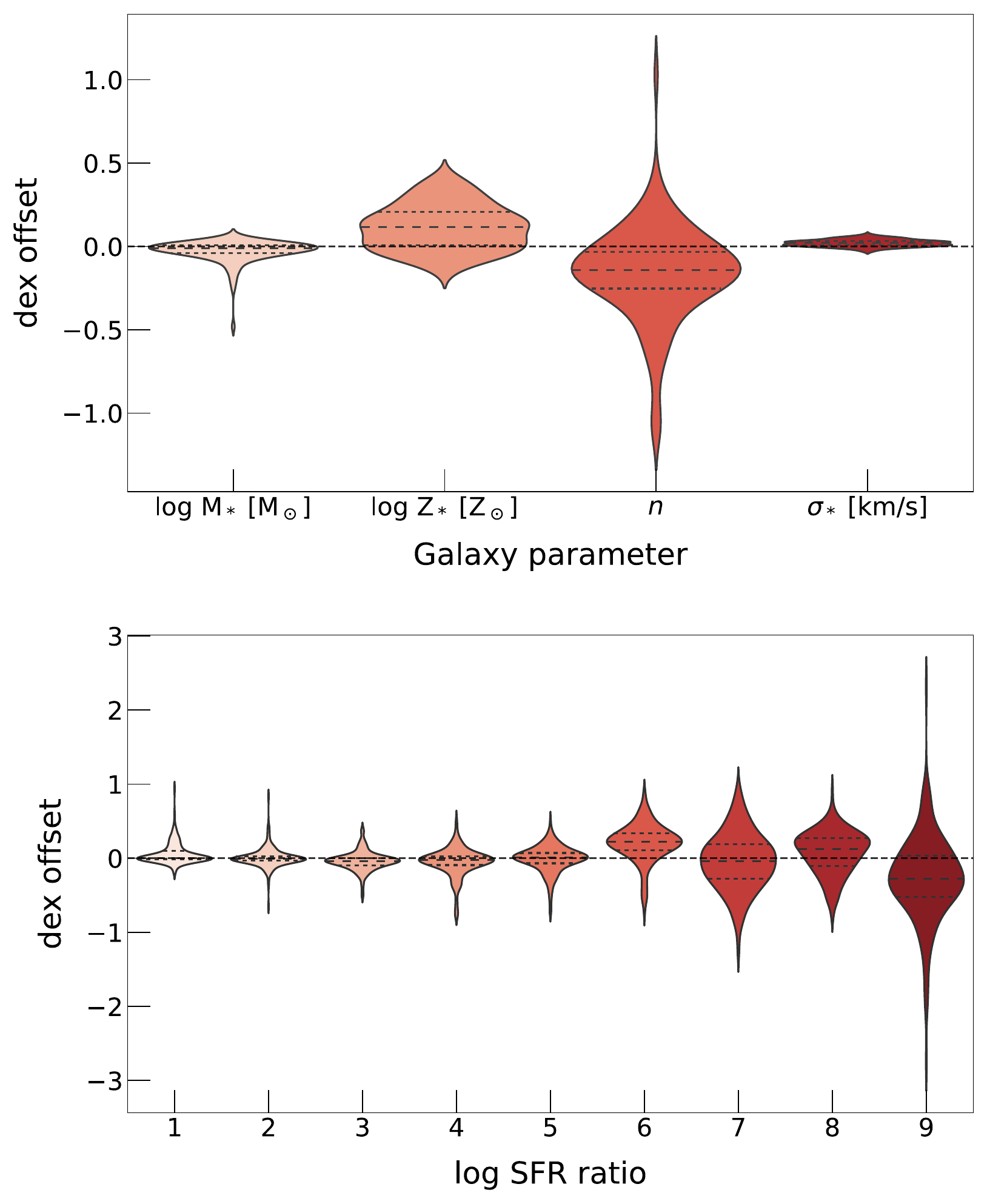}
    \caption{Histograms showing the dex offset between the median posterior recovered galaxy parameters (top) and log SFR ratios (bottom) and their true values for 100 mock quiescent galaxies. The stellar masses (log $\Mstar$) and stellar velocity dispersions ($\sigma_*$) are recovered with minimal systematic offset and reasonable scatter. However, the the stellar metallicities (log Z$_*$) are biased high and the diffuse dust optical depth  ($\hat{\tau}_{\rm{dust, 2}}$) are biased low. There is also significant scatter in the recovered dust indices. The log SFR ratios are generally well-recovered, but ratio 6 is systematically over-estimated and ratio 9 is under-estimated.}
    \label{fig:quiescent params}
\end{figure}

The top panel of Figure \ref{fig:quiescent params} shows how well the basic parameters of the mock quiescent galaxies described in Section \ref{sec:quiescent sample} are recovered by {\typewriter Prospector} using the stochastic SFH model. The stellar mass is well-recovered, with no significant bias and a scatter of 0.06 dex. The stellar velocity dispersion is also constrained well, with both a systematic offset and scatter of $\sim 0.02$ dex. On the other hand, the metallicity and dust index are less well-recovered. The metallicity values recovered by the stochastic model are biased high by $\sim 0.14$ dex and have a scatter of 0.14 dex. The dust index values are biased low by 0.14 with a scatter of 0.28. 

The bottom panel of Figure \ref{fig:quiescent params} shows the recovery efficacy for the log SFR ratios of the quiescent mock sample. The majority of the SFR ratios are recovered with minimal bias (systematic offsets of $\leq 0.07$ dex) and reasonable dispersion. We find that the scatter in the difference between model and input log SFR ratio values tends to increase with lookback time (i.e. log SFR ratio 1 has a scatter of 0.14 dex and log SFR ratio 9 has a scatter of 0.57 dex). However, we do find that the recovered log SFR ratio 6 values are systematically biased high by $\sim 0.20$ dex and the ratio 9 values are biased low by $\sim 0.28$ dex.

\begin{figure}
    \centering
    \includegraphics[width=0.48\textwidth]{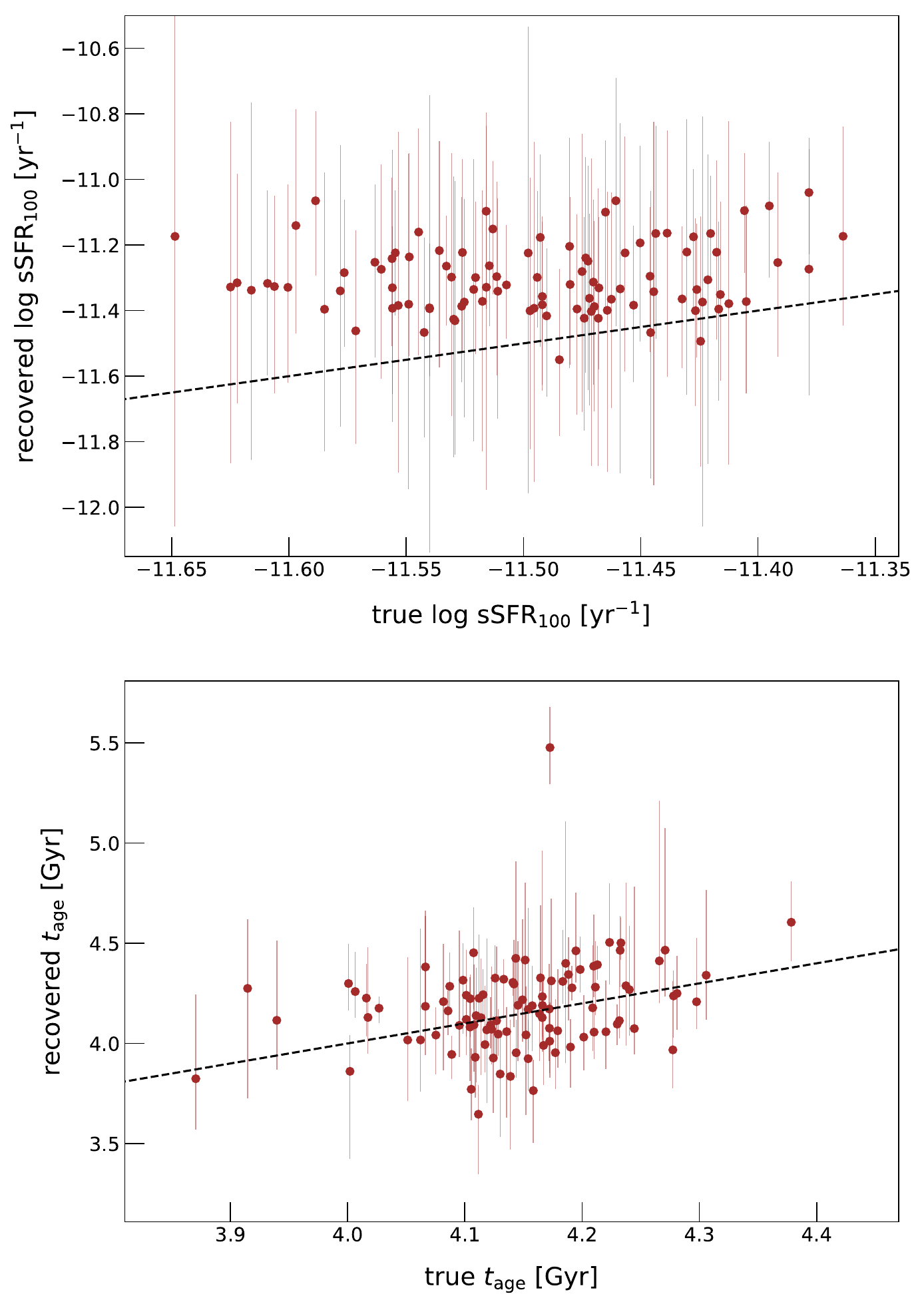}
    \caption{Recovered vs. true recent sSFR (log dSFR$_{100}$; top panel) and mass-weighted age ($t_{\rm{age}}$; bottom panel) for the mock sample of 100 quiescent galaxies. log sSFR$_{100}$ hits a floor at $\sim$-11.4~$\mathrm{yr}^{-1}$, although the true sSFR values do lie within the 1$\sigma$ errors of the recovered values. $t_{\rm{age}}$ also tends to be over-estimated in these galaxies. The perfect, one-to-one recovery scenario is denoted with a black dashed line.}
    \label{fig:quiescent sfr_mwa}
\end{figure}

We also show the ability for the stochastic model to recover the recent specific star-formation rate (log sSFR$_{100}$) and mass-weighted age ($t_{\rm{age}}$) of mock quiescent galaxies in Figure \ref{fig:quiescent sfr_mwa}. The model $t_{\rm{age}}$ values are generally consistent with the input values. The offsets between the fit and input $t_{\rm{age}}$ values has a $\sim0.22$ Gyr scatter and no significant bias. The recovered log sSFR$_{100}$ values are also consistent with the truth to within $1\sigma$. However, log SFR$_{100}$ is systematically over-estimated in these systems (biased high by $\sim0.19$ dex) and seems to hit a floor around a value of $-11.4~\mathrm{yr}^{-1}$, suggesting that the stochastic SFH prior struggles to reproduce low ongoing SFR activity in galaxies.

\begin{figure}
    \centering
    \includegraphics[width=0.48\textwidth]{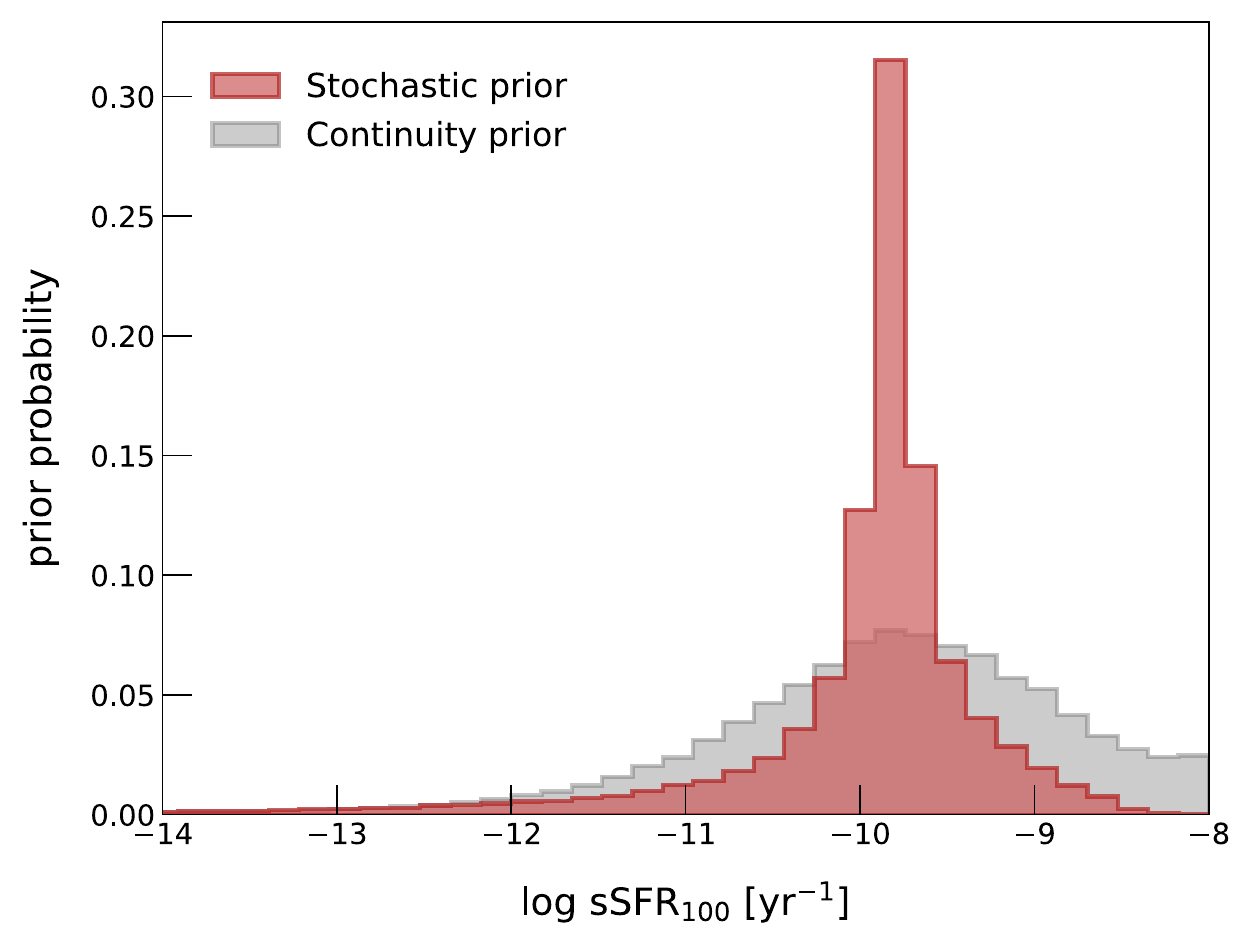}
    \caption{Histograms of the specific star-formation rate (sSFR = SFR/$\Mstar$) obtained from 100,000 draws of the prior distribution (with fixed $z = 0.8$) for the stochastic SFH model, which was used to fit the mock galaxies, and the continuity SFH model for comparison. While the stochastic prior has a long tail towards low sSFRs like the continuity prior, a smaller fraction of its prior probability exists at these low values, making it more difficult to obtain accurate constraints of quiescent galaxies' SFRs.}
    \label{fig:ssfr prior stoch vs cont}
\end{figure}

To investigate why this limit exists, we examine the sSFR prior distribution. We take 100,000 draws from the stochastic model prior distributions (with redshift fixed to $z = 0.8$) and calculate $\mathrm{sSFR = SFR/\Mstar}$ for each prior draw. The histogram of the resulting sSFR distribution is shown in red in Figure \ref{fig:ssfr prior stoch vs cont}. 
We do the same for the continuity prior and display the distribution in gray. 

We can see that the stochastic prior gives way to a rather tight sSFR distribution, with the 16th$-$84th percentiles spanning $10^{-10.4} - 10^{-9.6}~\mathrm{yr}^{-1}$. The distribution does have a long tail towards low sSFR values, similar to what is seen in the continuity prior, demonstrating that a non-negligible fraction of the stochastic model's prior probability exists these low sSFR values. However, the prior probability assigned to these low sSFRs is less than that of the continuity prior. This is likely the reason why the true sSFR values lie within the 1$\sigma$ errors of the recovered values, despite the median values being biased high. The model does explore the space of very low ongoing SFRs and cannot rule out such a solution, but because the prior probability associated with that space is small, it is much more difficult to obtain accurate constraints. So although the stochastic model should, in principle, be able to model very low ongoing star-formation activity, it struggles to do so in practice. Thus, this serves as a caution: when using the stochastic SFH model, any sSFR measured to be below $\sim 10^{-11.4}~\mathrm{yr}^{-1}$ should be treated as an upper limit rather than an absolute value.

The inaccurate recovery of low SFRs by the stochastic SFH model likely also plays a role in the biases we see in the recovery of the stellar metallicity and dust parameters. Because the effects that SFH, metallicity, and dust attenuation have on a galaxy SED are often degenerate, an inaccurate SFH will naturally lead to less accurate values for the metallicity and dust parameters as well. This challenge, however, is not unique to our SFH model. Across the board, SED-fitting models struggle to reproduce the low SFRs found in massive, quiescent galaxies \citep[e.g.][]{Fumagalli2014, Utomo2014} -- it is inherently difficult to distinguish the effect that a change in SFR from, e.g. 1 M$_\odot$ yr$^{-1}$ to 0.1 M$_\odot$ yr$^{-1}$ has on a galaxy's spectrum from noise, especially in lower S/N data. Therefore, one should always be careful when making claims about the SFRs of quiescent galaxies.

\section{Combining PSD parameter posterior distributions}
\label{sec:combining_posteriors}

In Section \ref{sec:recovery PSD}, the posterior distributions presented for the PSD parameter recovery test (Figure \ref{fig:psd params}) were obtained by merging the individual {\typewriter dynesty} runs of each mock galaxy (i.e. effectively adding together their prior volumes and associated likelihoods). Using this method, $\sigreg$ was recovered to within $1\sigma$ of its true value across all four mock regimes. The remaining three PSD parameters -- $\taueq$, $\sigdyn$, and $\taudyn$ -- had posterior distributions which were broad, unconstraining, and strongly prior-dominated. The caveat here is, because the likelihood function of every mock galaxy is unique to that galaxy, this method of combining their posteriors is not entirely correct. 

Consider we have $N$ independent sets of information $d_1, \dots, d_N$ (e.g. galaxy SEDs) to assess some parameter $\theta$ (e.g. each of the PSD parameters). We also have a model to estimate the individual posterior probabilities $P(\theta | d_1), \dots, P(\theta | d_N)$. Then, the combined posterior distribution on $\theta$ obtained from the independent sets of data $d_1,\dots,d_N$ is
\begin{equation}
    P(\theta | d_1,\dots,d_N) = \alpha \frac{\prod^{n=N}_{n=1} P(\theta | d_n)}{P(\theta)^{N-1}},
    \label{eq:probcomb}
\end{equation}
where $P(\theta)$ is the prior on $\theta$ and $\alpha$ is the normalization constant. In other words, instead of taking the weighted sum of the posterior distributions of our mock galaxies, we should really be taking the product.

\begin{figure}
    \centering
    \begin{subfigure}[b]{0.48\textwidth}
        \centering
        \includegraphics[width=\textwidth]{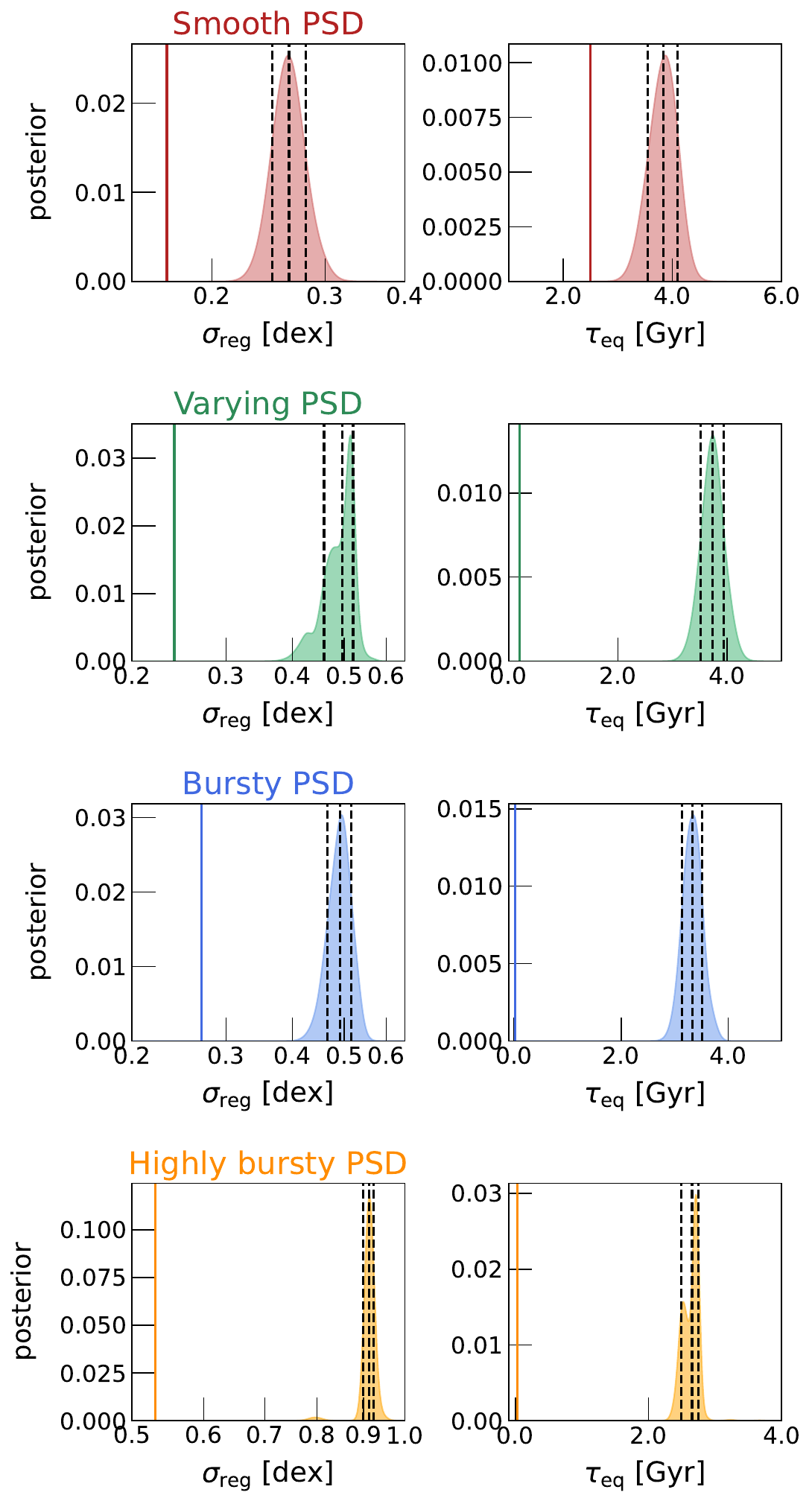}
    \end{subfigure}
    \caption{The combined posterior distributions of $\sigreg$ and $\taueq$ for each mock galaxy regime in Section \ref{sec:legac recovery} calculated according to Equation \ref{eq:probcomb}. The dashed black lines represent the [16, 50, 84]-th percentiles of the posteriors, and the solid colored line denotes the ``truth'', or input value. We find that in all cases, the combined posteriors are $\gtrsim10\times$ narrower than those presented in Figure \ref{fig:psd params}, and the true $\sigreg$ and $\taueq$ values are $\gtrsim2\sigma$ outside of their median estimated values.}
    \label{fig:correct combined psd params}
\end{figure}

We combine the $\sigreg$ and $\taueq$ posteriors of our mock $z$ = 0.7 galaxies according Equation \ref{eq:probcomb}, using the {\typewriter scipy.stats.gaussian\_kde} function to estimate their probability density functions. The result is shown in Figure \ref{fig:correct combined psd params}. We notice immediately that the posterior distributions obtained are significantly more sharply-peaked than those derived via the weighted sum method. In fact, the $1\sigma$ widths of both the $\sigreg$ and $\taudyn$ posterior distributions in Figure \ref{fig:correct combined psd params} are $\gtrsim10\times$ narrower than the distributions in Figure \ref{fig:psd params}. Additionally, while the true $\sigreg$ values across all four mock galaxy regimes were previously recovered within $1\sigma$ of their posterior distributions, they are all now well outside of the $2-3\sigma$ limit. The trend in $\sigreg$ across the four PSD regimes, however, is still reproduced (i.e. we still recover the fact that $\sigreg$ is smallest in the smooth PSD regime, middling in the varying and bursty regimes, and largest in the highly bursty regime).

The reason that $\sigreg$ seems to be well-recovered and the $\taueq$ posteriors mimic their priors in Section \ref{sec:recovery PSD}, whereas here, both the $\sigreg$ and $\taueq$ posteriors are sharply peaked around the wrong values is largely due to our sampling method. {\typewriter dynesty} systematically under-samples the edges of the prior distributions. This effect becomes magnified when we combine our posteriors, as we are exponentially suppressing the wings by taking the product of 100 of these distributions for each mock galaxy regime.

\begin{figure}
    \centering
    \begin{subfigure}[b]{0.4\textwidth}
        \centering
        \includegraphics[width=\textwidth]{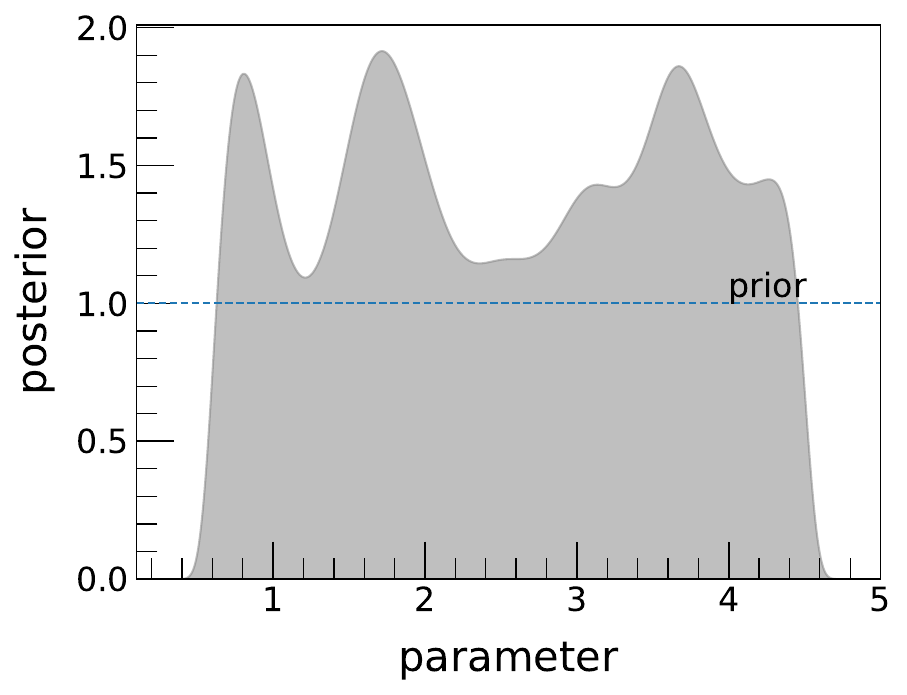}
    \end{subfigure}
    \caption{Combined posterior distribution of 100 {\typewriter dynesty} runs sampling from a uniform prior between 0 and 5. The prior is marked by the dashed blue line. We see that the edges of the prior are, in fact, under-sampled, causing the wings of the posterior distribution to be highly suppressed, similar to the effect afflicting the combined PSD parameter posteriors in Figure \ref{fig:correct combined psd params}.}
    \label{fig:example dynesty}
\end{figure}

We test that such under-sampling is occurring by independently sampling from a uniform distribution between 0 and 5 a hundred times using {\typewriter dynesty}, and then combining the results according to Equation \ref{eq:probcomb}, the outcome of which is displayed in Figure \ref{fig:example dynesty}.  We see that the edges of the prior are, in fact, under-sampled, causing the wings of the combined posterior to be highly suppressed. Because of this, for the time being, we merge the individual {\typewriter dynesty} runs of each mock galaxy to obtain the combined posterior distributions for each PSD parameter, rather than take their prior-weighted product. However, this issue implicates the need for both Bayesian hierarchical modelling and a better sampling method to improve this SFH modelling method in the future.

\bsp	
\label{lastpage}
\end{document}